\documentclass[a4paper]{article}
\newlength{\mpwidth}
\usepackage{graphicx}
\usepackage{multirow}
\usepackage{rotating}
\usepackage{amsthm}

\newtheorem{Definition}{Definition}

\begin{document}
\title{Validation of a PETSc based software implementing a 4DVAR Data Assimilation algorithm: a case study related with an Oceanic Model based on Shallow Water equation}
\author{Luisa Carracciuolo, Emil M. Constantinescu, Luisa D'Amore}
\maketitle

\section{Definition of the DA problem and description of the PETSc based software implementation}
The considered software intends to solve the problem defined on a suitable domain decomposition $$DD(\Delta\times \Omega)=\left\{\Delta_{j}\times \Omega_{h}\right\}_{j;h}$$ of 
time-space domain $\Delta\times \Omega$ as described in Definition \ref{DADef} (all the needed notations can be founded in \cite{DD-DA-Journal}).

\begin{Definition}[The 4D-VAR DA problem defined on the domain decomposition $DD(\Delta\times \Omega)$ - the 4D-VAR DD-DA problem]
\label{DADef}
%\begin{Definition}[Local 4D-VAR regularization functional]
The 4D Variational DD-DA problem consists in computing the vector ${\mathbf{\tilde{x}}}_{jh}^{DA}$ such that
\begin{eqnarray}
{\mathbf{\tilde{x}}}^{DA} & = & \sum_{j;h}{EO_{jh}\left({\mathbf{x}}_{jh}^{DA}\right)} 
\end{eqnarray}

\noindent{where}

\begin{eqnarray}
{\mathbf{x}}_{jh}^{DA} & = & argmin_{\mathbf{x}_{{jh}}}{J}_{jh}(\mathbf{u}_{{jh}}),\label{solloc}
\end{eqnarray}

\noindent{where the operator ${J}_{jh}$ (the local 4D-VAR regularization functional) is defined as follows}:

\begin{eqnarray}
J_{jh}(\mathbf{x}_{{jh}}) & = & RO_{jh}[J]+\mu \ {O}_{(ih)(jk)}[\mathbf{x}_{{(ih)(jk)}}] 
\end{eqnarray}

\noindent {and where $\mathcal{O}_{(ih)(jk)}$ is a suitably defined operator on the overlapped domain $\Delta_{jk}\times \Omega_{ih}$}.
Parameter $\mu$ is a regularization parameter. The 4D-VAR regularization functional $J$ is defined as:

\begin{equation}\label{min3sec}
 J(\mathbf{x})= \| \mathbf{x}- \mathbf{x}^{b}\|_{{\bf B}^{-1}}^{2}    + \lambda \sum_{k=0}^{nt_{obs}-1} \| \mathcal{H}_{t_k} ( \mathcal{M}_{t_0\rightarrow t_k}[  \mathbf{x}])-\mathbf{v}_k \|_{{\bf R}_k^{-1}}^{2}
\end{equation}

where  $\lambda$ is a regularization parameter,  ${\bf B}$ and ${\bf R}_k$ ($\forall k= 0,\dots, nt_{obs}-1$) are the covariance matrices of the errors on the 
background and the observations respectively, while  $\| \cdot \|_{{\bf B}^{-1}}$ and $\| \cdot \|_{{\bf R}_k^{-1}}$ denote the weighted euclidean norm. 
% \flushright{$\spadesuit$}
\end{Definition}

The 4D-VAR DD-DA problem solution is computed performing the following steps 
on each subdomains $\Delta_{j}\times \Omega_{h}$ (the so called {\em 4D-VAR DD-DA algorithm}): %, the software executes the following steps:
\begin{itemize}
\item {\bf Locally} compute all the parameters that define the local 4D-VAR regularization functional ${J}_{jh}$
\item {\bf Locally} compute the minimum ${\mathbf{x}}_{jh}^{DA}$ (needed values for overlapping regions are obtained when necessary - 
      i.e., for the model local evolution)
\item {\bf Globally} contribute to computation of ${\mathbf{x}}_{jh}^{DA}$
\end{itemize}

In order to compute the minimum of all the functionals $J_{jh}$, 
the DD-DA algorithm 
has to face with some issues.
In more details, we  have to address:
\begin{itemize}
\item the linearization of the operator $\mathcal{M}_{t-\Delta t\to t}$, let us say $\mathbf{M}_{t-\Delta t}$,  used
for the evaluations of $J_{jh}$ required by the minimisation algorithm;
\item the evaluation of the adjoint operator of $\mathbf{M}_{t-\Delta t}$, let us say ${\bf M}_{t-\Delta t}^{\star}$, used
for the evaluation of $\nabla J_{jh}$ required by the minimisation algorithm;
%\item[(d)] the computation of the preconditioner $\widetilde{RO_{jh}\left[B\right]}$ of the error covariance matrice $RO_{jh}\left[B\right]$: 
%in the proposed approach we choose to define $\widetilde{RO_{jh}\left[B\right]}$ as a Truncated SVD of $RO_{jh}\left[B\right]$;
\end{itemize}

Both the points above should require the computation of the discretization of the Jacobian of $\mathcal{M}_{t-\Delta t\rightarrow t}$
$$
\nabla \mathbf{M}_{t-\Delta t\rightarrow t}
$$

\begin{figure}
\centerline{\fbox{\includegraphics[scale=0.50]{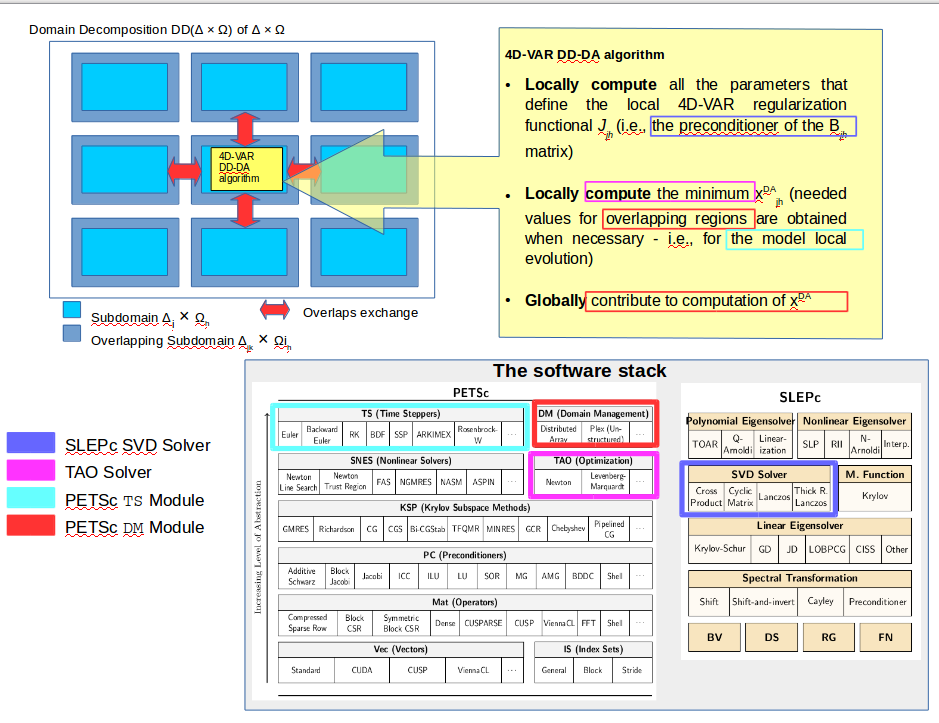}}}
\caption{Representation of the Software stack and algorithm implementation} \label{SoftStackPlusAlgo}
\end{figure}

Following some details about software implementation in  PETSc (Portable, Extensible Toolkit for Scientific Computation)\cite{PETScManual} environment.
To implement the entire algorithm we plan to use:

\begin{enumerate}
\item the PETSc time steppers {\tt TS} module for solving time-dependent (nonlinear) PDEs, including the computation of adjoint;
\item the PETSc {\tt DM} module wich is a powerfull tool for the managment of all mesh data related with domain decomposition;
\item
The TAO software library \cite{TAOLibrary} for the computation of (\ref{solloc}). The Toolkit for Advanced Optimization (TAO) is aimed at the solution of large-scale optimization problems on high-performance architectures.  TAO is suitable for both single-processor and
massively-parallel architectures. The current version of TAO has algorithms for unconstrained and bound-constrained optimization.
\item
The SLEPc software library \cite{SLEPcLibrary} for the computation of spectral decomposition usefull to compute a preconditioner
of the error covariance matrices (i.e., see approach used in \cite{DD-DA-Journal}). The Scalable Library for Eigenvalue Problem Computations
(SLEPc) is a software library for the solution of large scale sparse eigenvalue problems on parallel computers.
It can also be used for computing a partial SVD of a large, sparse, rectangular matrix, and to solve nonlinear eigenvalue problems.
\end{enumerate}
All  the abobe mentioned software are integrated or based on PETSc (see figure \ref{SoftStackPlusAlgo} for a representation of the Software stack 
and algorithm implementation).

To represent the Jacobian of ${\mathcal{M}}_{t-\Delta t  \rightarrow t}$ in PETSc we decided to follow a {``\em matrix-free approach''}:
we used a  {\tt MATHSHELL} type for PETSc {\tt Mat} object  to represent $\nabla \mathbf{M}_{t-\Delta t\rightarrow t}$  just defining its way of operating

%\begin{figure}
%\centerline{\fbox{\includegraphics[scale=1.00]{IMMAGINI/{ECEx-InitialGuess.bin.SMALL}.png}}}
%\caption{Initial guess $x_0$ for the model} \label{InitialGuess}
%\end{figure}

\section{Case study description}

The case study is based on the Shallow Water Equations (SWEs) on the sphere.
The SWE have been used extensively as a simple model of the atmosphere or ocean circulation since they
contain the essential wave propagation mechanisms found in general circulation models \cite{ShallowWater}. The SWEs in spherical coordinates are:

\begin{eqnarray}
\frac{\partial u}{\partial t} & = & - \frac{1}{a \cos{\theta}} \left( u \frac{\partial u}{\partial \lambda} + v \cos{\theta}\frac{\partial u}{\partial\theta}\right)
                                    + \left( f + \frac{u\tan{\theta}}{a}\right) v
                                    - \frac{g}{a \cos{\theta}} \frac{\partial h}{\partial \lambda} \label{EquU}\\
\frac{\partial v}{\partial t} & = & - \frac{1}{a \cos{\theta}} \left( u \frac{\partial v}{\partial \lambda} + v \cos{\theta}\frac{\partial v}{\partial\theta}\right)
                                    + \left( f + \frac{u\tan{\theta}}{a}\right) u
                                    - \frac{g}{a} \frac{\partial h}{\partial \theta} \label{EquV}\\
\frac{\partial h}{\partial t} & = & - \frac{1}{a \cos{\theta}} \left(\frac{\partial\left(hu\right)}{\partial \lambda} + \frac{\partial\left(hu\cos{\theta}\right)}{\partial \theta}\right) \label{EquH}
\end{eqnarray}

Here $f$ is the Coriolis parameter given by $f = 2\Omega\sin{\theta}$, where $\Omega$ is the angular speed of the rotation of the Earth,
$h$ is the height of the homogeneous atmosphere (or of the free ocean surface), $u$ and $v$ are the zonal and meridional wind (or the ocean velocity)
components, respectively, $\theta$ and $\lambda$ are the latitudinal and longitudinal directions, respectively, $a$ is the
radius of the earth and $g$ is the gravitational constant.

We  express the system of equations (\ref{EquU})-(\ref{EquH}) using a compact form, i.e.:

\begin{equation}
\frac{\partial{\mathbf{Z} }}{\partial t} = \mathcal{M}_{t-\Delta t\rightarrow t}\left({\mathbf{Z}} \right) \label{ShallowWaterEqu}
\end{equation}

where
\begin{equation}
{\mathbf{Z}}=\left(
\begin{array}{c}        
       u \\ v \\ h
\end{array}     
  \right)
\end{equation}
and

\begin{eqnarray}
\mathcal{M}_{t-\Delta t\rightarrow t}\left({\mathbf{Z}} \right) & = & \left(
                        \begin{array}{l}
- \frac{1}{a \cos{\theta}} \left( u \frac{\partial u}{\partial \lambda} + v \cos{\theta}\frac{\partial u}{\partial\theta}\right)
                                    + \left( f + \frac{u\tan{\theta}}{a}\right) v
                                    - \frac{g}{a \cos{\theta}} \frac{\partial h}{\partial \lambda} \\
- \frac{1}{a \cos{\theta}} \left( u \frac{\partial v}{\partial \lambda} + v \cos{\theta}\frac{\partial v}{\partial\theta}\right)
                                    + \left( f + \frac{u\tan{\theta}}{a}\right) u
                                    - \frac{g}{a} \frac{\partial h}{\partial \theta} \\
- \frac{1}{a \cos{\theta}} \left(\frac{\partial\left(hu\right)}{\partial \lambda} + \frac{\partial\left(hu\cos{\theta}\right)}{\partial \theta}\right)
                        \end{array}
                         \right) \nonumber \\
                      & = & \left(
                       \begin{array}{l}
                             F_1 \\
                             F_2 \\
                             F_3
                       \end{array}
                         \right)
\end{eqnarray}

We discretize (\ref{ShallowWaterEqu}) just in space using an un-staggered Turkel-Zwas scheme \cite{Turkel-Zwas1,Turkel-Zwas2}, and
we obtain:
\begin{equation}
\frac{\partial{\mathbf{Z}_{disc}}}{\partial t} = \mathcal{M}^{t-\Delta t\rightarrow t}_{disc}\left({\mathbf{Z}_{disc}} \right) \label{ShallowWaterSemiDiscrete}
\end{equation}
where
\begin{equation}
{ \mathbf{Z}_{disc}}=\left(
\begin{array}{c}        
       \left(u_{i,j}\right)_{i=0,\ldots,nlon-1;j=0,\ldots,nlat-1} \\
       \left(v_{i,j}\right)_{i=0,\ldots,nlon-1;j=0,\ldots,nlat-1} \\
       \left(h_{i,j}\right)_{i=0,\ldots,nlon-1;j=0,\ldots,nlat-1} \\
\end{array}     
  \right)
\end{equation}

and

\begin{equation}
\mathcal{M}^{t-\Delta t\rightarrow t}_{disc}\left({\mathbf{Z}_{disc}} \right) = \left(
                        \begin{array}{l}
       \left(U_{i,j}\right)_{i=0,\ldots,nlon-1;j=0,\ldots,nlat-1} \\
       \left(V_{i,j}\right)_{i=0,\ldots,nlon-1;j=0,\ldots,nlat-1} \\
       \left(H_{i,j}\right)_{i=0,\ldots,nlon-1;j=0,\ldots,nlat-1} \\
                        \end{array}
                         \right)
\end{equation}

\begin{eqnarray*}
U_{i,j}  & = & - \sigma_{lon} \frac{u_{i,j}}{\cos{\theta_j}} \left( u_{i+1,j} - u_{i-1,j}\right)  \\
         &   & - \sigma_{lat} \ {v_{i,j}}                \left( u_{i,j+1} - u_{i,j-1}\right) \\
         &   & - \sigma_{lon} \frac{g}{p \cos{\theta_j}} \left( h_{i+p,j} - h_{i-p,j}\right) \\
         &   & + 2 \left[ \left(1-\alpha\right) \left( 2\Omega\sin{\theta_j}+ \frac{u_{i,j}}{a}\tan\theta_j\right)v_{i,j}     \right. \\
         &   & +   \left. \frac{\alpha}{2}      \left( 2\Omega\sin{\theta_j}+ \frac{u_{i+p,j}}{a}\tan\theta_j\right)v_{i+p,j} \right. \\
         &   & +   \left. \frac{\alpha}{2}      \left( 2\Omega\sin{\theta_j}+ \frac{u_{i-p,j}}{a}\tan\theta_j\right)v_{i-p,j} \right] \\
V_{i,j}  & = & - \sigma_{lon} \frac{u_{i,j}}{\cos{\theta_j}} \left( v_{i+1,j} - v_{i-1,j}\right)  \\
         &   & - \sigma_{lat} \ {v_{i,j}} \left( u_{i,j+1} - u_{i,j-1}\right) \\
         &   & - \sigma_{lat} \frac{g}{q} \left( h_{i,j+q} - h_{i,j-q}\right) \\
         &   & - 2 \left[ \left(1-\alpha\right) \left( 2\Omega\sin{\theta_j}+ \frac{u_{i,j}}{a}\tan\theta_j\right)u_{i,j}             \right. \\
         &   & +   \left. \frac{\alpha}{2}      \left( 2\Omega\sin{\theta_{j+q}}+ \frac{u_{i,j+q}}{a}\tan\theta_{j+q}\right)u_{i,j+q} \right. \\
         &   & +   \left. \frac{\alpha}{2}      \left( 2\Omega\sin{\theta_{j-q}}+ \frac{u_{i,j-q}}{a}\tan\theta_{j-q}\right)u_{i,j-q} \right] \\
H_{i,j}  & = & - \alpha \left\{ \frac{u_{i,j}}{\cos{\theta_j}}  \left( h_{i+1,j} - h_{i-1,j}\right) \right.\\
         &   & + \ {v_{i,j}} \left( h_{i,j+1} - h_{i,j-1}\right) \\
         &   & + \frac{h_{i,j}}{\cos{\theta_j}} \left[ \left(1-\alpha\right) \left( u_{i+p,j} - u_{i-p,j}\right)  \right. \\
         &   & + \left. \frac{\alpha}{2} \left(u_{i+p,j+q} - u_{i-p,j+q} + u_{i+p,j-q} - u_{i-p,j-q}\right)\right]\frac{1}{p} \\
         &   & + \left[ \left(1-\alpha\right) \left( v_{i,j+q}\cos{\theta_{j+q}} - v_{i,j-q}\cos{\theta_{j-q}}\right) \right. \\
         &   & + \frac{\alpha}{2} \left( v_{i+p,j+q}\cos{\theta_{j+q}} - v_{i+p,j-q}\cos{\theta_{j-q}} \right)\\
         &   &\left.\left.+\frac{\alpha}{2} \left( v_{i-p,j+q}\cos{\theta_{j+q}} - v_{i-p,j-q}\cos{\theta_{j-q}} \right)\right]\frac{1}{q} \right\}
\end{eqnarray*}

%Here $\sigma_{lat}=\frac{1}{a\Delta \theta}$ and $\sigma_{lon}=\frac{1}{a\Delta\lambda}$; 
%$p$ and $q$ are parameters of the coarse mesh ratio to the fine
%mesh and $\alpha$ is a real value such that $0\le \alpha < 1$ where

%\begin{eqnarray*}
%\Delta \theta  & = & \frac{\pi}{nlat-1} \\
%\Delta \lambda & = &\frac{2\pi}{nlon-1}
%\end{eqnarray*}
%and
%\begin{eqnarray*}
%\theta_j  & = & j\Delta\theta,\ j=0,\ldots,nlat-1 \\
%\lambda_i & = & i\Delta\lambda,\ i=0,\ldots,nlon-1
%\end{eqnarray*}

If we define $\mathbf{M}_{\Delta t}^{0,M_{steps}}\left(\cdot\right)$ as follows:

\begin{equation}
\mathbf{M}_{\Delta t}^{0,M_{steps}}\left(\cdot\right) = \underbrace{\mathcal{M}^{t-\Delta t\rightarrow t}_{disc}\left(\mathcal{M}^{t-\Delta t\rightarrow t}_{disc}\left( \cdots \mathcal{M}^{t-\Delta t\rightarrow t}_{disc}\left({\cdot} \right) \right)\right)}_{M_{steps}},
\end{equation}

we note that the symbol $\mathbf{M}_{\Delta t}^{0,M_{steps}}\left(\cdot\right)$  represents the model $\mathcal{M}^{t-\Delta t\rightarrow t}_{disc}\left(\cdot\right)$ 
{\em ``applied''} $M_{steps}$ times.

We also note that the numerical model is 
defined by the following parameters:
\begin{itemize}
\item $\Delta t$ discretization step in time domain, 
\item $\Delta \lambda$, $\Delta \theta$ discretization step in space domain,
\item $\alpha$ parameter of the Turkel-Zwas schema,  
\item $p$, $q$ parameters of the Turkel-Zwas schema.
\end{itemize}

To verify the correct operation of the software module which implements the model, we tested the computed values of 
$\mathbf{M}_{\Delta t}^{0,M_{steps}=30}\left(x_0\right)$, when $\left|DD(\Delta\times \Omega)\right|=1$ (i.e., when domain $\Delta\times \Omega$ is not decomposed), where:

\begin{enumerate}
\item $\Delta t=50.0, 100.0, 150.0, 200.0$, \label{parametroDT}
\item $\alpha=\frac{1}{3}$, \label{parametroALFA}
\item $p=4$ and $q=2$, \label{parametroPQ}
\item $x_0$ is a syntetic vector containing all the considered fields: the see-level field $h$ is generated by a Gaussian stochastic process; both velocity fields 
$v$ and $u$ are set to zero,   %as provided by Emil C.,
\item $\Delta \lambda$ e $\Delta \theta$ 
defined on the basis of discretization grid used by data available
at repository {\em Ocean Synthesis/Reanalysis Directory} of Hamburg University (see \cite{Dati}).
\end{enumerate}

We note that the values for parameters at above mentioned points \ref{parametroDT}, \ref{parametroALFA} and \ref{parametroPQ}
were chosen on the basis of the considerations and results described in \cite{AnalysisTurkelZwas}.

\begin{figure}
\begin{center}
\begin{tabular}{cc}
\multicolumn{2}{c}{\fbox{\includegraphics[scale=1.00]{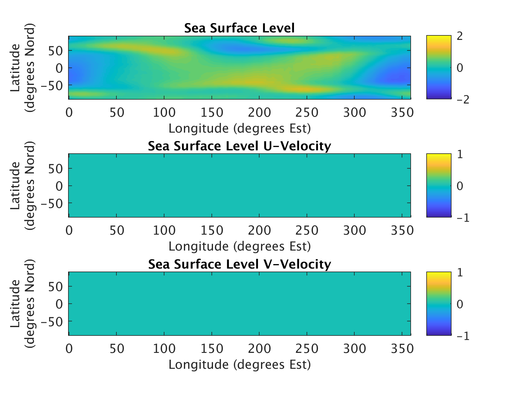}}} \\[1ex]
\multicolumn{2}{c}{(a) - Initial guess $x_0$ for the model} \\[2ex]
%\caption{Initial guess $x_0$ for the model} \label{InitialGuess}

%\centerline{
\fbox{\includegraphics[scale=1.00]{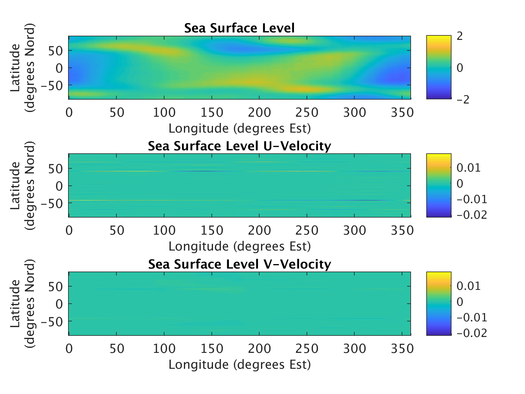}}%}
%\caption{$\mathbf{M}_{\Delta t}^{0,M_{steps}=30}\left(x_0\right)$ ($\Delta t=50.0$)} \label{ModelSolution.dt-50.000}
%\end{figure}
&
%\begin{figure}
%\centerline{
\fbox{\includegraphics[scale=1.00]{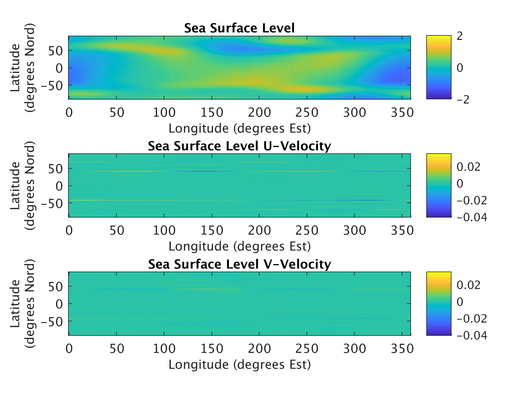}}%}
%\caption{$\mathbf{M}_{\Delta t}^{0,M_{steps}=30}\left(x_0\right)$ ($\Delta t=100.0$)} \label{ModelSolution.dt-100.000}
%\end{figure}
\\[1ex]
(b) - $\Delta t=50.0$ & (c) - $\Delta t=100.0$ \\[2ex]
%\begin{figure}
%\centerline{
\fbox{\includegraphics[scale=1.00]{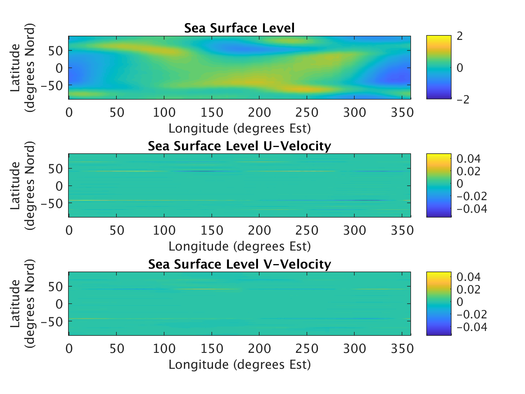}}%}
%\caption{$\mathbf{M}_{\Delta t}^{0,M_{steps}=30}\left(x_0\right)$ ($\Delta t=150.0$)} \label{ModelSolution.dt-150.000}
%\end{figure}
&
%\begin{figure}
%\centerline{
\fbox{\includegraphics[scale=1.00]{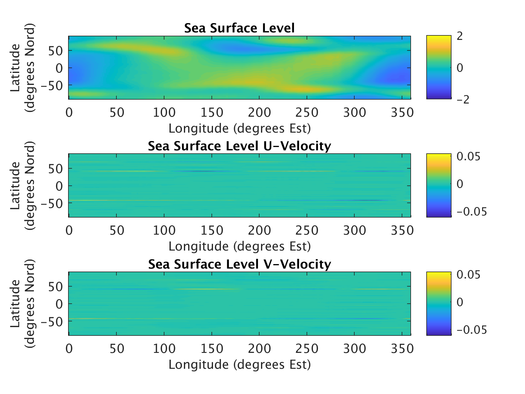}}%}
\\[1ex]
$\Delta t=150.0$ & $\Delta t=200.0$ \\[2ex]
%\caption{$\mathbf{M}_{\Delta t}^{0,M_{steps}=30}\left(x_0\right)$ ($\Delta t=200.0$)} \label{ModelSolution.dt-200.000}
\end{tabular}
\end{center}
\caption{$\mathbf{M}_{\Delta t}^{0,M_{steps}=30}\left(x_0\right)$ ($\Delta t=200.0, 150.0, 100.0, 50.0$)} \label{ModelSolution}
\end{figure}

%In figure \ref{InitialGuess} $x_0$ is represented.
In figure \ref{ModelSolution}-(a) $x_0$ is represented.
In figure \ref{ModelSolution}-(b)-(d) % \ref{ModelSolution.dt-50.000}, \ref{ModelSolution.dt-100.000}, \ref{ModelSolution.dt-150.000} and  \ref{ModelSolution.dt-200.000} 
are represented respectively $\mathbf{M}_{\Delta t}^{0,M_{steps}=30}\left(x_0\right)$ 
where $\Delta t=50.0, 100.0, 150.0, 200.0$.
We used different values for $\Delta t$ with the aim to empirically determine the {\em ``best value''} for $\Delta t$. The considered values for $\Delta t$ were chosen taking into account the considerations about CFL condition for Turkel-Zwas methods  described in \cite{AnalysisTurkelZwas}.

To give a measure of how $\mathbf{M}_{\Delta t}^{0,M_{steps}=30}\left(x_0\right)$ differs from $x_0$ depending from $\Delta t$, in table \ref{TabellaDifferenzeXsolXorig}
we show $$\|\mathbf{M}_{\Delta t}^{0,M_{steps}=30}\left(x_0\right)-x_0\|_{2}/\|x_0\|_{2}$$ for the considered values of $\Delta t$. 

\begin{table}
\begin{center}
\begin{tabular}{c|c}
$\Delta t$ & $\|\mathbf{M}_{\Delta t}^{0,M_{steps}=30}\left(x_0\right)-x_0\|_{2}/\|x_0\|_{2}$ \\ \hline
50.0  &	2.493588e-03 \\
100.0 &	4.846299e-03 \\
150.0 &	6.949851e-03 \\
200.0 &	8.737334e-03 \\
\end{tabular}
\end{center}
\caption{The values of $\|\mathbf{M}_{\Delta t}^{0,M_{steps}=30}\left(x_0\right)-x_0\|_{2}/\|x_0\|_{2}$ as function of $\Delta t$.} \label{TabellaDifferenzeXsolXorig}
\end{table}

%Following these tests, it can be concluded that the considered values of $\Delta t$ do not seem to determine significant differences in the values of 
%$\mathbf{M}_{\Delta t}^{0,M_{steps}=30}\left(x_0\right)$.

\section{Test results related with the DA software module operation}
To verify the correct operation of the software module which implements the DA process % on the basis of the approach described in \cite{4DVAR}, 
we tested the computed values of 
$x^{\Delta t}_{DA}$, when $\left|DD(\Delta\times \Omega)\right|=1$ (i.e., when domain $\Delta\times \Omega$ is not decomposed), starting:
\begin{itemize}
\item from the ${\bf R}_k={\bf R}$ ($\forall k= 0,\dots, nt_{obs}-1$) diagonal matrix representing the covariance matrix of the errors on all the observations vectors 
\item from the $H$ diagonal matrix representing the observational operator and
\item from the backgroud $x^{\Delta t}_b$ and
\item from the set of $nt_{obs}$ observations vectors $\left\{_{\Delta t}x^{n}_{obs}\right\}_{n=1,\ldots,nt_{obs}}$, 
\item and by using, as preconditioner $\tilde{B^{\Delta t}}_{nSVs}$ of covariance matrix $B^{\Delta t}$, its Truncated SVD (the first $nSVs$ singular values 
$\left(S_i\right)_{i=0,\ldots,nSVs-1}$ of $B^{\Delta t}$ are considered),
\end{itemize} 
where
\begin{enumerate}
\item $x^{\Delta t}_b = \mathbf{M}_{\Delta t}^{0,M_{steps}=30-nt_{obs}+1}\left(x_0\right)$,
\item \label{ProblemDefinition}
\fbox{
\begin{minipage}[c]{\mpwidth}
\begin{description}
\item{\bf Problem 1} For each $n=1,\ldots,nt_{obs}$,
\begin{eqnarray*}
_{\Delta t}x^{n}_{obs} &=&\frac{1}{10^2}\left[ 10^2*\mathbf{M}_{\Delta t}^{0,M_{steps}=30-nt_{obs}+n}\left(x_0\right)\right], 
\end{eqnarray*}
(each elements of $_{\Delta t}x^{n}_{obs}$ were obtained by rounding, on the third significant digit, the respective 
elements of $\mathbf{M}_{\Delta t}^{0,M_{steps}=30-nt_{obs}+n}\left(x_0\right)$).
$H$ is the identity matrix.
\item{\bf Problem 2} For each $n=1,\ldots,nt_{obs}$,
if $i$ is a multiple of $STEP=5$
\begin{eqnarray*}
\left(_{\Delta t}x^{n}_{obs}\right)_i & = & \left(\mathbf{M}_{\Delta t}^{0,M_{steps}=30-nt_{obs}+n}\left(x_0\right)\right)_i + 0.01*randn, \\
diag(H)_i & = & 1,
\end{eqnarray*}
else
\begin{eqnarray*}
\left(_{\Delta t}x^{n}_{obs}\right)_i & = &0, \\
diag(H)_i & = &0, 
\end{eqnarray*}
($_{\Delta t}x^{n}_{obs}$ are sparse vectors whose lenght is $3*nlat*nlon$ and whose non-zero elements (the $20\%$ of the total) were obtained by adding a scaled number  
$0.01*randn$ from normal distribution to 
the respective elements of $\mathbf{M}_{\Delta t}^{0,M_{steps}=30-nt_{obs}+n}\left(x_0\right)$).
\item{\bf Problem 3} For each $n=1,\ldots,nt_{obs}$,
\begin{eqnarray*}
_{\Delta t}x^{n}_{obs} &=&\frac{1}{10^1}\left[ 10^1*\mathbf{M}_{\Delta t}^{0,M_{steps}=30-nt_{obs}+n}\left(x_0\right)\right], 
\end{eqnarray*}
(each elements of $_{\Delta t}x^{n}_{obs}$ were obtained by rounding, on the second significant digit, the respective 
elements of $\mathbf{M}_{\Delta t}^{0,M_{steps}=30-nt_{obs}+n}\left(x_0\right)$).
$H$ is the identity matrix.
\item{\bf Problem 4} For each $n=1,\ldots,nt_{obs}$,
if $i$ is a multiple of $STEP=5$
\begin{eqnarray*}
\left(_{\Delta t}x^{n}_{obs}\right)_i & = & \left(\mathbf{M}_{\Delta t}^{0,M_{steps}=30-nt_{obs}+n}\left(x_0\right)\right)_i + 0.01*O\left(\left(\mathbf{M}_{\Delta t}^{0,M_{steps}=30-nt_{obs}+n}\left(x_0\right)\right)_i\right)*randn, \\
diag(H)_i & = & 1,
\end{eqnarray*}
else
\begin{eqnarray*}
\left(_{\Delta t}x^{n}_{obs}\right)_i & = &0, \\
diag(H)_i & = &0, 
\end{eqnarray*}
($_{\Delta t}x^{n}_{obs}$ are sparse vectors whose lenght is $3*nlat*nlon$ and whose non-zero elements (the $20\%$ of the total) were obtained by adding,
to the respective elements of $\mathbf{M}_{\Delta t}^{0,M_{steps}=30-nt_{obs}+n}\left(x_0\right)$, a number 
from normal distribution scaled by a factor related with order of magnitude of each elements 
$O\left(\left(\mathbf{M}_{\Delta t}^{0,M_{steps}=30-nt_{obs}+n}\left(x_0\right)\right)_i\right)$).
\end{description}
\end{minipage}}
\item $nt_{obs}=1,2,4,6,8,10$,
\item $nSVs=4,6,8,10,12$.
\item $\left({\bf R}^{-1}\right)_i=\left\{\begin{array}{cl}
				\frac{1}{1e+6} & \mbox{if\ } i=0,\ldots,nlat*nlon-1 \\
				1 & \mbox{if\ } i=nlat*nlon,\ldots,3*nlat*nlon \\
                                   \end{array}\right.$.
\item
$B^{\Delta t}=\left(x^{\Delta t}_{err}\right) \left(x^{\Delta t}_{err}\right)^T$, where 
$$
\left(x^{\Delta t}_{err}\right)_i=\left(x^{\Delta t}_{b}\right)_i-\nu^{\Delta t}
$$
and where
$$
\nu^{\Delta t}=\frac{\sum_{j=0,ldots,3*nlat*nlon-1}{\left(x^{\Delta t}_{b}\right)_j}}{3*nlat*nlon}
$$
\end{enumerate}

\begin{figure}
\centerline{\framebox{\includegraphics[scale=0.70]{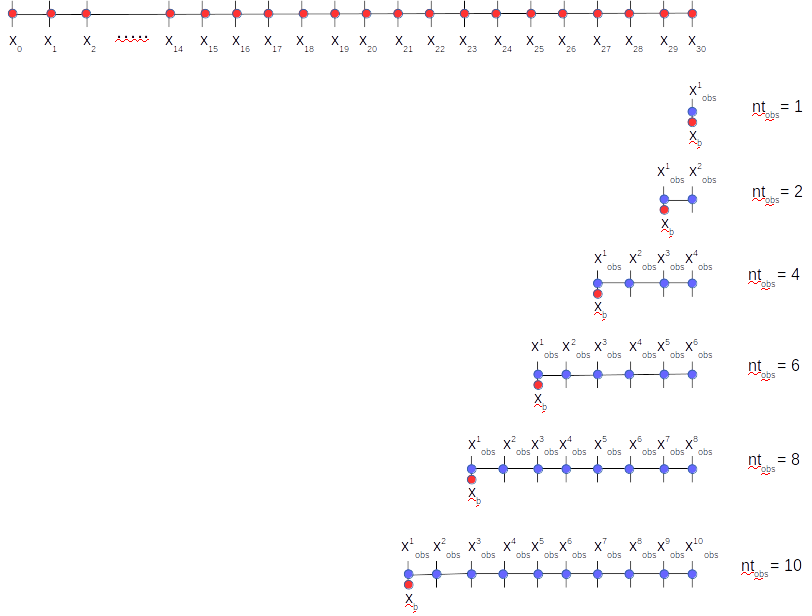}}}
\caption{How $x^{\Delta t}_b$ and $_{\Delta t}x^{n}_{obs}$ were chosen/built from $\mathbf{M}_{\Delta t}^{0,M_{steps}=m}{\left(x_0\right)}$ data} 
\label{CaseStudyDataGeneration}
\end{figure}

Figure \ref{CaseStudyDataGeneration} shows how the background $x^{\Delta t}_b$ (the red circle in the image) and observations $_{\Delta t}x^{n}_{obs}$ 
(the blue circle in the image) 
were chosen/built from 
$x^{\Delta t}_m=\mathbf{M}_{\Delta t}^{0,M_{steps}=m}{\left(x_0\right)}$ data: in particular, for each values of $nt_{obs}=1,2,4,6,8,10$, the image intends to show which subset 
of $\left\{x^{\Delta t}_m\right\}_{m=0,\ldots,30}$ is considered to generate background and observations.
In particular, the procedure used to build the input data for DA problem performs the following steps:
\begin{enumerate}
\item
{\em ``application''} of the model $\mathbf{M}_{\Delta t}^{0,M_{steps}=m}\left(\cdot\right)$ to the starting point $x_0$ to obtain 
the vector $x^{\Delta t}_m$ where $m=30-nt_{obs}+1$; \label{puntoXB}
\item 
from the vector $x^{\Delta t}_m$ computed at above point \ref{puntoXB} we obtain both 
the background vector $x^{\Delta t}_b$ and, by using one of the definitions for {Problem 1} or {Problem 2}, the 
first observation vector $_{\Delta t}x^{1}_{obs}$;
\item
further {\em ``application''} of the model to compute the set of vectors 
$\left\{x^{\Delta t}_m\right\}_{m=30-nt_{obs}+2,\ldots,30}$
from which obtain, by using one of the definitions for {Problem 1} or {Problem 2}, the remaining $nt_{obs}-1$ observation vectors 
$_{\Delta t}x^{i}_{obs}, i=2,..,nt_{obs}$.
\end{enumerate}

Then, the {\em ``assimilation window''} $AW^{nt_{obs}}_{\Delta t}$ in time domain, when the value of ${\Delta t}$ is fixed, is 
the intervall defined as:
\begin{displaymath}
AW^{nt_{obs}}_{\Delta t} = \left[ \left(30-nt_{obs}+1\right)\Delta t,30\Delta t\right].
\end{displaymath}

Three sets of tests are performed: 
the first set intends to evaluate how different values for $nt_{obs}$ and $nSVs$ influence the behavior of DA software module;
the second and third sets 
of tests intend to give elements to evaluate how Data Assimilation used during model application (i.e., $\mathbf{M}_{\Delta t}^{0,M_{steps}=m}\left(x^{\Delta t}_{DA}\right)$)
 improve the quality of the model computed values without DA (i.e., $\mathbf{M}_{\Delta t}^{0,M_{steps}=m}\left(x^{\Delta t}_b\right)$) 
with respect to observations
\footnote{We note that the symbols 
$\mathbf{M}^{0,M_{steps}=n}\left(x^{\Delta t}_{b}\right)$ 
and 
$\mathbf{M}_{\Delta t}^{0,M_{steps}=n}\left(x^{\Delta t}_{DA}\right)$ 
represent the model {\em ``applied''} $n$ times to the background $x^{\Delta t}_{b}$ and to the solution of DA problem $x^{\Delta t}_{DA}$ respectively}.

\begin{description}
\item{\bf Tests Set 1}
In order to evaluate how different values for $nt_{obs}$ and $nSVs$ influence the behavior of DA software module, 
in tables \ref{TabellaERRDA-Problem1} (for Problem 1),  \ref{TabellaERRDA-Problem2} (for Problem 2), \ref{TabellaERRDA-Problem3} (for Problem 3) and \ref{TabellaERRDA-Problem4} (for Problem 4) the values of $err^{\Delta t}_{b}$ and $err^{\Delta t}_{DA}$ are showed for the above listed values 
of $nt_{obs}$ and $nSVs$ and for the four considered problems where
%\ref{TabellaERRDA.dt-50.000}, \ref{TabellaERRDA.dt-100.000}, \ref{TabellaERRDA.dt-150.000} e \ref{TabellaERRDA.dt-200.000} 

\begin{eqnarray}
err^{\Delta t}_{b}  & = & \left\|H x^{\Delta t}_b     - _{\Delta t}x^{n=1}_{obs}\right\|_{2}/\left\|_{\Delta t}x^{n=1}_{obs}\right\|_{2} \\
err^{\Delta t}_{DA} & = & \left\|H x^{\Delta t}_{DA}  - _{\Delta t}x^{n=1}_{obs}\right\|_{2}/\left\|_{\Delta t}x^{n=1}_{obs}\right\|_{2}
\end{eqnarray}

and where $\Delta t=50.0, 100.0, 150.0, 200.0$.
We note that values of $err^{\Delta t}_{DA}$ are not reported when the algorithm failed (i.e., when the Truncated SVD computation failed).
The table \ref{SingularValues} show, as examples, the values of $\left(S_i\right)_{i=0,\ldots,nSVs-1}$ when $nt_{obs}=2$ and $\Delta t=50.0,100.0,150.0,200.0$.

\begin{table}
\begin{center}
\begin{tabular}{|c|c|c|c|c|} \hline
    & \multicolumn{4}{c|}{$\left(S_i\right)_{i=0,\ldots,nSVs-1}$}  \\\cline{2-5}
    & \multicolumn{4}{c|}{$\Delta t$}                            \\\cline{2-5}                               
$i$ &  50.0        & 100.0        & 150.0        &  200.0      \\\hline 
00  & 1.489436e+05 & 1.489436e+05 & 1.489436e+05 & 1.489436e+05 \\
01  & 5.922436e-12 & 5.918084e-12 & 1.421180e-11 & 1.518229e-07 \\
02  & 1.098639e-18 & 3.745277e-12 & 1.058226e-11 & 3.045052e-11 \\
03  & 2.717045e-22 & 2.956363e-12 & 5.891872e-12 & 2.154711e-11 \\
04  & 7.077415e-23 & 2.450420e-12 & 5.477584e-12 & 1.815192e-11 \\
05  & 1.984375e-23 & 2.034103e-12 & 4.955457e-12 & 1.770864e-11 \\
06  & 1.308422e-25 & 1.248364e-12 & 4.426257e-12 & 1.447040e-11 \\
07  & 6.804022e-26 & 9.572109e-13 & 4.043577e-12 & 1.247191e-11 \\
08  & 6.736329e-26 & 9.029654e-13 & 3.518900e-12 & 1.176448e-11 \\
09  & 6.642977e-26 & 7.345570e-13 & 2.587594e-12 & 1.014551e-11 \\
10  & 6.614099e-26 & 7.164269e-13 & 2.226629e-12 & 9.308843e-12 \\
11  & 5.977310e-28 & 5.974983e-13 & 1.710227e-12 & 5.925403e-12 \\\hline
\end{tabular}
\end{center}
\caption{The first $nSVs$ singular values $\left(S_i\right)_{i=0,\ldots,nSVs-1}$ of matrix $B^{\Delta t}$ when $nt_{obs}=2$ 
($\Delta t=50.0,100.0,150.0,200.0$).} \label{SingularValues}
\end{table}

We also note that: 
\begin{enumerate}
\item
the smaller values for $nSVs$ is (i.e., where $nSVs=4$), more the DA software module is able to effectively compute the solution of the DA problem;
\item
the larger values for $nSVs$ is, more accurate is the solution of the DA problem computed by DA software module (if it is successful).
%This consideration is mostly applicable in the case of problem 1: in the case of problem 2, the use of the highest values for $nSVs$ seems to lead to accuracy degradation.
\end{enumerate}

This behavior could be explained considering that:
\begin{enumerate}
\item the smaller the value of $nSVs$ is, better conditioned is the matrix $\tilde{B}_{nSVs}$, but also 
\item the larger values for $nSVs$ is, {\em ``closer''} the matrices $\tilde{B}_{nSVs}$ and $B$ are.
\end{enumerate}

%\begin{table}
%\hspace{-2cm}
\begin{sidewaystable}
\begin{minipage}[t]{\textwidth}\small
\begin{tabular}{c||c||cccccc}
                     &                                                 & \multicolumn{6}{|c}{$nt_{obs}$} \\[1ex]\cline{3-8}
$\Delta t=50.0$      &                                                 & $1$                &         $2$        &         $4$        &         $6$        &         $8$        &         $10$ \\\hline\hline
\multicolumn{2}{c||}{$err^{\Delta t}_{b}$}       & 5.923000468033734e-03 & 5.918233049175874e-03 & 5.909744367520904e-03 & 5.901306860071136e-03 & 5.893094645517731e-03 & 5.884752741767643e-03 \\\hline\hline 
\multirow{6}{*}{$err^{\Delta t}_{DA}$}   & $nSVs$&                       &                       &                       &                       &                       &                       \\\hline\hline
                              & $4$   & 5.923000468033728e-03 & 5.918233049175878e-03 & 5.909744367520882e-03 & 5.901306860071130e-03 & 5.893094645517735e-03 & 5.884752741767650e-03      \\
                              & $6$   & 5.923000468033734e-03 & 5.918233049175892e-03 & $-$                   & $-$                   & 5.893094645517728e-03 & 5.884752741767642e-03      \\
                              & $8$   & 5.923000468033734e-03 & 5.918233049175873e-03 & 5.909744367520906e-03 & 5.901306860071137e-03 & 5.893094645517726e-03 & 5.884752724034771e-03      \\
                              & $10$  & 5.923000468033739e-03 & 5.918233049175869e-03 & 5.909744367520902e-03 & 5.901306860071123e-03 & 5.893094612088676e-03 & 5.884752741767648e-03      \\
                              & $12$  & $-$                   & 5.918233049175874e-03 & 5.909744377775199e-03 & 5.901306706186256e-03 & $-$                   & $-$                       \\                           \\ \hline\hline
\end{tabular}
\end{minipage}
\\[3ex]
%\centerline{(a)}
%\caption{$err_{b}$ and $err_{DA}$ as function of $nt_{obs}$ and $nSVs$ ($\Delta t=50.0$)} 
%\label{TabellaERRDA.dt-50.000}
%%\end{table}
%\end{sidewaystable}
%
%\begin{sidewaystable}
%%\begin{table}
%%\hspace{-2cm}
\begin{minipage}[t]{\textwidth}\small
\begin{tabular}{c||c||cccccc}
                      &                                                 & \multicolumn{6}{|c}{$nt_{obs}$} \\[1ex]\cline{3-8}
$\Delta t=100.0$      &                                                 & $1$                &         $2$        &         $4$        &         $6$        &         $8$        &         $10$ \\\hline\hline
\multicolumn{2}{c||}{$err^{\Delta t}_{b}$}       & 6.084454299095919e-03 & 6.073321261976583e-03 & 6.052230870989613e-03 & 6.028625075904369e-03 & 6.006605050609536e-03 & 5.984624179759802e-03  \\\hline\hline\\\
\multirow{6}{*}{$err^{\Delta t}_{DA}$}   & $nSVs$&                       &                       &                       &                       &                       &                       \\\hline\hline
                              & $4$   & 6.084454299095923e-03 & 6.073321261976588e-03 & 6.052230870989617e-03 & 6.028625075904372e-03 & 6.006605050609541e-03 & 5.984624179759798e-03            \\
                              & $6$   & 6.084454299095920e-03 & 6.073321261976588e-03 & 6.052230870989613e-03 & $-$                   & 6.006605050609541e-03 & 5.984624179759808e-03            \\
                              & $8$   & 6.084454298277842e-03 & 6.073321261976589e-03 & 6.052230870989612e-03 & 6.028625162320382e-03 & 6.006594251561000e-03 & 5.984624177098236e-03            \\
                              & $10$  & 6.084454340791144e-03 & $-$                   & 6.052230870989642e-03 & 6.028625075904334e-03 & $-$                   & $-$                              \\
                              & $12$  & 6.084454255839374e-03 & 6.073321261494072e-03 & 6.052230870989607e-03 & 6.028625075885917e-03 & $-$                   & $-$                              \\ \hline\hline
\end{tabular}
\end{minipage}
\\[3ex]
%\centerline{(b)}
%\caption{$err_{b}$ and $err_{DA}$ as function of $nt_{obs}$ and $nSVs$ ($\Delta t=100.0$)} 
%\label{TabellaERRDA.dt-100.000}
%%\end{table}
%\end{sidewaystable}
%
%\begin{sidewaystable}
%%\begin{table}
%%\hspace{-2cm}
\begin{minipage}[t]{\textwidth}\small
\begin{tabular}{c||c||cccccc}
                      &                                                 & \multicolumn{6}{|c}{$nt_{obs}$} \\[1ex]\cline{3-8}
$\Delta t=150.0$      &                                                 & $1$                &         $2$        &         $4$        &         $6$        &         $8$        &         $10$ \\\hline\hline
\multicolumn{2}{c||}{$err^{\Delta t}_{b}$}       & 6.280754455601023e-03 & 6.263201134463331e-03 & 6.225669222102765e-03 & 6.184831399882742e-03 & 6.143615055490961e-03 & 6.102356895266559e-03 \\\hline\hline
\multirow{6}{*}{$err^{\Delta t}_{DA}$}   & $nSVs$&                    &                    &                    &                    &                    &                                      \\\hline\hline
                              & $4$   & 6.280754455601023e-03 & 6.263201134463330e-03 & 6.225669222102762e-03 & 6.184831399882739e-03 & 6.143615055490959e-03 & 6.102356895266554e-03           \\
                              & $6$   & 6.280754455601023e-03 & 6.263201134463330e-03 & 6.225669222102764e-03 & 6.184831399882742e-03 & $-$                   & 6.102356895266563e-03           \\
                              & $8$   & 6.280754455601023e-03 & 6.263201064275920e-03 & $-$                   & 6.184831399882738e-03 & 6.143615055490968e-03 & 6.102356895266562e-03           \\
                              & $10$  & 6.280754446106453e-03 & 6.263201055314105e-03 & 6.225669243979252e-03 & $-$                   & 6.143615055490962e-03 & 6.102356790788719e-03           \\
                              & $12$  & 6.280754387072876e-03 & 6.263201117930033e-03 & 6.225668830117284e-03 & $-$                   & 6.143615138972201e-03 & 6.102356811770067e-03  \\ \hline\hline
\end{tabular}
\end{minipage}
\\[3ex]
%\centerline{(c)}
%\caption{$err_{b}$ and $err_{DA}$ as function of $nt_{obs}$ and $nSVs$ ($\Delta t=150.0$)} 
%\label{TabellaERRDA.dt-150.000}
%%\end{table}
%\end{sidewaystable}

%\begin{sidewaystable}
%\begin{table}
%\hspace{-2cm}
\begin{minipage}[t]{\textwidth}\small
\begin{tabular}{c||c||cccccc}
                      &                                                 & \multicolumn{6}{|c}{$nt_{obs}$} \\[1ex]\cline{3-8}
$\Delta t=200.0$      &                                                 & $1$                &         $2$        &         $4$        &         $6$        &         $8$        &         $10$ \\\hline\hline
\multicolumn{2}{c||}{$err^{\Delta t}_{b}$}       & 6.462142195081026e-03 & 6.437375271097644e-03 & 6.392026891087157e-03 & 6.340783651666366e-03 & 6.292150917739605e-03 & 6.244798377227290e-03  \\\hline\hline
\multirow{6}{*}{$err^{\Delta t}_{DA}$}   & $nSVs$&                    &                    &                    &                    &                    &                                              \\\hline\hline
                              & $4$   & 6.462142195081031e-03 & 6.437375271097648e-03 & 6.392026891087095e-03 & 6.340783651666372e-03 & 6.292150917739616e-03 & 6.244798377227289e-03  \\
                              & $6$   & 6.462142195081022e-03 & $-$                   & 6.392026891087161e-03 & 6.340783652553978e-03 & 6.292150917739606e-03 & 6.244798377227294e-03  \\
                              & $8$   & $-$                   & 6.437375271097721e-03 & 6.392026891087158e-03 & 6.340783669355231e-03 & 6.292150917739599e-03 & 6.244798377227293e-03  \\
                              & $10$  & $-$                   & 6.437375294652374e-03 & $-$                   & $-$                   & 6.292150917739615e-03 & $-$                    \\
                              & $12$  & $-$                   & 6.437375306111083e-03 & 6.392026896528792e-03 & 6.340783515516571e-03 & $-$                   & $-$                    \\ \hline\hline
\end{tabular}
\end{minipage}
%\centerline{(d)}
%\caption{$err_{b}$ and $err_{DA}$ as function of $nt_{obs}$ and $nSVs$ ($\Delta t=200.0$)} 
%\label{TabellaERRDA.dt-200.000}
\caption{$err^{\Delta t}_{b}$ and $err^{\Delta t}_{DA}$ as function of $nt_{obs}$ and $nSVs$ ($\Delta t=50.0,100.0,150.0,200.0$) - Problem 1} 
\label{TabellaERRDA-Problem1}
%%\end{table}
\end{sidewaystable}

\begin{sidewaystable}
\begin{minipage}[t]{\textwidth}\small
\begin{tabular}{c||c||cccccc}
                     &                                                 & \multicolumn{6}{|c}{$nt_{obs}$} \\[1ex]\cline{3-8}
$\Delta t=50.0$      &                                                 & $1$                &         $2$        &         $4$        &         $6$        &         $8$        &         $10$ \\\hline\hline
\multicolumn{2}{c||}{$err^{\Delta t}_{b}$}       & 3.453467033333584e-02 & 3.496046426389585e-02 & 3.461814248792758e-02 & 3.496047407097717e-02 & 3.461815048663301e-02 & 3.496048249133105e-02 \\\hline\hline
\multirow{6}{*}{$err^{\Delta t}_{DA}$}   & $nSVs$&                       &                       &                       &                       &                       &                       \\\hline\hline
                              & $4$   & 3.453467033333583e-02 & 3.496046426389589e-02 & 3.461814248792760e-02 & 3.496047407097717e-02 & 3.461815048663298e-02 & 3.496048249133105e-02  \\
                              & $6$   & 3.453467033333584e-02 & 3.496046426389582e-02 & $-$                   & $-$                   & 3.461815048663301e-02 & 3.496048249133105e-02  \\
                              & $8$   & 3.453467033333584e-02 & 3.496046426389584e-02 & 3.461814248792756e-02 & 3.496047407097716e-02 & 3.461815048663301e-02 & 3.496048237755320e-02  \\
                              & $10$  & 3.453467033333585e-02 & 3.496046426389584e-02 & 3.461814248792758e-02 & 3.496047407097717e-02 & 3.461815029375775e-02 & 3.496048249133106e-02  \\
                              & $12$  & $-$                   & 3.496046426389585e-02 & 3.461814250679108e-02 & 3.496047385347141e-02 & $-$                   & $-$   \\ \hline\hline
\end{tabular}
\end{minipage}
\\[3ex]
%\centerline{(a)}
%\caption{$err_{b}$ and $err_{DA}$ as function of $nt_{obs}$ and $nSVs$ ($\Delta t=50.0$)} 
%\label{TabellaERRDA.dt-50.000}
%%\end{table}
%\end{sidewaystable}
%
%\begin{sidewaystable}
%%\begin{table}
%%\hspace{-2cm}
\begin{minipage}[t]{\textwidth}\small
\begin{tabular}{c||c||cccccc}
                      &                                                 & \multicolumn{6}{|c}{$nt_{obs}$} \\[1ex]\cline{3-8}
$\Delta t=100.0$      &                                                 & $1$                &         $2$        &         $4$        &         $6$        &         $8$        &         $10$ \\\hline\hline
\multicolumn{2}{c||}{$err^{\Delta t}_{b}$}       & 3.453457238151639e-02 & 3.496035264638818e-02 & 3.461805630410621e-02 & 3.496039039880471e-02 & 3.461808760666546e-02 & 3.496042291664354e-02                     \\\hline\hline
\multirow{6}{*}{$err^{\Delta t}_{DA}$}   & $nSVs$&                       &                       &                       &                       &                       &                       \\\hline\hline
                              & $4$   & 3.453457238151639e-02 & 3.496035264638819e-02 & 3.461805630410621e-02 & 3.496039039880478e-02 & 3.461808760666545e-02 & 3.496042291664356e-02   \\
                              & $6$   & 3.453457238151640e-02 & 3.496035264638819e-02 & 3.461805630410621e-02 & $-$                   & 3.461808760666545e-02 & 3.496042291664352e-02   \\
                              & $8$   & 3.453457249154981e-02 & 3.496035264638817e-02 & 3.461805630410621e-02 & 3.496039035077939e-02 & 3.461809290479863e-02 & 3.496042290325619e-02   \\
                              & $10$  & 3.453457274084655e-02 & $-$                   & 3.461805630410599e-02 & 3.496039039880472e-02 & $-$                   & $-$                   \\
                              & $12$  & 3.453457237819704e-02 & 3.496035261803864e-02 & 3.461805630410619e-02 & 3.496039039846557e-02 & $-$                   & $-$                   \\ \hline\hline
\end{tabular}
\end{minipage}
\\[3ex]
%\centerline{(b)}
%\caption{$err_{b}$ and $err_{DA}$ as function of $nt_{obs}$ and $nSVs$ ($\Delta t=100.0$)} 
%\label{TabellaERRDA.dt-100.000}
%%\end{table}
%\end{sidewaystable}
%
%\begin{sidewaystable}
%%\begin{table}
%%\hspace{-2cm}
\begin{minipage}[t]{\textwidth}\small
\begin{tabular}{c||c||cccccc}
                      &                                                 & \multicolumn{6}{|c}{$nt_{obs}$} \\[1ex]\cline{3-8}
$\Delta t=150.0$      &                                                 & $1$                &         $2$        &         $4$        &         $6$        &         $8$        &         $10$ \\\hline\hline
\multicolumn{2}{c||}{$err^{\Delta t}_{b}$}       & 3.453440979970871e-02 & 3.496017506248329e-02 & 3.488429733019234e-02 & 3.483606155320147e-02 & 3.461798505691022e-02 & 3.496032646744604e-02                     \\\hline\hline
\multirow{6}{*}{$err^{\Delta t}_{DA}$}   & $nSVs$&                    &                    &                    &                    &                    &                                      \\\hline\hline
                              & $4$  & 3.453440979970870e-02 & 3.496017506248329e-02 & 3.488429733019234e-02 & 3.483606155320147e-02 & 3.461798505691021e-02 & 3.496032646744603e-02 \\
                              & $6$  & 3.453440979970870e-02 & 3.496017506248329e-02 & 3.488429733019234e-02 & 3.483606155320146e-02 & $-$                   & 3.496032646744604e-02 \\
                              & $8$  & 3.453440979970870e-02 & 3.496017499975523e-02 & $-$                   & 3.483606155320146e-02 & 3.461798505691016e-02 & 3.496032646744601e-02 \\
                              & $10$ & 3.453440976085218e-02 & 3.496017512611298e-02 & 3.488429724470573e-02 & $-$                   & 3.461798505691022e-02 & 3.496032641471299e-02 \\
                              & $12$ & 3.453440988054077e-02 & 3.496017506439134e-02 & 3.488429726668655e-02 & $-$                   & 3.461798464694495e-02 & 3.496032654977042e-02 \\ \hline\hline
\end{tabular}
\end{minipage}
\\[3ex]
%\centerline{(c)}
%\caption{$err_{b}$ and $err_{DA}$ as function of $nt_{obs}$ and $nSVs$ ($\Delta t=150.0$)} 
%\label{TabellaERRDA.dt-150.000}
%%\end{table}
%\end{sidewaystable}

%\begin{sidewaystable}
%\begin{table}
%\hspace{-2cm}
\begin{minipage}[t]{\textwidth}\small
\begin{tabular}{c||c||cccccc}
                      &                                                 & \multicolumn{6}{|c}{$nt_{obs}$} \\[1ex]\cline{3-8}
$\Delta t=200.0$      &                                                 & $1$                &         $2$        &         $4$        &         $6$        &         $8$        &         $10$ \\\hline\hline
\multicolumn{2}{c||}{$err^{\Delta t}_{b}$}       & 3.453419647204346e-02 & 3.495994493038772e-02 & 3.461773231566689e-02 & 3.496007727466858e-02 & 3.496013862728978e-02 & 3.496019616952864e-02 \\\hline\hline
\multirow{6}{*}{$err^{\Delta t}_{DA}$}   & $nSVs$&                    &                    &                    &                    &                    &                                              \\\hline\hline
                              & $4$  & 3.453419647204347e-02 & 3.495994493038771e-02 & 3.461773231566698e-02 & 3.496007727466857e-02 & 3.496013862728979e-02 & 3.496019616952864e-02 \\
                              & $6$  & 3.453419647204343e-02 & $-$                   & 3.461773231566688e-02 & 3.496007727559594e-02 & 3.496013862728978e-02 & 3.496019616952863e-02 \\
                              & $8$  & $-$                   & 3.495994493038770e-02 & 3.461773231566689e-02 & 3.496007721824784e-02 & 3.496013862728979e-02 & 3.496019616952864e-02 \\
                              & $10$ & $-$                   & 3.495994489860085e-02 & $-$                   & $-$                   & 3.496013862728979e-02 & $-$                   \\
                              & $12$ & $-$                   & 3.495994485406865e-02 & 3.461773231104990e-02 & 3.496007753154848e-02 & $-$                   & $-$                   \\
\hline\hline

\end{tabular}
\end{minipage}
%\centerline{(d)}
%\caption{$err_{b}$ and $err_{DA}$ as function of $nt_{obs}$ and $nSVs$ ($\Delta t=200.0$)} 
%\label{TabellaERRDA.dt-200.000}
\caption{$err^{\Delta t}_{b}$ and $err^{\Delta t}_{DA}$ as function of $nt_{obs}$ and $nSVs$ ($\Delta t=50.0,100.0,150.0,200.0$) - Problem 2}
\label{TabellaERRDA-Problem2}
%%\end{table}
\end{sidewaystable}

\begin{sidewaystable}
\begin{minipage}[t]{\textwidth}\small
\begin{tabular}{c||c||cccccc}
                     &                                                 & \multicolumn{6}{|c}{$nt_{obs}$} \\[1ex]\cline{3-8}
$\Delta t=50.0$      &                                                 & $1$                &         $2$        &         $4$        &         $6$        &         $8$        &         $10$ \\\hline\hline
\multicolumn{2}{c||}{$err^{\Delta t}_{b}$}       & 5.773468954018757e-02 & 5.773118330947371e-02 & 5.772450392218931e-02 & 5.771827398570289e-02 & 5.771249955261612e-02 & 5.770718624499813e-02 \\\hline\hline
\multirow{6}{*}{$err^{\Delta t}_{DA}$}   & $nSVs$&                       &                       &                       &                       &                       &                       \\\hline\hline
                              & $4$  & 5.773468954018757e-02 & 5.773118330947368e-02 & 5.772450392218937e-02 & 5.771827398570288e-02 & 5.771249955261611e-02 & 5.770718624499812e-02 \\
                              & $6$  & 5.773468954018758e-02 & 5.773118330947370e-02 & $-$                   & $-$                   & 5.771249955261611e-02 & 5.770718624499811e-02 \\
                              & $8$  & 5.773468954018757e-02 & 5.773118330947373e-02 & 5.772450392218934e-02 & 5.771827398570289e-02 & 5.771249955261611e-02 & 5.770718629444516e-02 \\
                              & $10$  & 5.773468954018760e-02 & 5.773118330947369e-02 & 5.772450392218934e-02 & 5.771827398570289e-02 & 5.771249952331193e-02 & 5.770718624499811e-02 \\
                              & $12$  & $-$                   & 5.773118330947372e-02 & 5.772450392378905e-02 & 5.771827419320513e-02 & $-$                   & $-$                   \\
\hline\hline
\end{tabular}
\end{minipage}
\\[3ex]
%\centerline{(a)}
%\caption{$err_{b}$ and $err_{DA}$ as function of $nt_{obs}$ and $nSVs$ ($\Delta t=50.0$)} 
%\label{TabellaERRDA.dt-50.000}
%%\end{table}
%\end{sidewaystable}
%
%\begin{sidewaystable}
%%\begin{table}
%%\hspace{-2cm}
\begin{minipage}[t]{\textwidth}\small
\begin{tabular}{c||c||cccccc}
                      &                                                 & \multicolumn{6}{|c}{$nt_{obs}$} \\[1ex]\cline{3-8}
$\Delta t=100.0$      &                                                 & $1$                &         $2$        &         $4$        &         $6$        &         $8$        &         $10$ \\\hline\hline
\multicolumn{2}{c||}{$err^{\Delta t}_{b}$}       & 5.788632535497507e-02 & 5.787373829705302e-02 & 5.784950155085092e-02 & 5.782658849902559e-02 & 5.780508422056688e-02 & 5.778506929509217e-02  \\\hline\hline
\multirow{6}{*}{$err^{\Delta t}_{DA}$}   & $nSVs$&                       &                       &                       &                       &                       &                       \\\hline\hline
                              & $4$  & 5.788632535497507e-02 & 5.787373829705301e-02 & 5.784950155085092e-02 & 5.782658849902555e-02 & 5.780508422056686e-02 & 5.778506929509220e-02 \\
                              & $6$  & 5.788632535497510e-02 & 5.787373829705302e-02 & 5.784950155085092e-02 & $-$                   & 5.780508422056686e-02 & 5.778506929509220e-02 \\
                              & $8$  & 5.788632536157578e-02 & 5.787373829705302e-02 & 5.784950155085091e-02 & 5.782658848730357e-02 & 5.780506330666032e-02 & 5.778506929813398e-02 \\
                              & $10$  & 5.788632527723454e-02 & $-$                   & 5.784950155085083e-02 & 5.782658849902568e-02 & $-$                   & $-$                   \\
                              & $12$  & 5.788632541333875e-02 & 5.787373829803212e-02 & 5.784950155085090e-02 & 5.782658849917136e-02 & $-$                   & $-$                   \\
\hline\hline
\end{tabular}
\end{minipage}
\\[3ex]
%\centerline{(b)}
%\caption{$err_{b}$ and $err_{DA}$ as function of $nt_{obs}$ and $nSVs$ ($\Delta t=100.0$)} 
%\label{TabellaERRDA.dt-100.000}
%%\end{table}
%\end{sidewaystable}
%
%\begin{sidewaystable}
%%\begin{table}
%%\hspace{-2cm}
\begin{minipage}[t]{\textwidth}\small
\begin{tabular}{c||c||cccccc}
                      &                                                 & \multicolumn{6}{|c}{$nt_{obs}$} \\[1ex]\cline{3-8}
$\Delta t=150.0$      &                                                 & $1$                &         $2$        &         $4$        &         $6$        &         $8$        &         $10$ \\\hline\hline
\multicolumn{2}{c||}{$err^{\Delta t}_{b}$}       & 5.809599803382869e-02 & 5.807549989275792e-02 & 5.803325570359528e-02 & 5.798937823866770e-02 & 5.794647454306275e-02 & 5.790575933289238e-02  \\\hline\hline
\multirow{6}{*}{$err^{\Delta t}_{DA}$}   & $nSVs$&                    &                    &                    &                    &                    &                                      \\\hline\hline
                              & $4$  & 5.809599803382870e-02 & 5.807549989275791e-02 & 5.803325570359529e-02 & 5.798937823866770e-02 & 5.794647454306275e-02 & 5.790575933289238e-02 \\
                              & $6$  & 5.809599803382870e-02 & 5.807549989275791e-02 & 5.803325570359529e-02 & 5.798937823866768e-02 & $-$                   & 5.790575933289237e-02 \\
                              & $8$  & 5.809599803382870e-02 & 5.807549990183274e-02 & $-$                   & 5.798937823866769e-02 & 5.794647454306268e-02 & 5.790575933289235e-02 \\
                              & $10$  & 5.809599807145778e-02 & 5.807550000874059e-02 & 5.803325568836281e-02 & $-$                   & 5.794647454306276e-02 & 5.790575939801962e-02 \\
                              & $12$  & 5.809599805160266e-02 & 5.807549983114331e-02 & 5.803325538684955e-02 & $-$                   & 5.794647445626924e-02 & 5.790575931600109e-02 \\
\hline\hline
\end{tabular}
\end{minipage}
\\[3ex]
%\centerline{(c)}
%\caption{$err_{b}$ and $err_{DA}$ as function of $nt_{obs}$ and $nSVs$ ($\Delta t=150.0$)} 
%\label{TabellaERRDA.dt-150.000}
%%\end{table}
%\end{sidewaystable}

%\begin{sidewaystable}
%\begin{table}
%\hspace{-2cm}
\begin{minipage}[t]{\textwidth}\small
\begin{tabular}{c||c||cccccc}
                      &                                                 & \multicolumn{6}{|c}{$nt_{obs}$} \\[1ex]\cline{3-8}
$\Delta t=200.0$      &                                                 & $1$                &         $2$        &         $4$        &         $6$        &         $8$        &         $10$ \\\hline\hline
\multicolumn{2}{c||}{$err^{\Delta t}_{b}$}       & 5.830785511535787e-02 & 5.827839030475356e-02 & 5.822099874390471e-02 & 5.816499040667799e-02 & 5.810964168493156e-02 & 5.805455701855686e-02 \\\hline\hline
\multirow{6}{*}{$err^{\Delta t}_{DA}$}   & $nSVs$&                    &                    &                    &                    &                    &                                              \\\hline\hline
                              & $4$  & 5.830785511535792e-02 & 5.827839030475357e-02 & 5.822099874390501e-02 & 5.816499040667800e-02 & 5.810964168493161e-02 & 5.805455701855686e-02 \\
                              & $6$  & 5.830785511535794e-02 & $-$                   & 5.822099874390471e-02 & 5.816499040727242e-02 & 5.810964168493156e-02 & 5.805455701855683e-02 \\
                              & $8$  & $-$                   & 5.827839030475357e-02 & 5.822099874390473e-02 & 5.816499039611871e-02 & 5.810964168493157e-02 & 5.805455701855686e-02 \\
                              & $10$  & $-$                   & 5.827839030296449e-02 & $-$                   & $-$                   & 5.810964168493159e-02 & $-$                   \\
                              & $12$  & $-$                   & 5.827839029079828e-02 & 5.822099874092245e-02 & 5.816499015380254e-02 & $-$                   & $-$                   \\
\hline\hline
\end{tabular}
\end{minipage}
%\centerline{(d)}
%\caption{$err_{b}$ and $err_{DA}$ as function of $nt_{obs}$ and $nSVs$ ($\Delta t=200.0$)} 
%\label{TabellaERRDA.dt-200.000}
\caption{$err^{\Delta t}_{b}$ and $err^{\Delta t}_{DA}$ as function of $nt_{obs}$ and $nSVs$ ($\Delta t=50.0,100.0,150.0,200.0$) - Problem 3}
\label{TabellaERRDA-Problem3}
%%\end{table}
\end{sidewaystable}

\begin{sidewaystable}
\begin{minipage}[t]{\textwidth}\small
\begin{tabular}{c||c||cccccc}
                     &                                                 & \multicolumn{6}{|c}{$nt_{obs}$} \\[1ex]\cline{3-8}
$\Delta t=50.0$      &                                                 & $1$                &         $2$        &         $4$        &         $6$        &         $8$        &         $10$ \\\hline\hline
\multicolumn{2}{c||}{$err^{\Delta t}_{b}$}       & 3.212944523284356e-02 & 3.146374761827107e-02 & 2.712145713499754e-02 & 3.057501318054897e-02 & 2.712146361761048e-02 & 3.057501994577008e-02 \\\hline\hline
\multirow{6}{*}{$err^{\Delta t}_{DA}$}   & $nSVs$&                       &                       &                       &                       &                       &                       \\\hline\hline
                              & $4$  & 3.212944523284356e-02 & 3.146374761827107e-02 & 2.712145713499750e-02 & 3.057501318054896e-02 & 2.712146361761048e-02 & 3.057501994577008e-02 \\
                              & $6$  & 3.212944523284356e-02 & 3.146374761827107e-02 & $-$                   & $-$                   & 2.712146361761048e-02 & 3.057501994577008e-02 \\
                              & $8$  & 3.212944523284356e-02 & 3.146374761827107e-02 & 2.712145713499755e-02 & 3.057501318054896e-02 & 2.712146361761049e-02 & 3.057502001608988e-02 \\
                              & $10$  & 3.212944523284356e-02 & 3.146374761827108e-02 & 2.712145713499754e-02 & 3.057501318054895e-02 & 2.712146366967664e-02 & 3.057501994577007e-02 \\
                              & $12$  & $-$                   & 3.146374761827107e-02 & 2.712145708523746e-02 & 3.057501359781360e-02 & $-$                   & $-$                   \\
\hline\hline
\end{tabular}
\end{minipage}
\\[3ex]
%\centerline{(a)}
%\caption{$err_{b}$ and $err_{DA}$ as function of $nt_{obs}$ and $nSVs$ ($\Delta t=50.0$)} 
%\label{TabellaERRDA.dt-50.000}
%%\end{table}
%\end{sidewaystable}
%
%\begin{sidewaystable}
%%\begin{table}
%%\hspace{-2cm}
\begin{minipage}[t]{\textwidth}\small
\begin{tabular}{c||c||cccccc}
                      &                                                 & \multicolumn{6}{|c}{$nt_{obs}$} \\[1ex]\cline{3-8}
$\Delta t=100.0$      &                                                 & $1$                &         $2$        &         $4$        &         $6$        &         $8$        &         $10$ \\\hline\hline
\multicolumn{2}{c||}{$err^{\Delta t}_{b}$}       & 3.212934767372494e-02 & 3.146365882582471e-02 & 2.712139102136180e-02 & 3.057494708253965e-02 & 2.712141382551856e-02 & 3.057497214505261e-02 \\\hline\hline
\multirow{6}{*}{$err^{\Delta t}_{DA}$}   & $nSVs$&                       &                       &                       &                       &                       &                       \\\hline\hline
                              & $4$  & 3.212934767372495e-02 & 3.146365882582471e-02 & 2.712139102136180e-02 & 3.057494708253973e-02 & 2.712141382551856e-02 & 3.057497214505259e-02 \\
                              & $6$  & 3.212934767372496e-02 & 3.146365882582471e-02 & 2.712139102136180e-02 & $-$                   & 2.712141382551856e-02 & 3.057497214505262e-02 \\
                              & $8$  & 3.212934776080620e-02 & 3.146365882582471e-02 & 2.712139102136179e-02 & 3.057494706301675e-02 & 2.712144164470810e-02 & 3.057497214657582e-02 \\
                              & $10$  & 3.212934748665278e-02 & $-$                   & 2.712139102136160e-02 & 3.057494708253960e-02 & $-$                   & $-$                   \\
                              & $12$  & 3.212934758819253e-02 & 3.146365881283841e-02 & 2.712139102136179e-02 & 3.057494708234279e-02 & $-$                   & $-$                   \\
\hline\hline
\end{tabular}
\end{minipage}
\\[3ex]
%\centerline{(b)}
%\caption{$err_{b}$ and $err_{DA}$ as function of $nt_{obs}$ and $nSVs$ ($\Delta t=100.0$)} 
%\label{TabellaERRDA.dt-100.000}
%%\end{table}
%\end{sidewaystable}
%
%\begin{sidewaystable}
%%\begin{table}
%%\hspace{-2cm}
\begin{minipage}[t]{\textwidth}\small
\begin{tabular}{c||c||cccccc}
                      &                                                 & \multicolumn{6}{|c}{$nt_{obs}$} \\[1ex]\cline{3-8}
$\Delta t=150.0$      &                                                 & $1$                &         $2$        &         $4$        &         $6$        &         $8$        &         $10$ \\\hline\hline
\multicolumn{2}{c||}{$err^{\Delta t}_{b}$}       & 2.983871447550081e-02 & 2.728177997180958e-02 & 2.906807057721612e-02 & 3.146357791255923e-02 & 2.712133412057428e-02 & 3.057489714184252e-02 \\\hline\hline
\multirow{6}{*}{$err^{\Delta t}_{DA}$}   & $nSVs$&                    &                    &                    &                    &                    &                                      \\\hline\hline
                              & $4$  & 2.983871447550081e-02 & 2.728177997180958e-02 & 2.906807057721612e-02 & 3.146357791255922e-02 & 2.712133412057427e-02 & 3.057489714184253e-02 \\
                              & $6$  & 2.983871447550081e-02 & 2.728177997180958e-02 & 2.906807057721612e-02 & 3.146357791255922e-02 & $-$                   & 3.057489714184252e-02 \\
                              & $8$  & 2.983871447550081e-02 & 2.728177987323344e-02 & $-$                   & 3.146357791255923e-02 & 2.712133412057424e-02 & 3.057489714184251e-02 \\
                              & $10$  & 2.983871441410072e-02 & 2.728178024259760e-02 & 2.906807054678072e-02 & $-$                   & 2.712133412057428e-02 & 3.057489704777779e-02 \\
                              & $12$  & 2.983871463090574e-02 & 2.728178016178572e-02 & 2.906806976987517e-02 & $-$                   & 2.712133508936694e-02 & 3.057489710667452e-02 \\
\hline\hline
\end{tabular}
\end{minipage}
\\[3ex]
%\centerline{(c)}
%\caption{$err_{b}$ and $err_{DA}$ as function of $nt_{obs}$ and $nSVs$ ($\Delta t=150.0$)} 
%\label{TabellaERRDA.dt-150.000}
%%\end{table}
%\end{sidewaystable}

%\begin{sidewaystable}
%\begin{table}
%\hspace{-2cm}
\begin{minipage}[t]{\textwidth}\small
\begin{tabular}{c||c||cccccc}
                      &                                                 & \multicolumn{6}{|c}{$nt_{obs}$} \\[1ex]\cline{3-8}
$\Delta t=200.0$      &                                                 & $1$                &         $2$        &         $4$        &         $6$        &         $8$        &         $10$ \\\hline\hline
\multicolumn{2}{c||}{$err^{\Delta t}_{b}$}       & 3.212898155458577e-02 & 3.146332038022308e-02 & 2.712113568922558e-02 & 3.057469532393129e-02 & 3.057474522264301e-02 & 3.057479199381502e-02 \\\hline\hline
\multirow{6}{*}{$err^{\Delta t}_{DA}$}   & $nSVs$&                    &                    &                    &                    &                    &                            \\\hline\hline
                              & $4$  & 3.212898155458578e-02 & 3.146332038022308e-02 & 2.712113568922544e-02 & 3.057469532393129e-02 & 3.057474522264302e-02 & 3.057479199381503e-02 \\
                              & $6$  & 3.212898155458573e-02 & $-$                   & 2.712113568922558e-02 & 3.057469532852141e-02 & 3.057474522264301e-02 & 3.057479199381503e-02 \\
                              & $8$  & $-$                   & 3.146332038022306e-02 & 2.712113568922558e-02 & 3.057469537977613e-02 & 3.057474522264301e-02 & 3.057479199381503e-02 \\
                              & $10$  & $-$                   & 3.146332040002525e-02 & $-$                   & $-$                   & 3.057474522264301e-02 & $-$                   \\
                              & $12$  & $-$                   & 3.146332042802096e-02 & 2.712113568750191e-02 & 3.057469542330091e-02 & $-$                   & $-$                   \\
\hline\hline
\end{tabular}
\end{minipage}
%\centerline{(d)}
%\caption{$err_{b}$ and $err_{DA}$ as function of $nt_{obs}$ and $nSVs$ ($\Delta t=200.0$)} 
%\label{TabellaERRDA.dt-200.000}
\caption{$err^{\Delta t}_{b}$ and $err^{\Delta t}_{DA}$ as function of $nt_{obs}$ and $nSVs$ ($\Delta t=50.0,100.0,150.0,200.0$) - Problem 4}
\label{TabellaERRDA-Problem4}
%%\end{table}
\end{sidewaystable}

\item{\bf Tests Set 2}
In order to evaluate how the use of the Data Assimilation influences the model's performance, in figures  \ref{AndamentoErrADA.nSVs-04.nt_obs-10.dt-50.000},
\ref{AndamentoErrADA.nSVs-04.nt_obs-10.dt-100.000}, \ref{AndamentoErrADA.nSVs-04.nt_obs-10.dt-150.000} and \ref{AndamentoErrADA.nSVs-04.nt_obs-10.dt-200.000}
trends of  $$\| H \mathbf{M}_{\Delta t}^{0,M_{steps}=n}\left(x^{\Delta t}_{DA}\right) -_{\Delta t}x^{n+1}_{obs}\|_{2}/\left\|_{\Delta t}x^{n+1}_{obs}\right\|_{2}$$
and        $$\| H \mathbf{M}^{0,M_{steps}=n}\left(x^{\Delta t}_{b}\right) -_{\Delta t}x^{n+1}_{obs}\|_{2}/\left\|_{\Delta t}x^{n+1}_{obs}\right\|_{2}$$
 as functions of  $n=0,\ldots,nt_{obs}-1$ are showed 
(values of $nt_{obs}$ and $nSVs$ are fixed and their values are $nt_{obs}=10$ e $nSVs=4$, (a)-(d)-labeled plots are related with Problems 1-4 respectively).
It seems that the use of DA positively influences the model performance just for Problems 1 and 3: 
such influence is more significant if greater is the value of $\Delta t$.

\begin{figure}
\begin{center}
\begin{tabular}{cc}
\includegraphics[width=8cm]{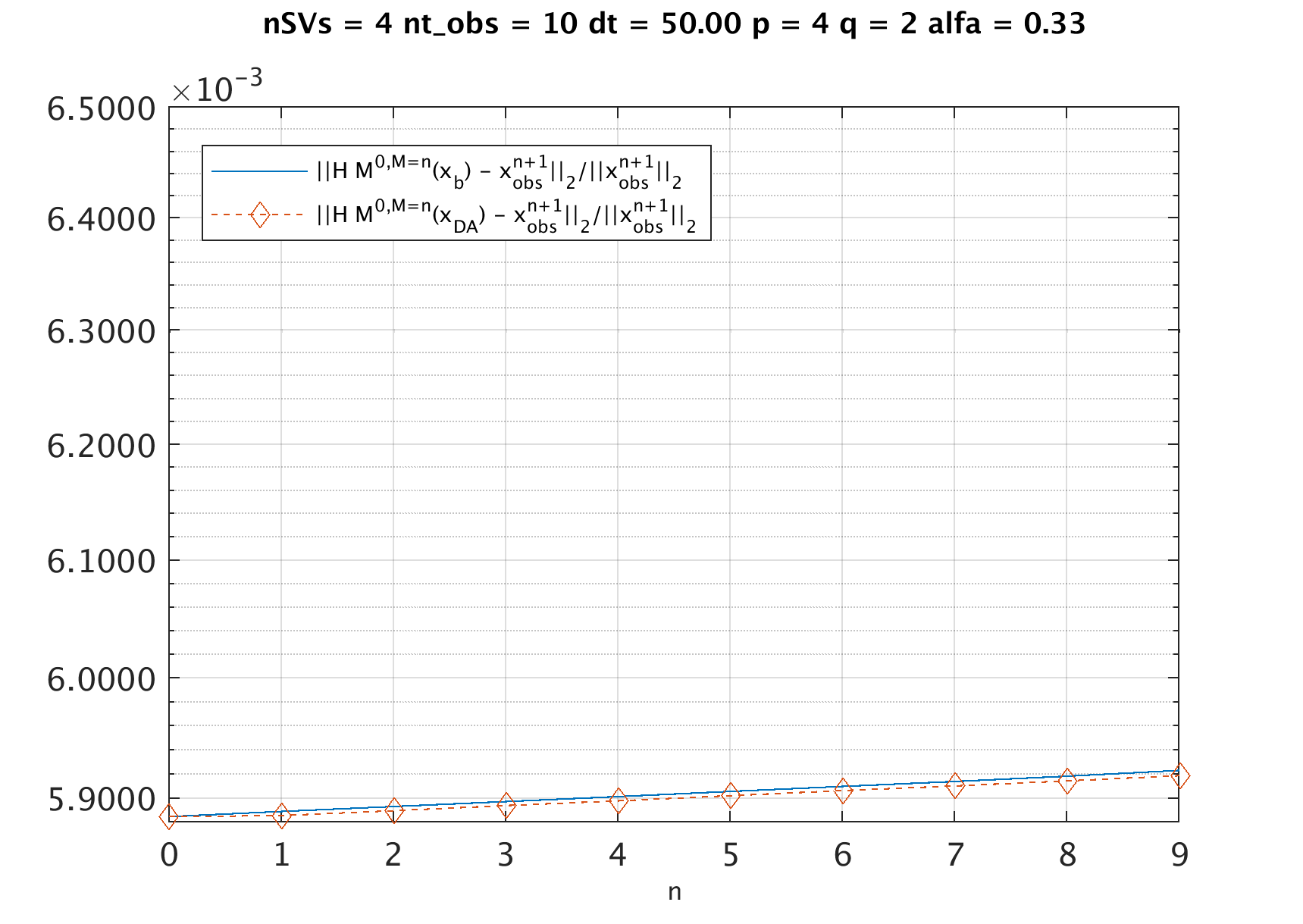}&
\includegraphics[width=8cm]{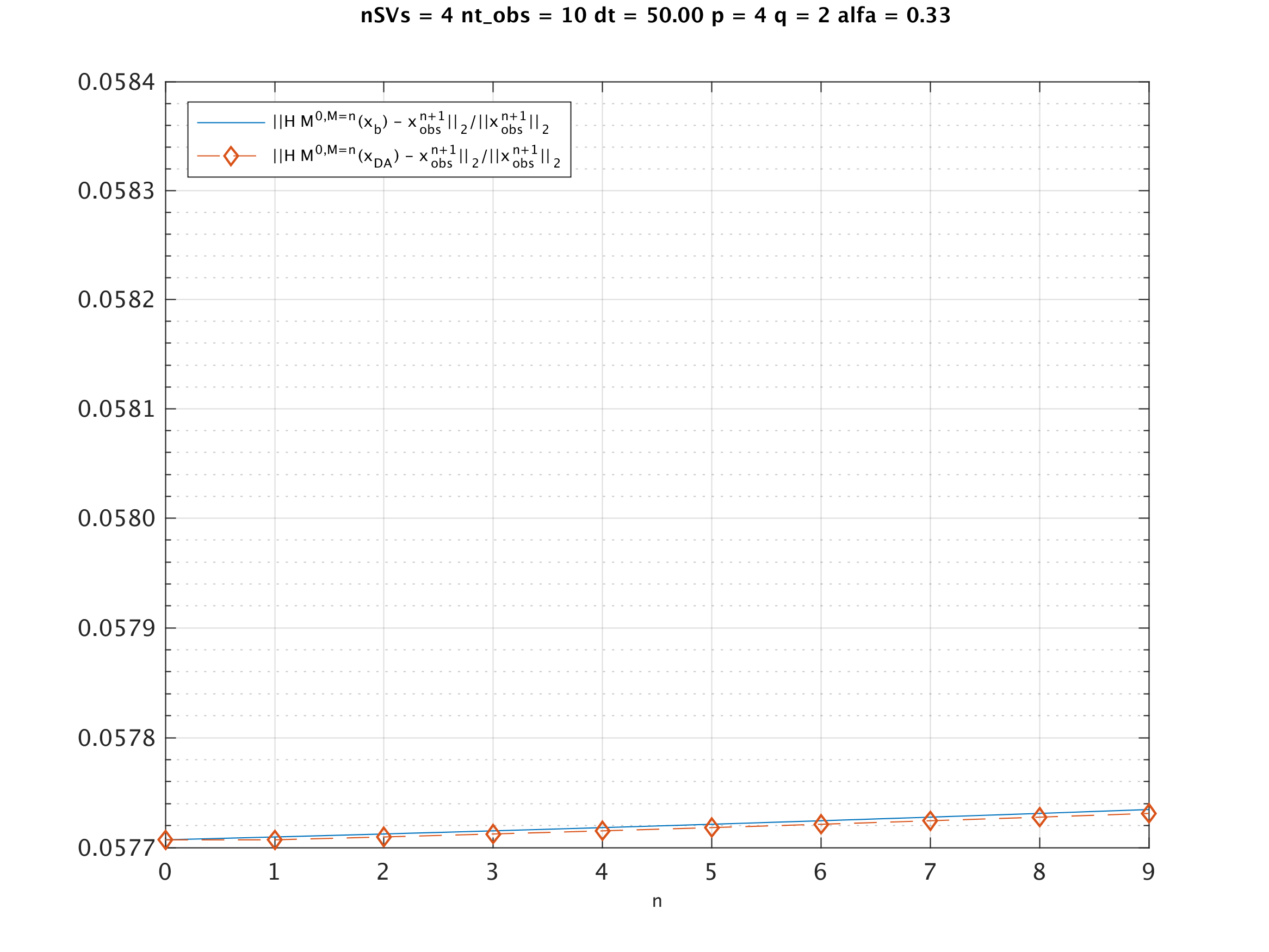} \\
(a) & (c) \\[4ex]
\includegraphics[width=8cm]{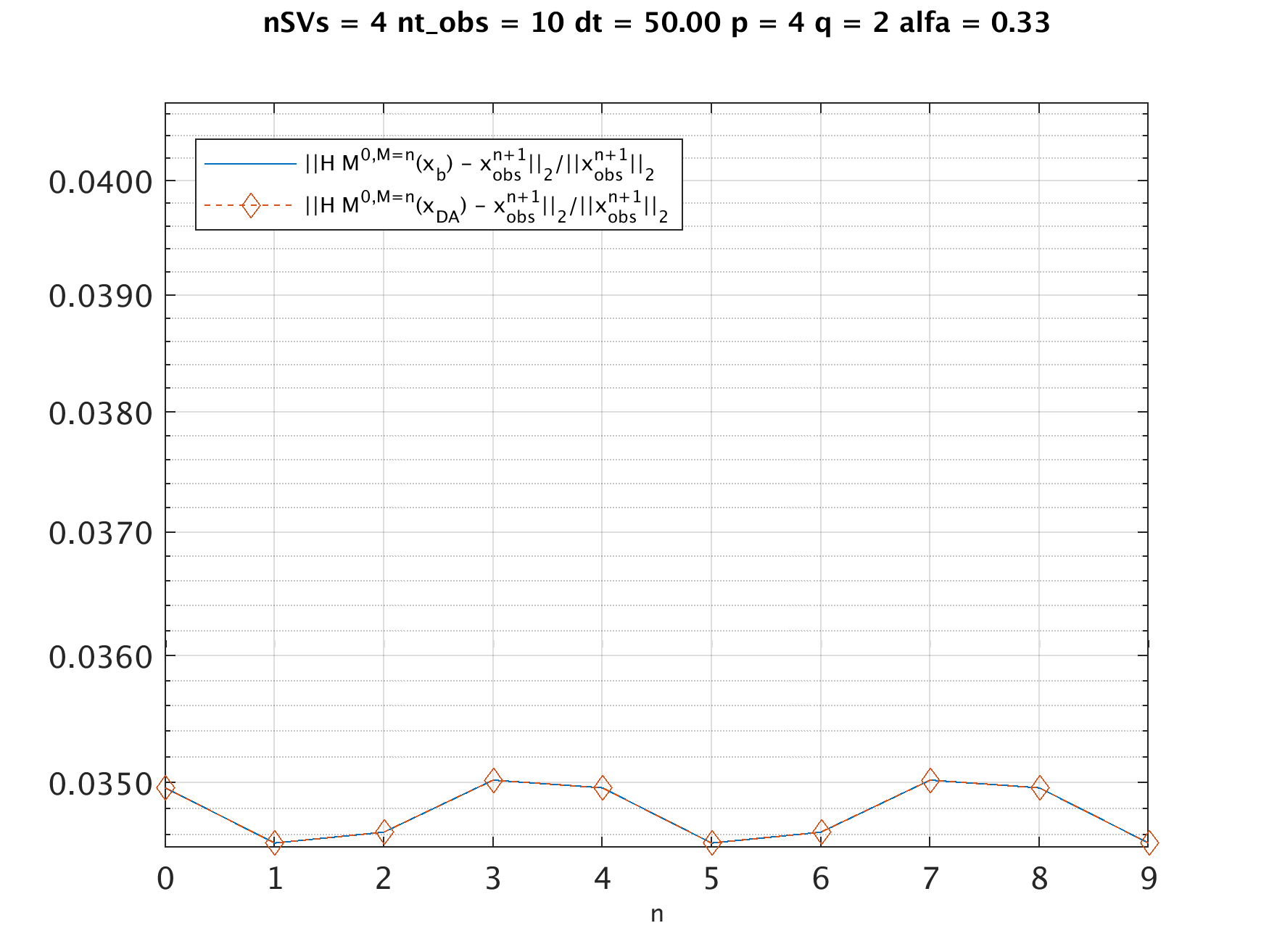} &
\includegraphics[width=8cm]{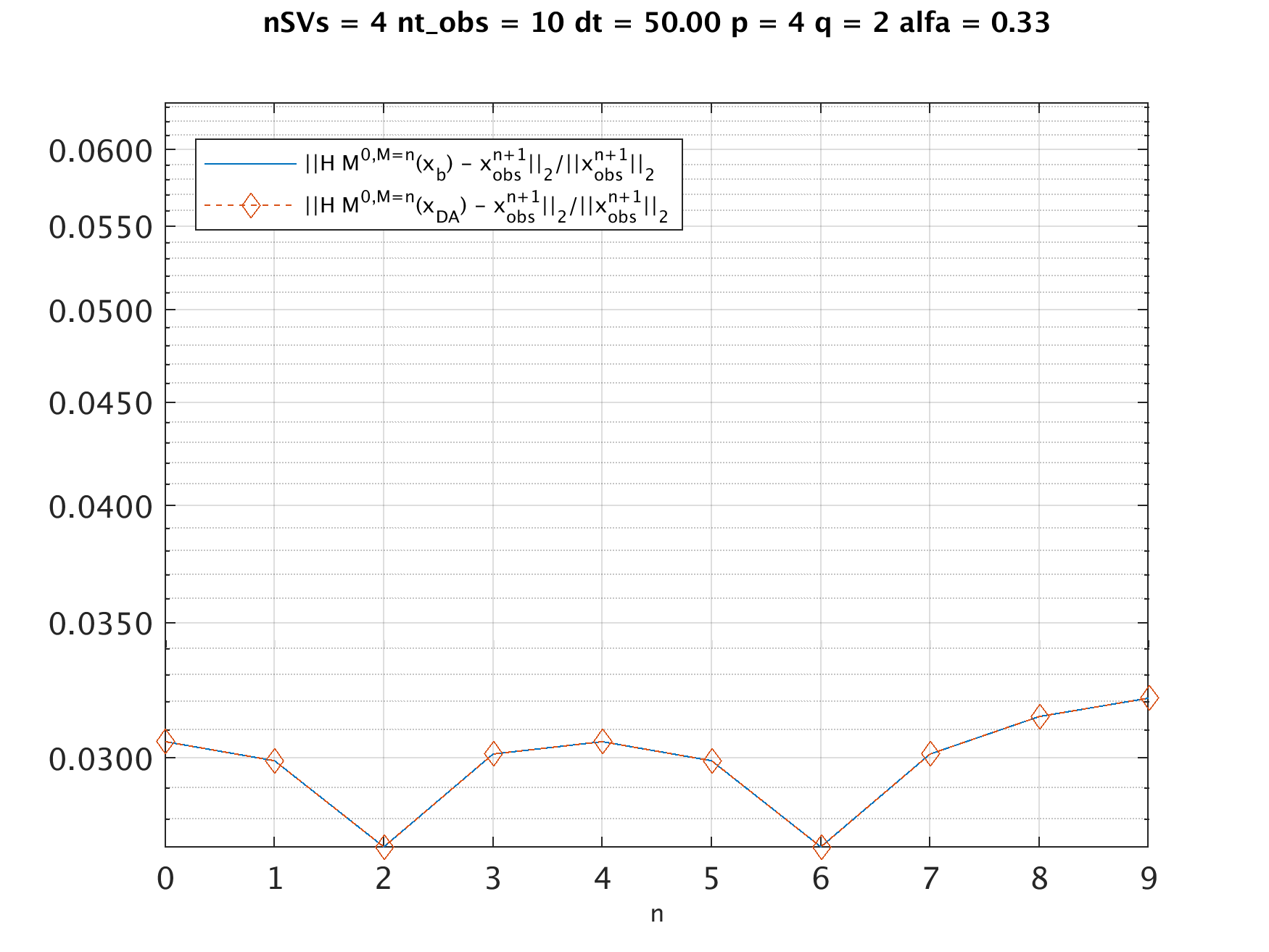} \\
(b) & (d) \\
\end{tabular}
\end{center}
\caption{\mbox{$\| H \mathbf{M}_{\Delta t}^{0,M_{steps}=n}\left(x^{\Delta t}_{DA}\right) -_{\Delta t}x^{n+1}_{obs}\|_{2}/\left\|_{\Delta t}x^{n+1}_{obs}\right\|_{2}$} 
and      \mbox{$\| H \mathbf{M}_{\Delta t}^{0,M_{steps}=n}\left(x^{\Delta t}_{b}\right)  -_{\Delta t}x^{n+1}_{obs}\|_{2}/\left\|_{\Delta t}x^{n+1}_{obs}\right\|_{2}$} 
as function of $n$ ($\Delta t=50.0$, $nt_{obs}=10$, $nSVs=4$, 
(a)-(d)-labeled plots are related with Problem 1-4 respectively)} 
\label{AndamentoErrADA.nSVs-04.nt_obs-10.dt-50.000}
\end{figure}

\begin{figure}
\begin{center}
\begin{tabular}{cc}
{\includegraphics[width=8cm]{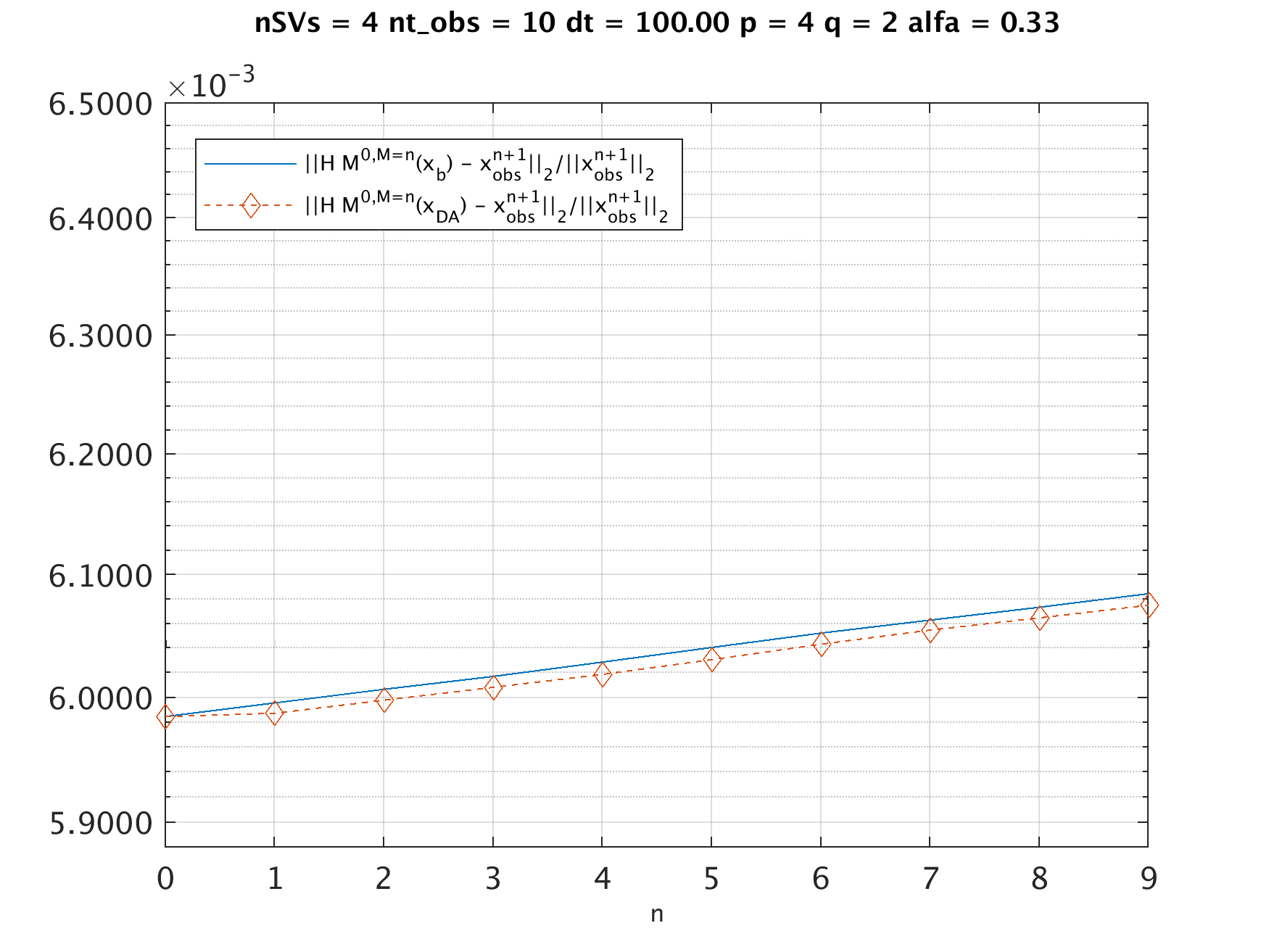}} &
{\includegraphics[width=8cm]{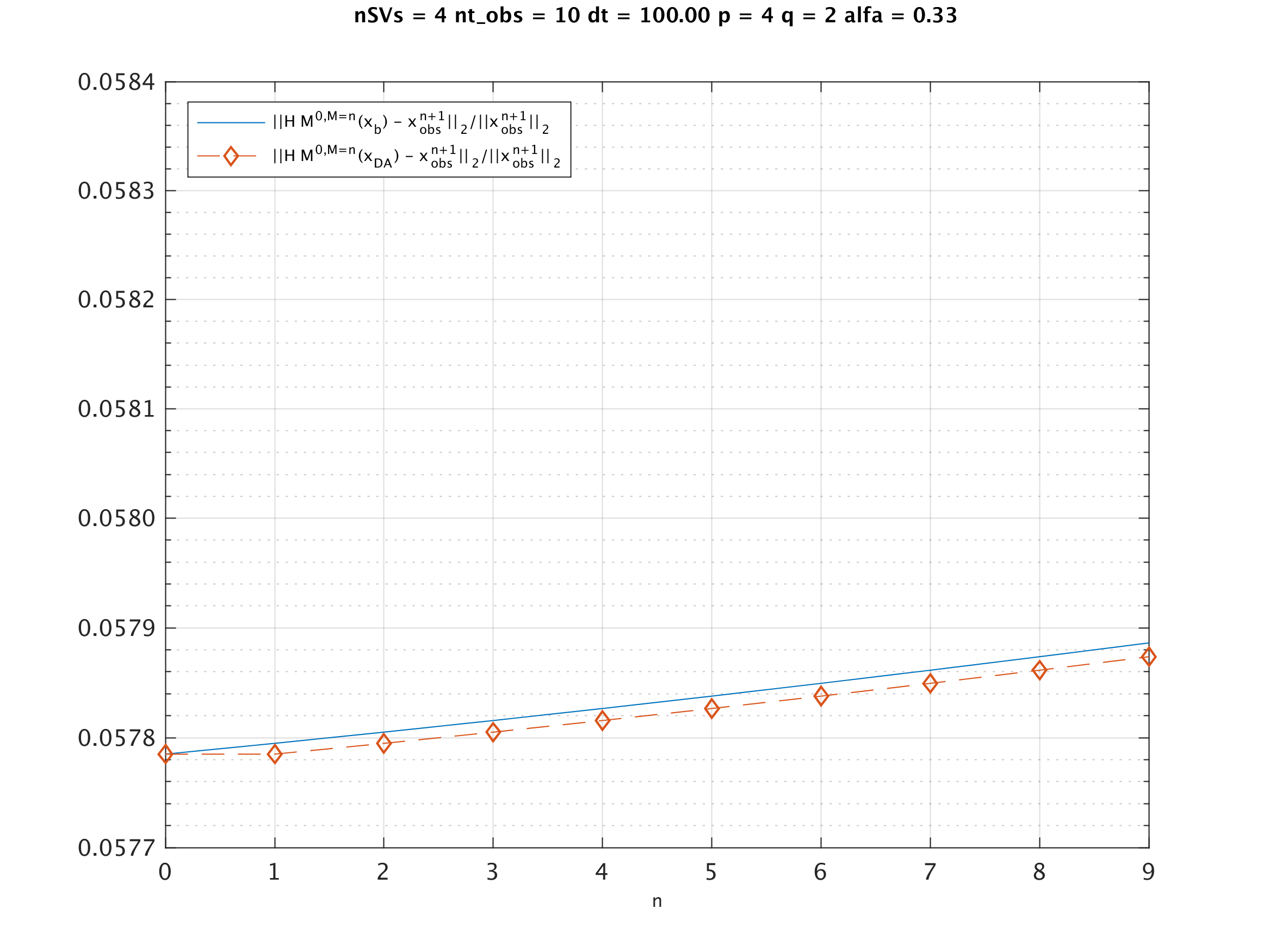}} \\
(a) & (c) \\[4ex]
{\includegraphics[width=8cm]{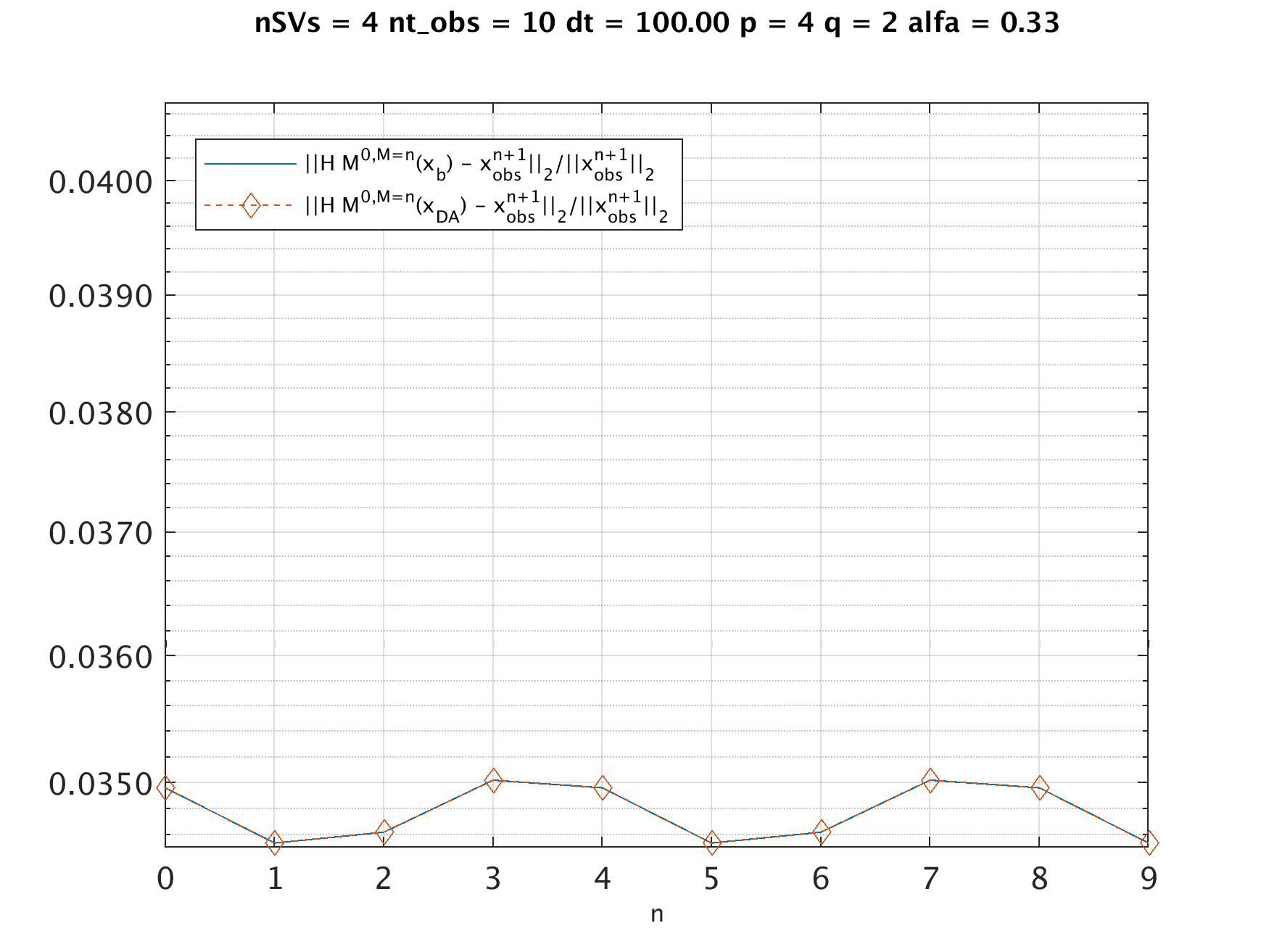}} &
{\includegraphics[width=8cm]{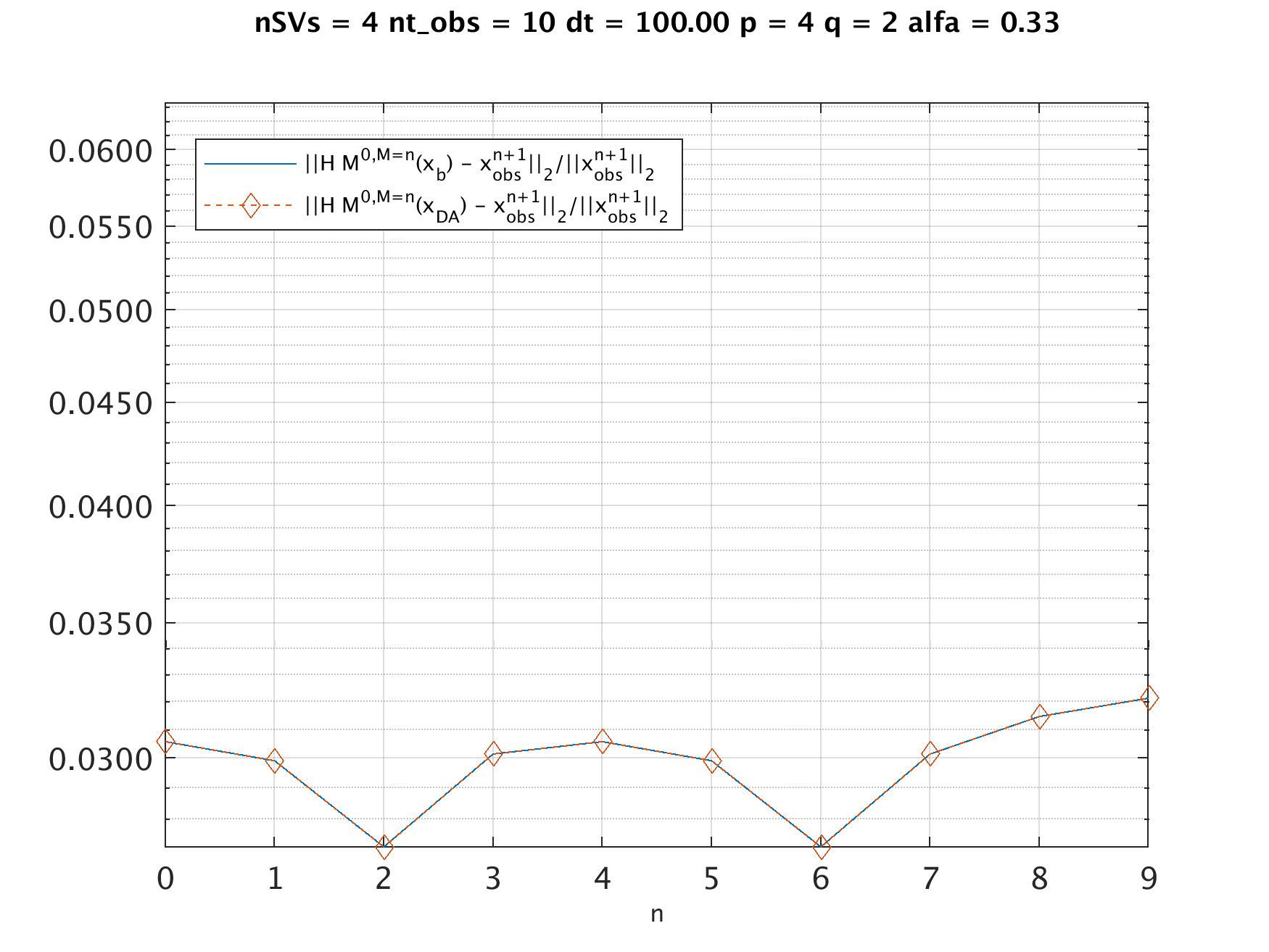}} \\
(b) & (d) \\
\end{tabular}
\end{center}
\caption{\mbox{$\| H \mathbf{M}_{\Delta t}^{0,M_{steps}=n}\left(x^{\Delta t}_{DA}\right) -_{\Delta t}x^{n+1}_{obs}\|_{2}/\left\|_{\Delta t}x^{n+1}_{obs}\right\|_{2}$} 
and      \mbox{$\| H \mathbf{M}_{\Delta t}^{0,M_{steps}=n}\left(x^{\Delta t}_{b}\right)  -_{\Delta t}x^{n+1}_{obs}\|_{2}/\left\|_{\Delta t}x^{n+1}_{obs}\right\|_{2}$} 
as function of $n$ ($\Delta t=100.0$, $nt_{obs}=10$, $nSVs=4$, 
(a)-(d)-labeled plots are related with Problem 1-4 respectively)} 
\label{AndamentoErrADA.nSVs-04.nt_obs-10.dt-100.000}
\end{figure}

\begin{figure}
\begin{center}
\begin{tabular}{cc}
{\includegraphics[width=8cm]{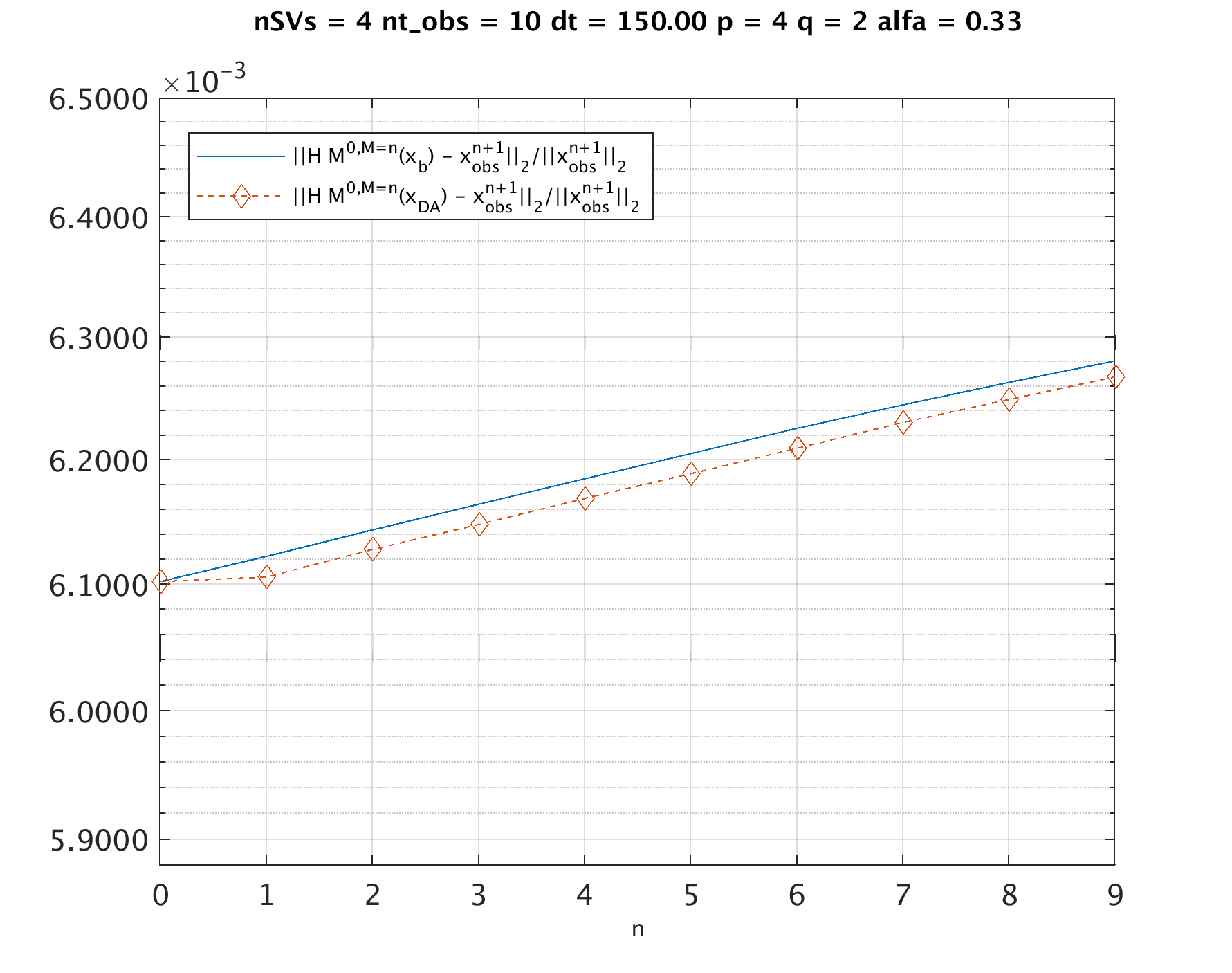}} &
{\includegraphics[width=8cm]{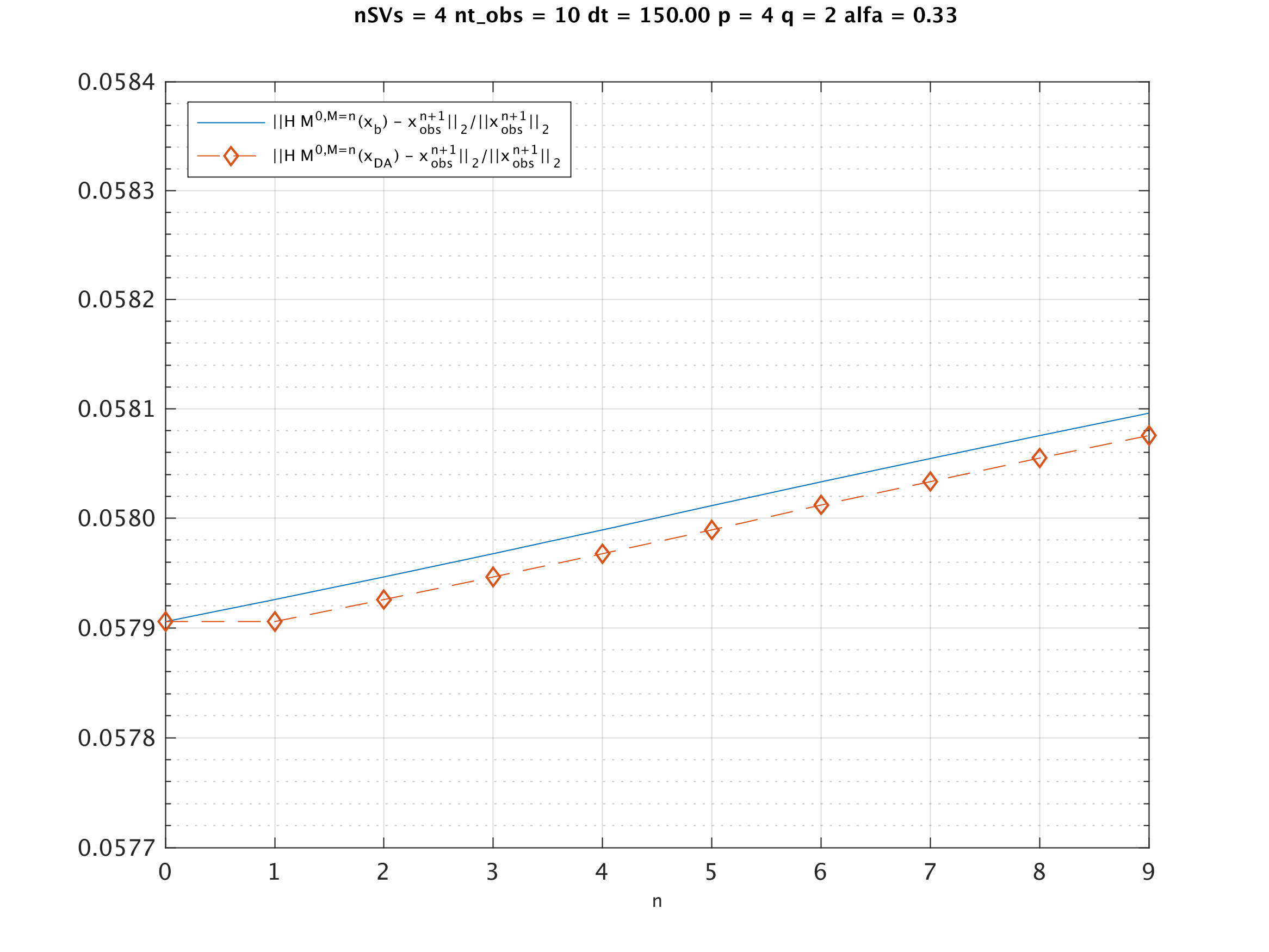}} \\
(a) & (c) \\[4ex]
{\includegraphics[width=8cm]{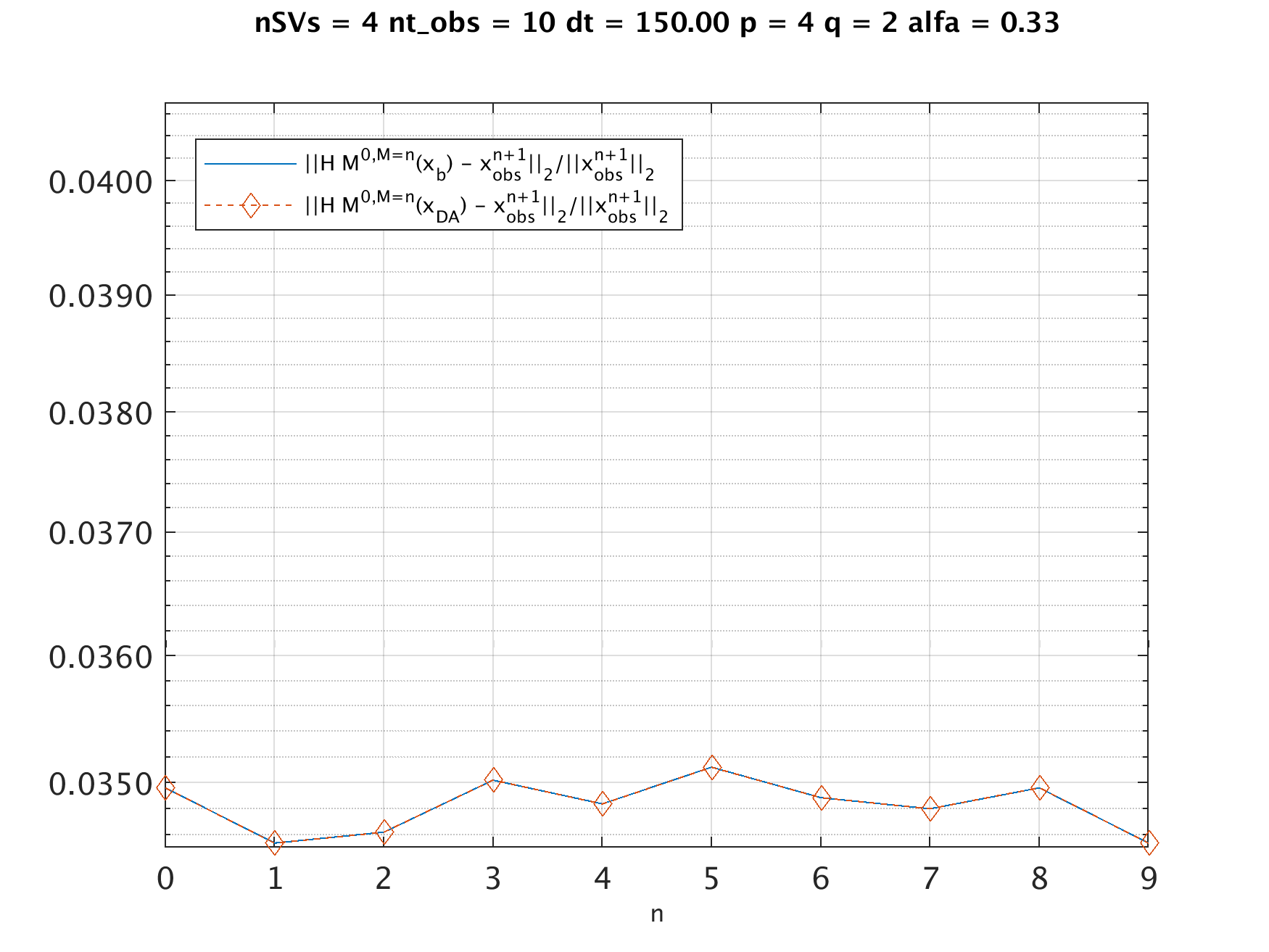}} &
{\includegraphics[width=8cm]{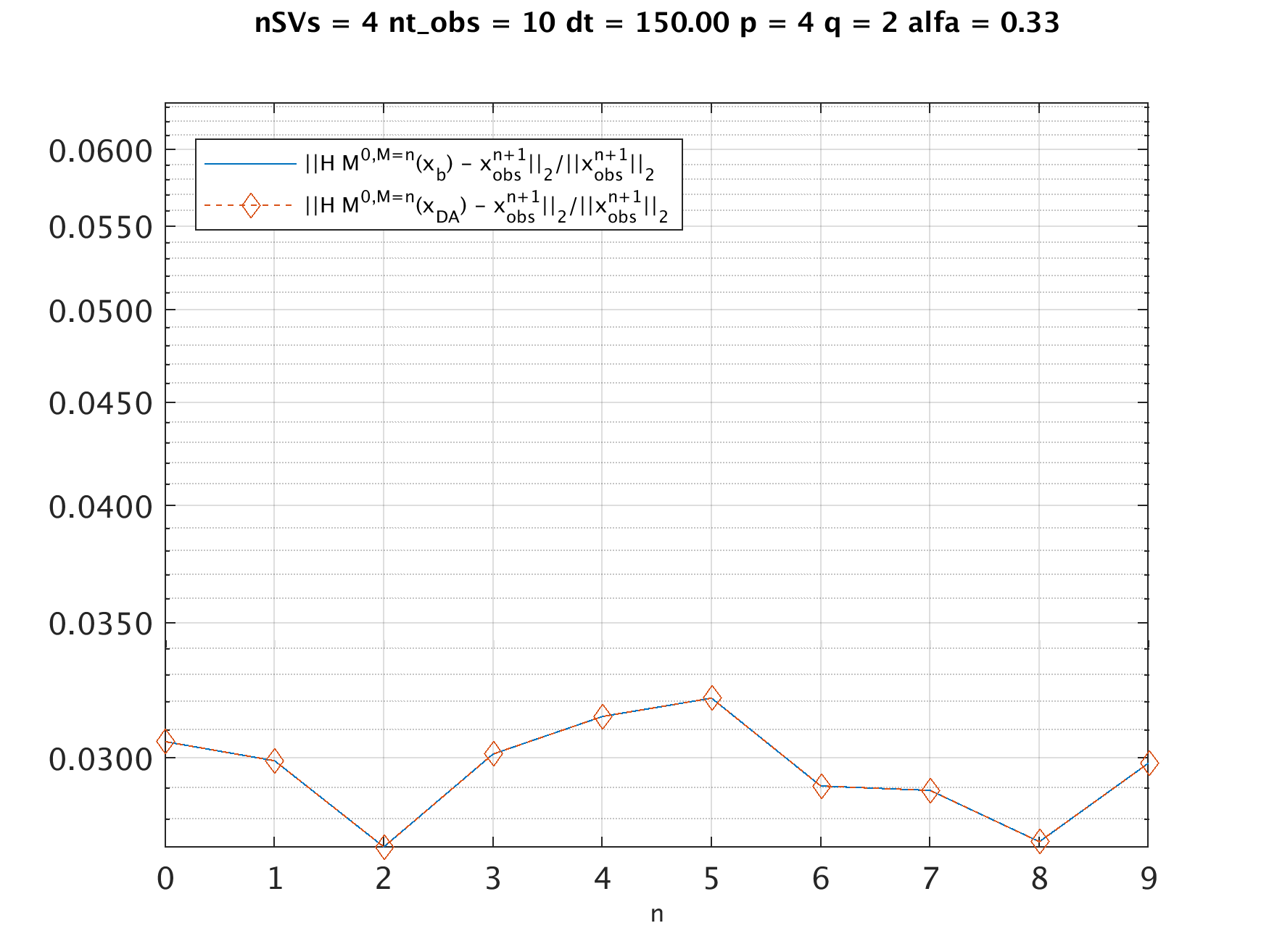}} \\
(b) & (d) \\
\end{tabular}
\end{center}
\caption{\mbox{$\| H \mathbf{M}_{\Delta t}^{0,M_{steps}=n}\left(x^{\Delta t}_{DA}\right) -_{\Delta t}x^{n+1}_{obs}\|_{2}/\left\|_{\Delta t}x^{n+1}_{obs}\right\|_{2}$} 
and      \mbox{$\| H \mathbf{M}_{\Delta t}^{0,M_{steps}=n}\left(x^{\Delta t}_{b}\right)  -_{\Delta t}x^{n+1}_{obs}\|_{2}/\left\|_{\Delta t}x^{n+1}_{obs}\right\|_{2}$} 
as function of $n$ ($\Delta t=150.0$, $nt_{obs}=10$, $nSVs=4$, 
(a)-(d)-labeled plots are related with Problem 1-4 respectively)} 
\label{AndamentoErrADA.nSVs-04.nt_obs-10.dt-150.000}
\end{figure}

\begin{figure}
\begin{center}
\begin{tabular}{cc}
{\includegraphics[width=8cm]{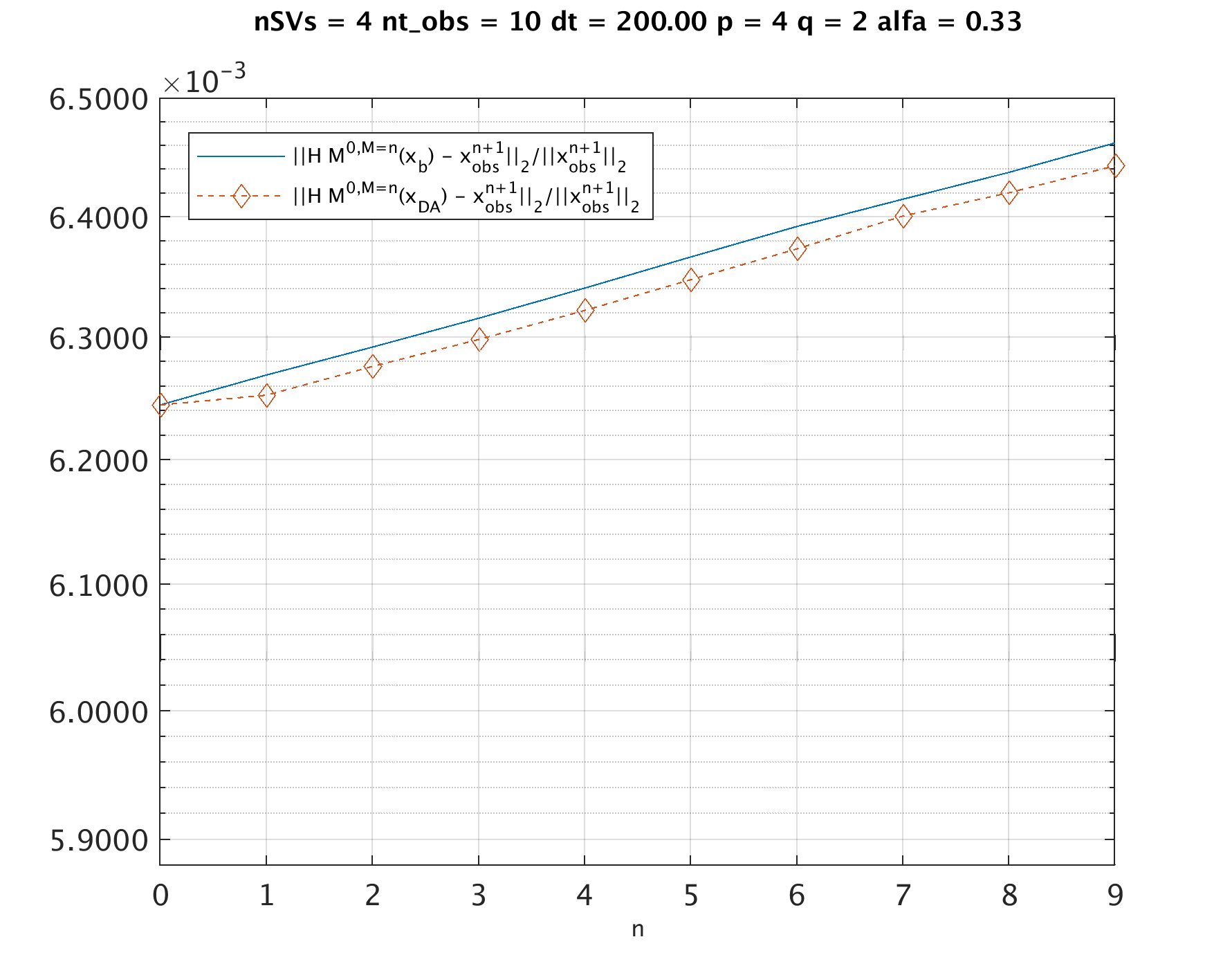}} &
{\includegraphics[width=8cm]{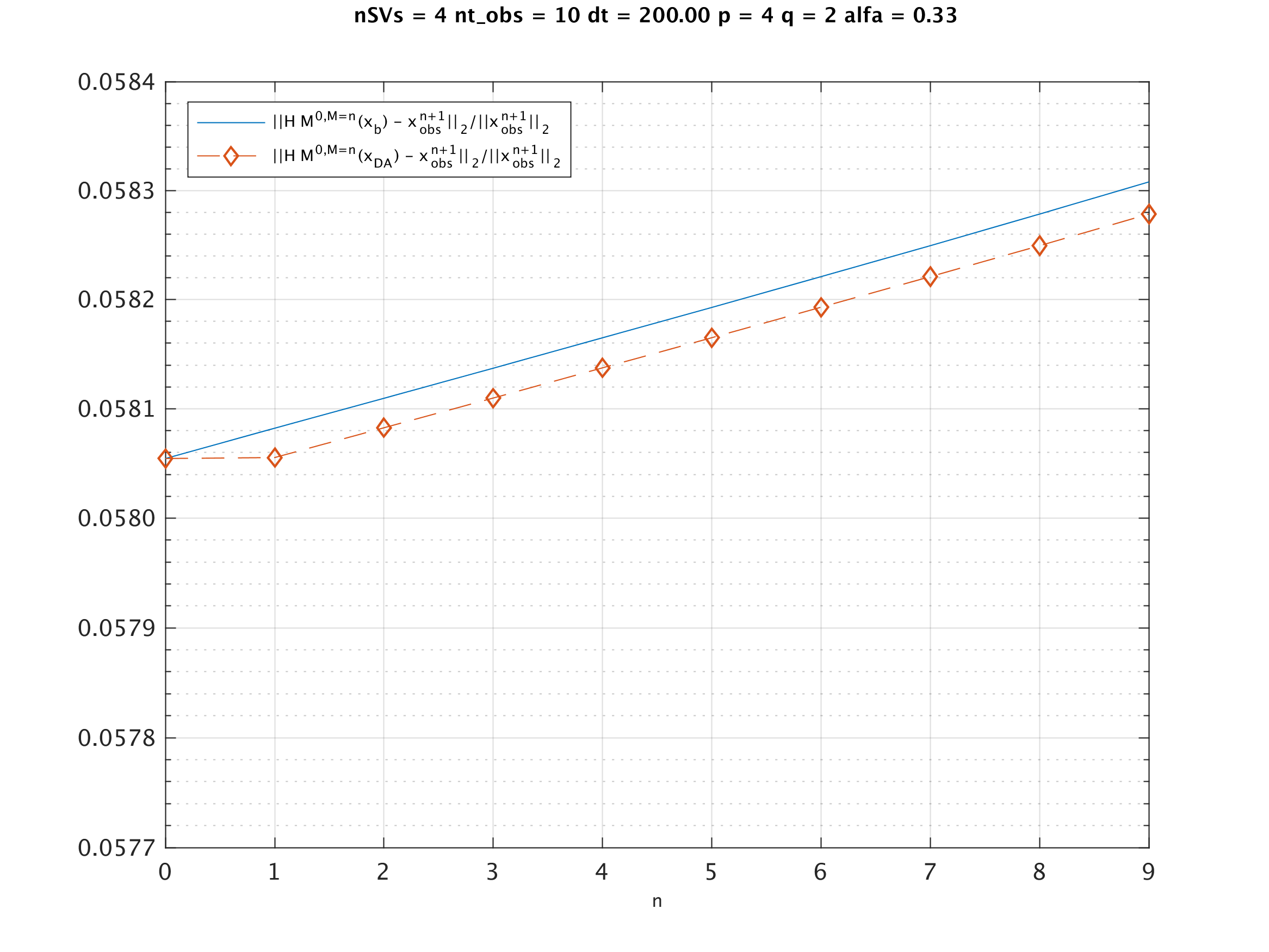}} \\
(a) & (c) \\[4ex]
{\includegraphics[width=8cm]{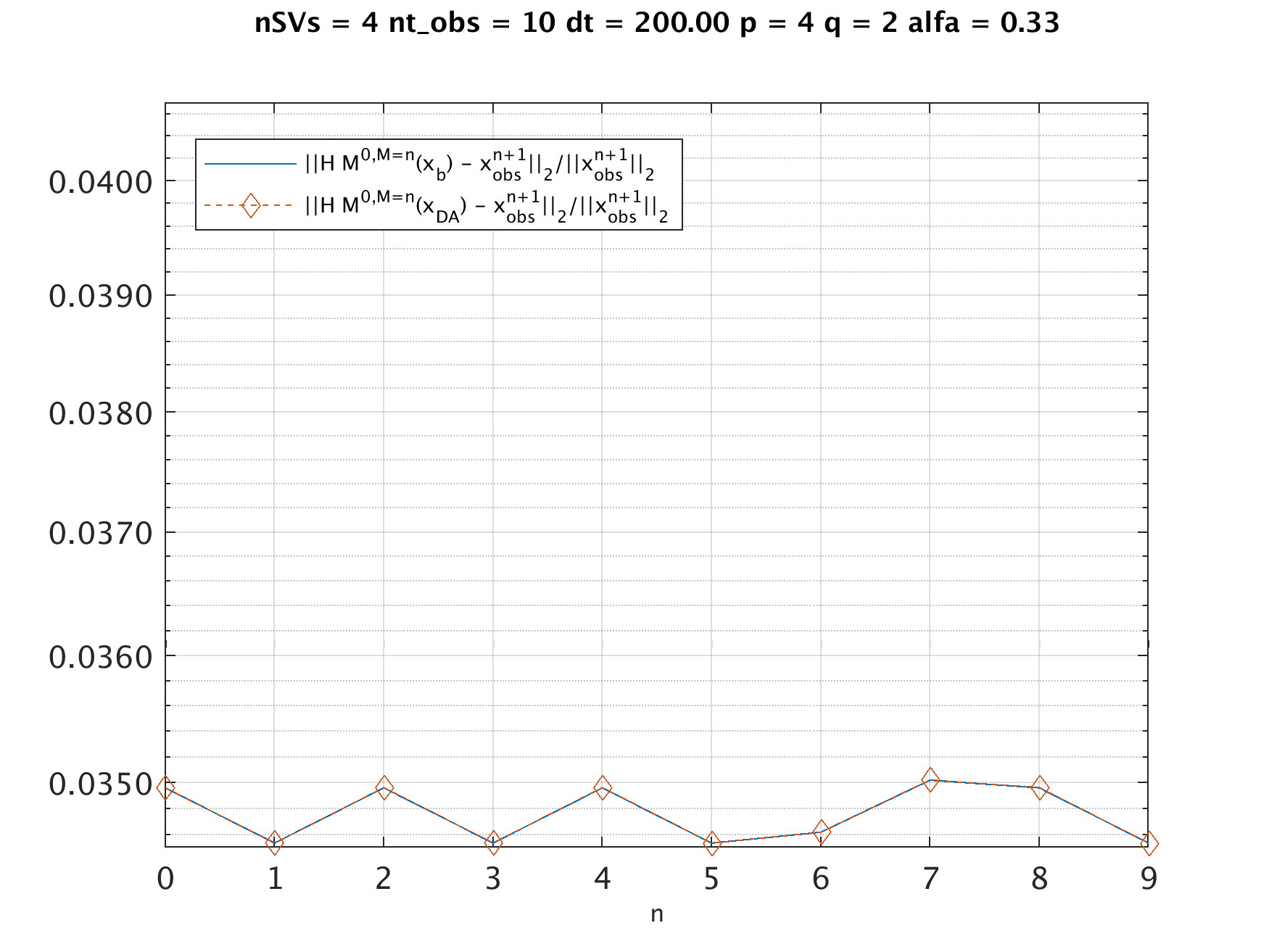}} &
{\includegraphics[width=8cm]{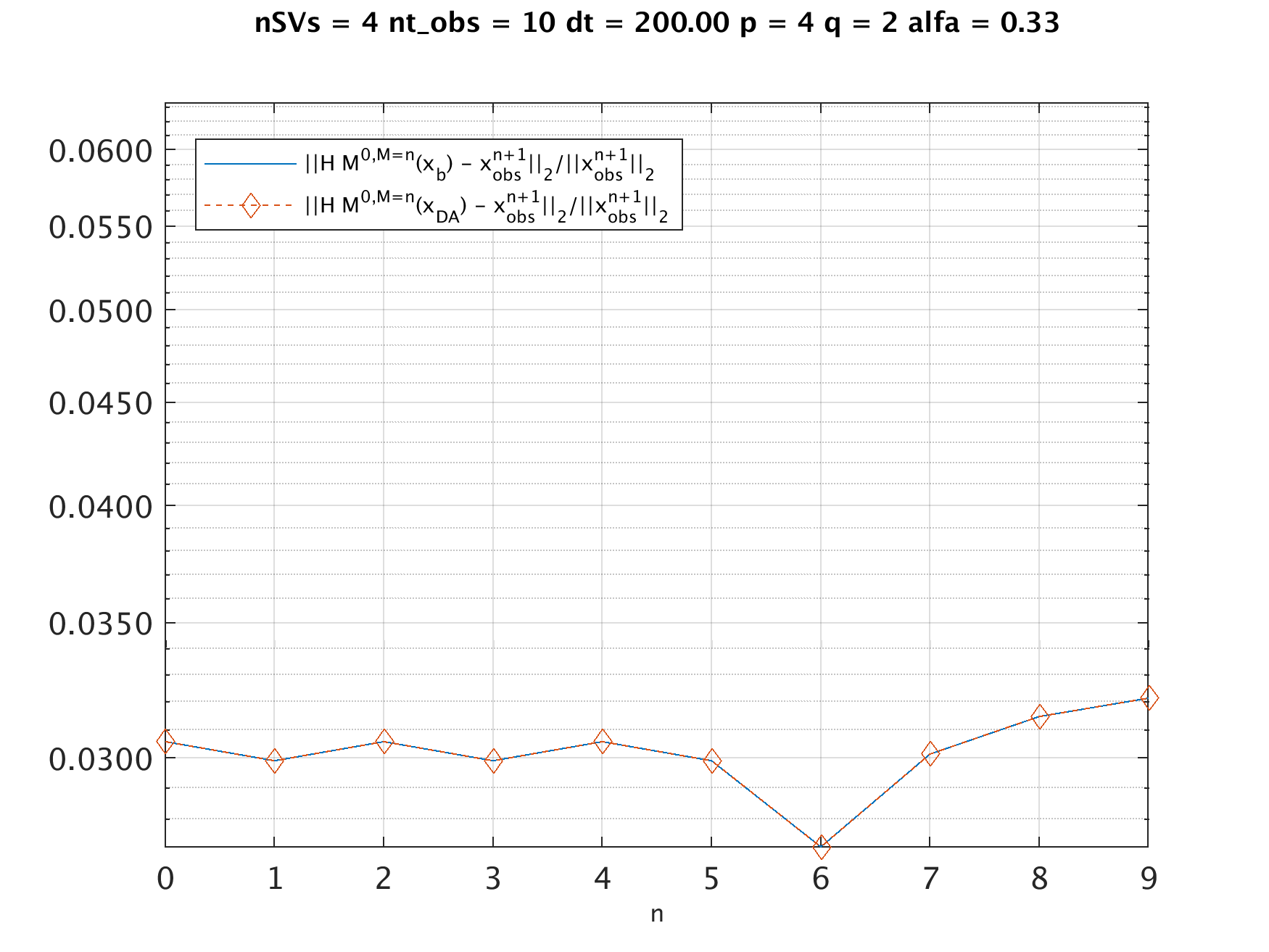}} \\
(b) & (d) \\
\end{tabular}
\end{center}
\caption{\mbox{$\| H \mathbf{M}_{\Delta t}^{0,M_{steps}=n}\left(x^{\Delta t}_{DA}\right) -_{\Delta t}x^{n+1}_{obs}\|_{2}/\left\|_{\Delta t}x^{n+1}_{obs}\right\|_{2}$} 
and      \mbox{$\| H \mathbf{M}_{\Delta t}^{0,M_{steps}=n}\left(x^{\Delta t}_{b}\right)  -_{\Delta t}x^{n+1}_{obs}\|_{2}/\left\|_{\Delta t}x^{n+1}_{obs}\right\|_{2}$} 
as function of $n$ ($\Delta t=200.0$, $nt_{obs}=10$, $nSVs=4$, 
(a)-(d)-labeled plots are related with Problem 1-4 respectively)} 
\label{AndamentoErrADA.nSVs-04.nt_obs-10.dt-200.000}
\end{figure}

\item{\bf Tests Set 3}
In order to evaluate how the use of 4DVAR approach (which use a number of observations vectors $nt_{obs}$ greater than $1$) influences the model's performance,
in figures \ref{ConfrontoErrADA.nSVs-04.dt-50.000},
\ref{ConfrontoErrADA.nSVs-04.dt-100.000}, \ref{ConfrontoErrADA.nSVs-04.dt-150.000} and \ref{ConfrontoErrADA.nSVs-04.dt-200.000},
trends of  $$\| H \mathbf{M}_{\Delta t}^{0,M_{steps}=n}\left(x^{\Delta t}_{DA}\right) -_{\Delta t}x^{n+1}_{obs}\|_{2}/\left\|_{\Delta t}x^{n+1}_{obs}\right\|_{2}$$ 
and $$\| H \mathbf{M}^{0,M_{steps}=n}\left(x^{\Delta t}_{b}\right) -_{\Delta t}x^{n+1}_{obs}\|_{2}/\left\|_{\Delta t}x^{n+1}_{obs}\right\|_{2}$$
as functions of $n=0,\ldots,nt_{obs}-1$ are compared when $nt_{obs}=2,6,10$ 
(the value of $nSVs$ is fixed and its value is $nSVs=4$, (a)-(d)-labeled plots are related with Problems 1-4 respectively). 
It seems that the use of a number of observations $nt_{obs}$ greater than $1$ doesn't significantly influence the model performance so it may not be convenient 
to use multiple observations (indeed, the computational cost of the DA algorithm increases with $nt_{obs}$).
\end{description}

\begin{figure}
\begin{center}
\begin{tabular}{cc}
{\includegraphics[width=8cm]{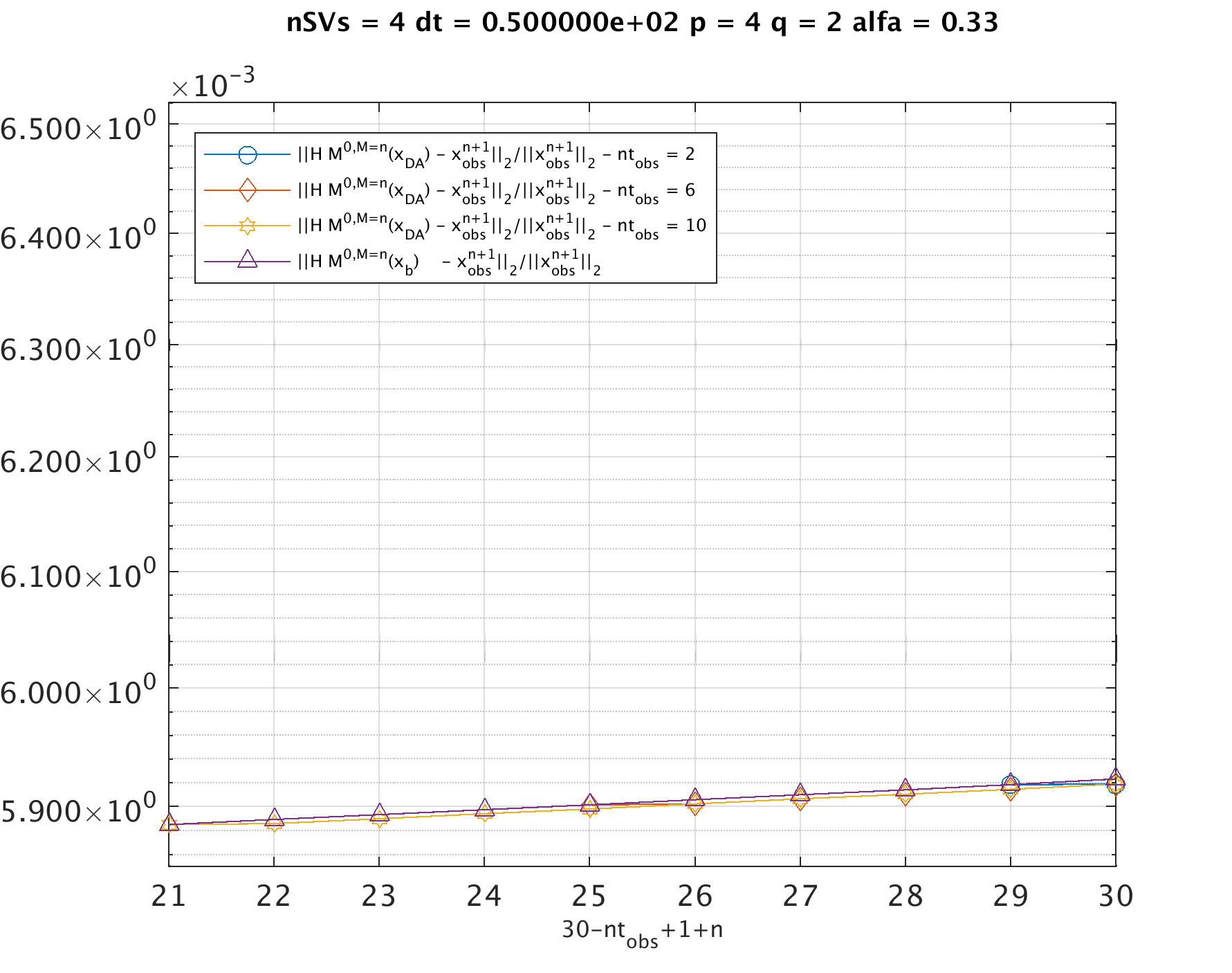}} &
{\includegraphics[width=8cm]{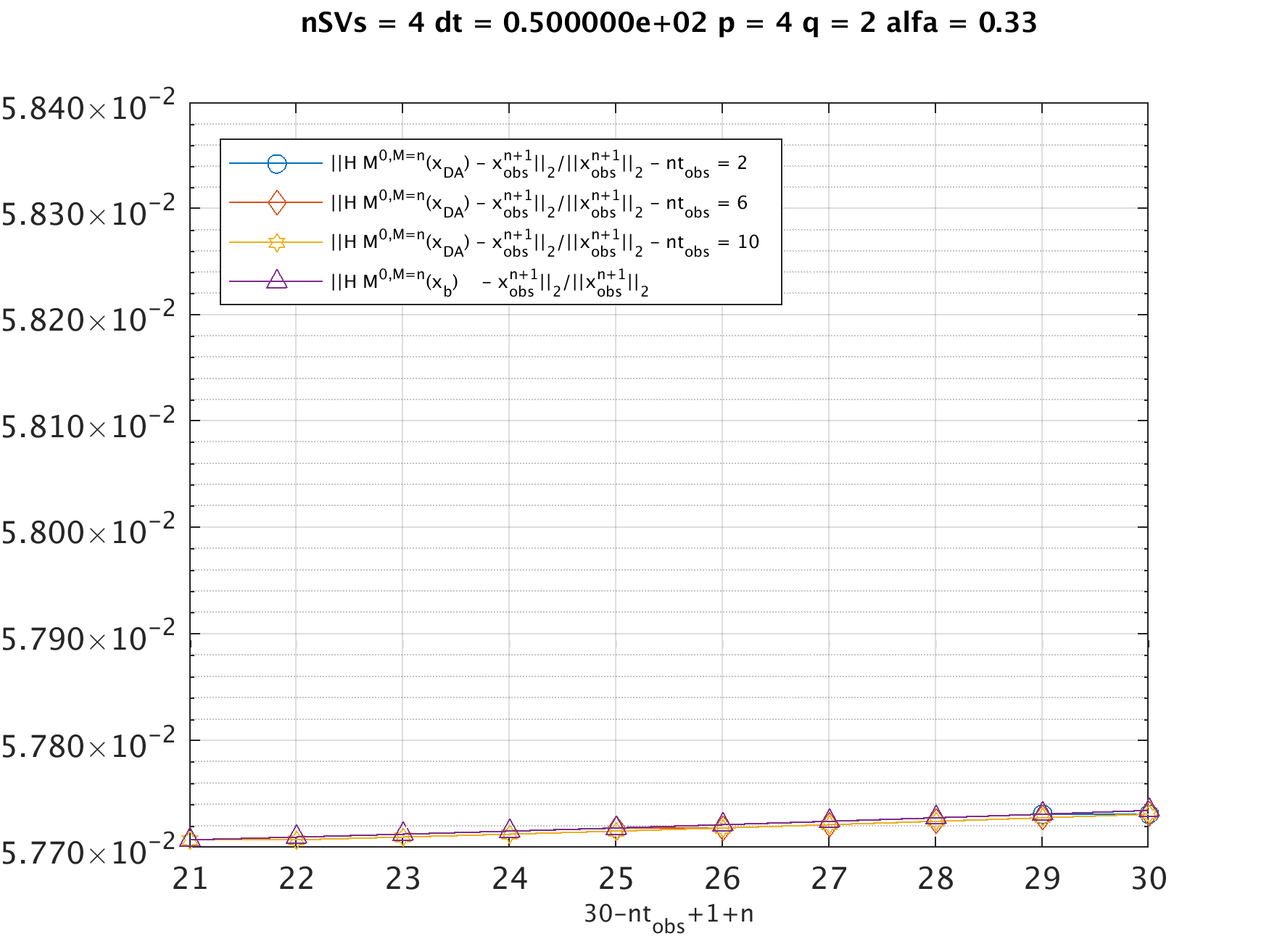}} \\
(a) & (c) \\[4ex]
{\includegraphics[width=8cm]{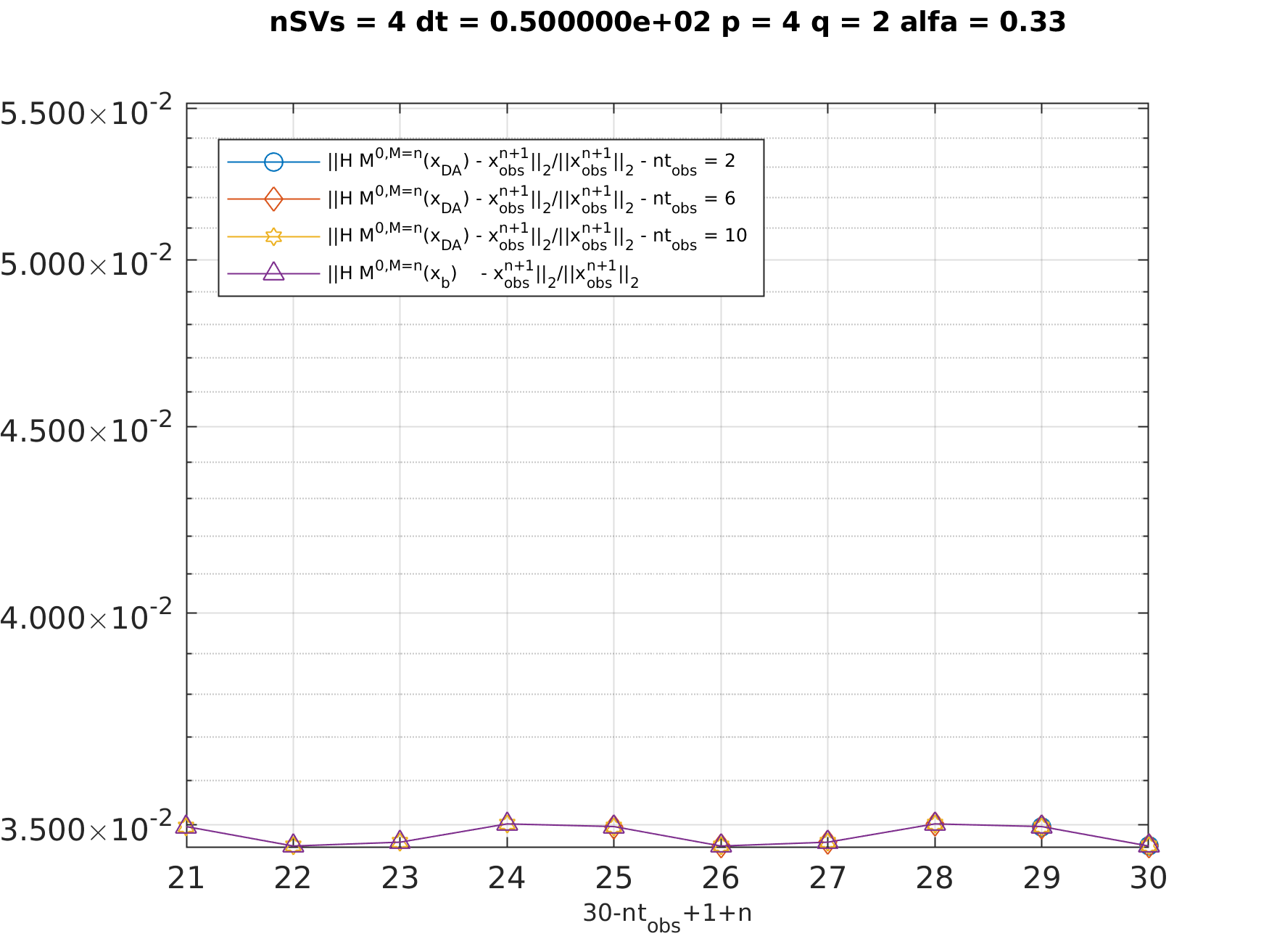}} &
{\includegraphics[width=8cm]{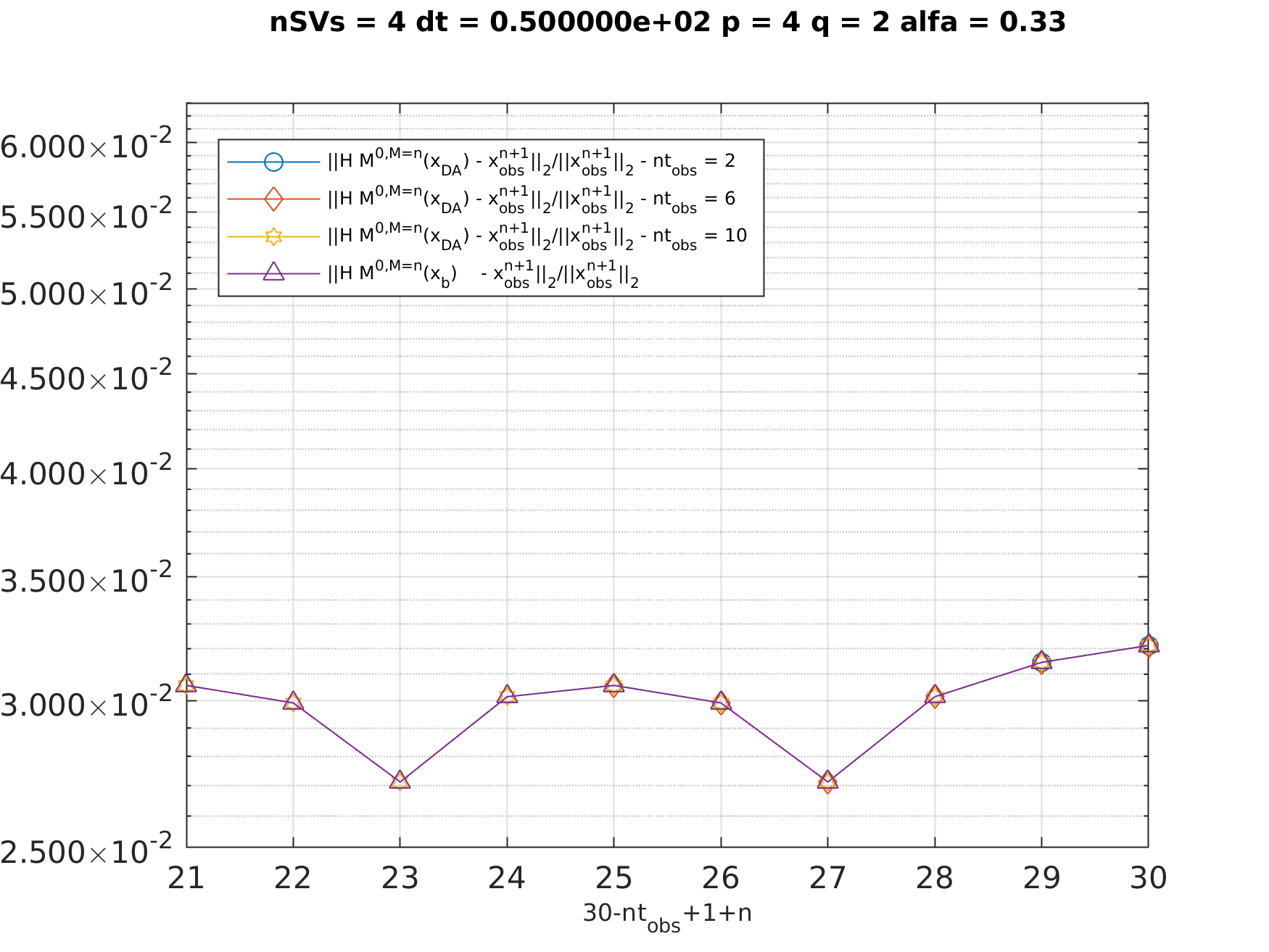}} \\
(b) & (d) \\
\end{tabular}
\end{center}
\caption{\mbox{$\| H \mathbf{M}_{\Delta t}^{0,M_{steps}=n}\left(x^{\Delta t}_{DA}\right) -_{\Delta t}x^{n+1}_{obs}\|_{2}/\left\|_{\Delta t}x^{n+1}_{obs}\right\|_{2}$}
and      \mbox{$\| H \mathbf{M}_{\Delta t}^{0,M_{steps}=n}\left(x^{\Delta t}_{b}\right)  -_{\Delta t}x^{n+1}_{obs}\|_{2}/\left\|_{\Delta t}x^{n+1}_{obs}\right\|_{2}$}
as function of $n=0,\ldots,nt_{obs}-1$ ($\Delta t=50.0$, $nt_{obs}=2,6,10$, $nSVs=4$, 
(a)-(b)-labeled plots are related with Problem 1-4 respectively)} 
\label{ConfrontoErrADA.nSVs-04.dt-50.000}
\end{figure}

\begin{figure}
\begin{center}
\begin{tabular}{cc}
{\includegraphics[width=8cm]{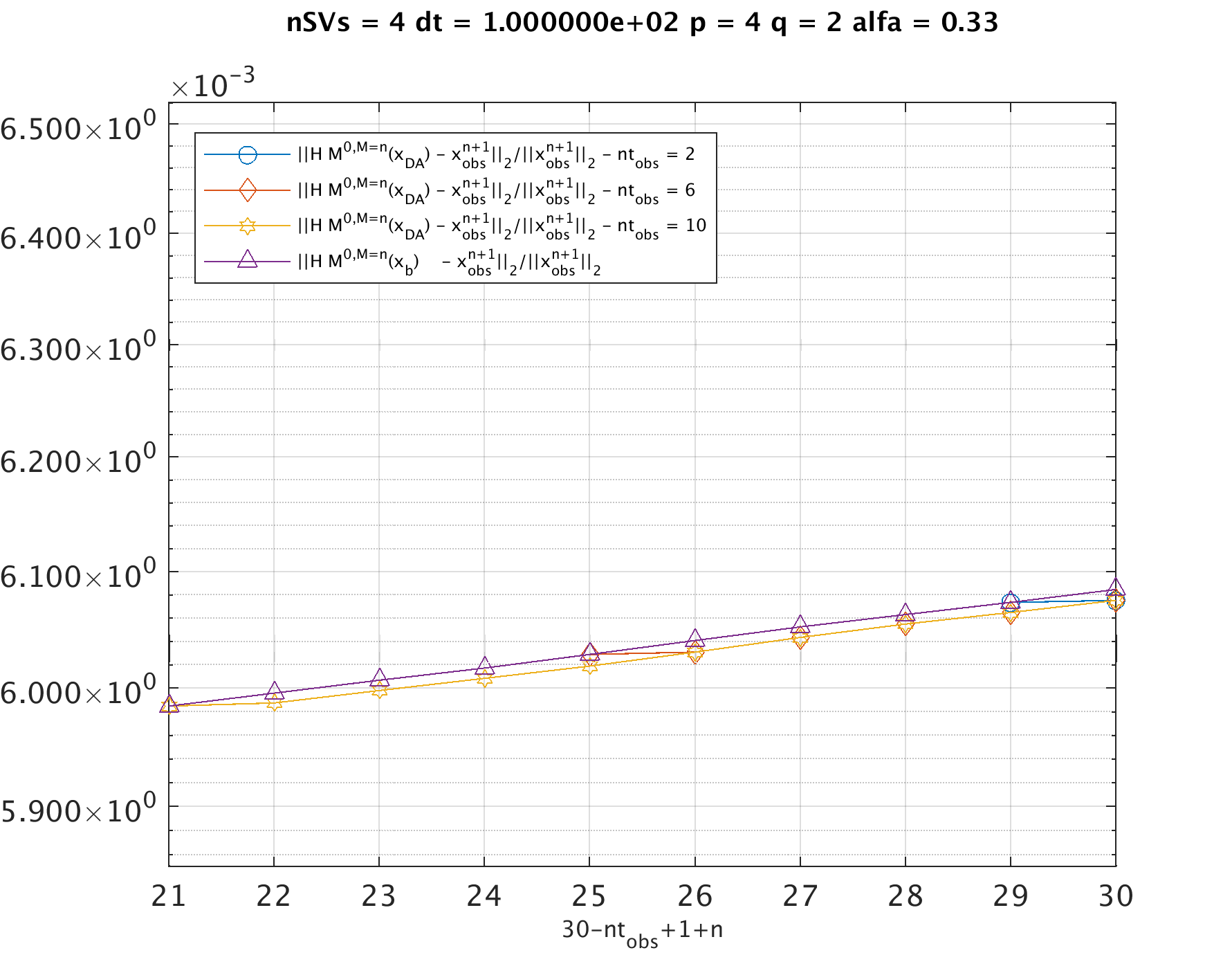}} &
{\includegraphics[width=8cm]{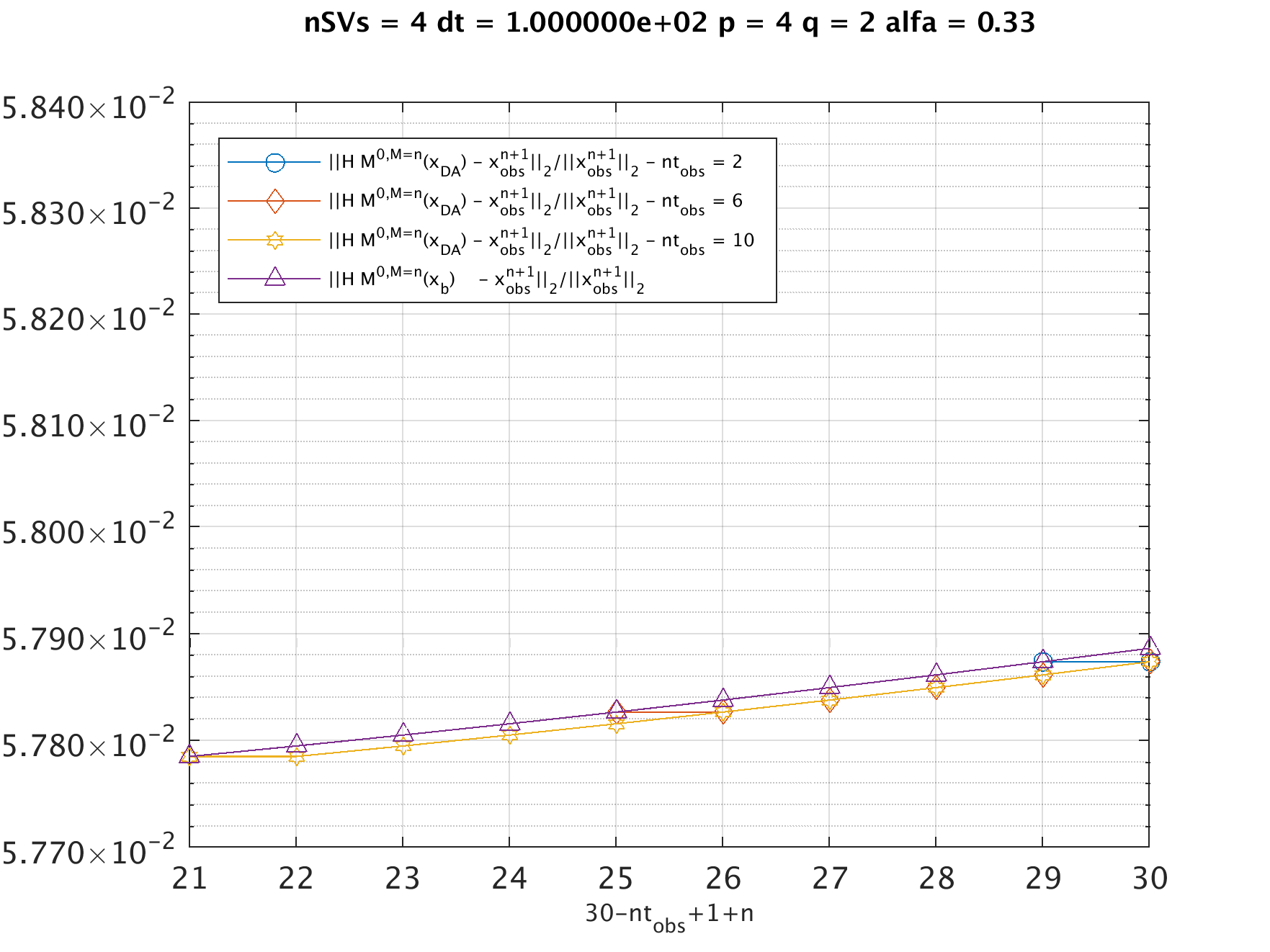}} \\
(a) & (c) \\[4ex]
{\includegraphics[width=8cm]{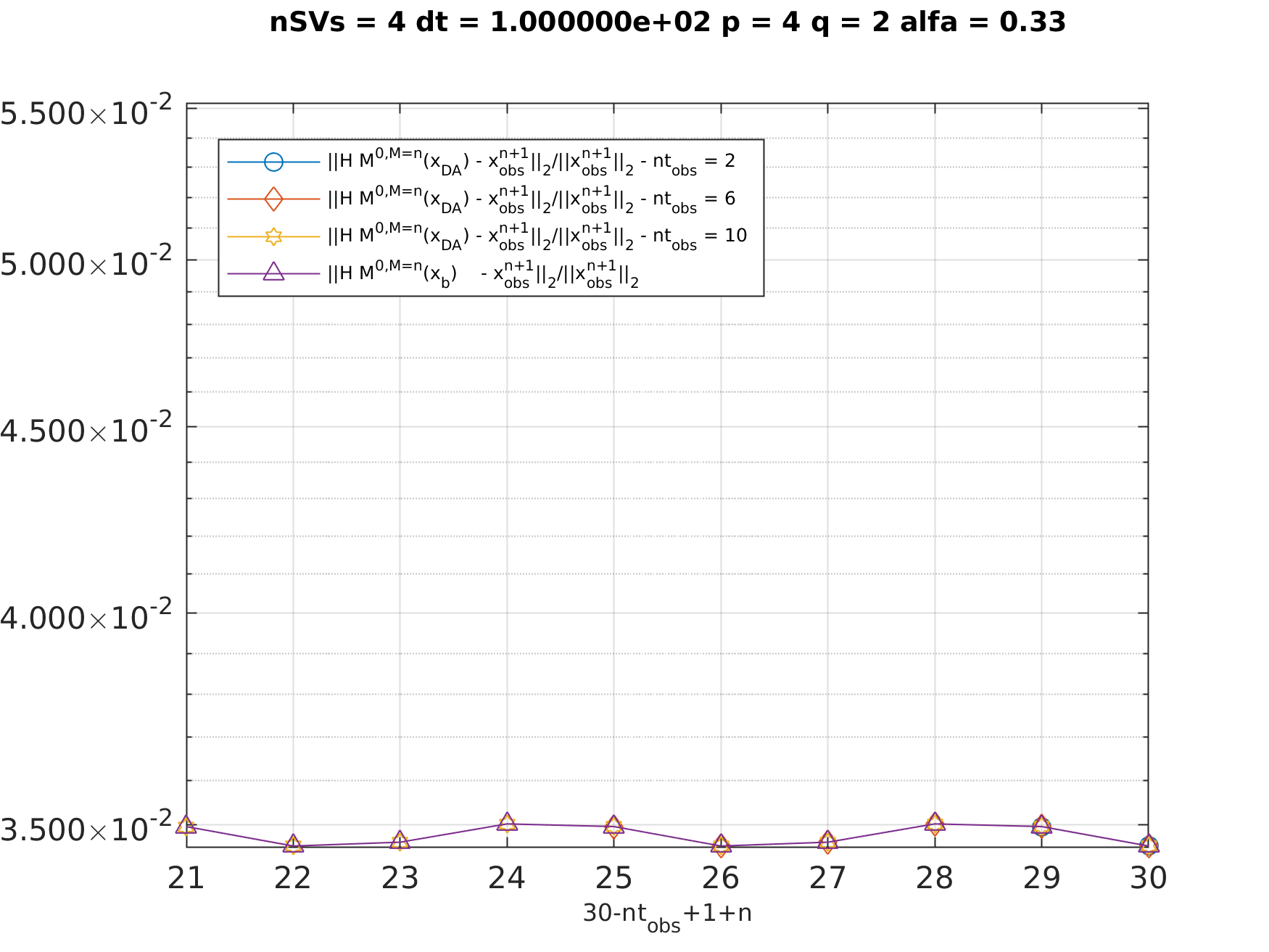}} &
{\includegraphics[width=8cm]{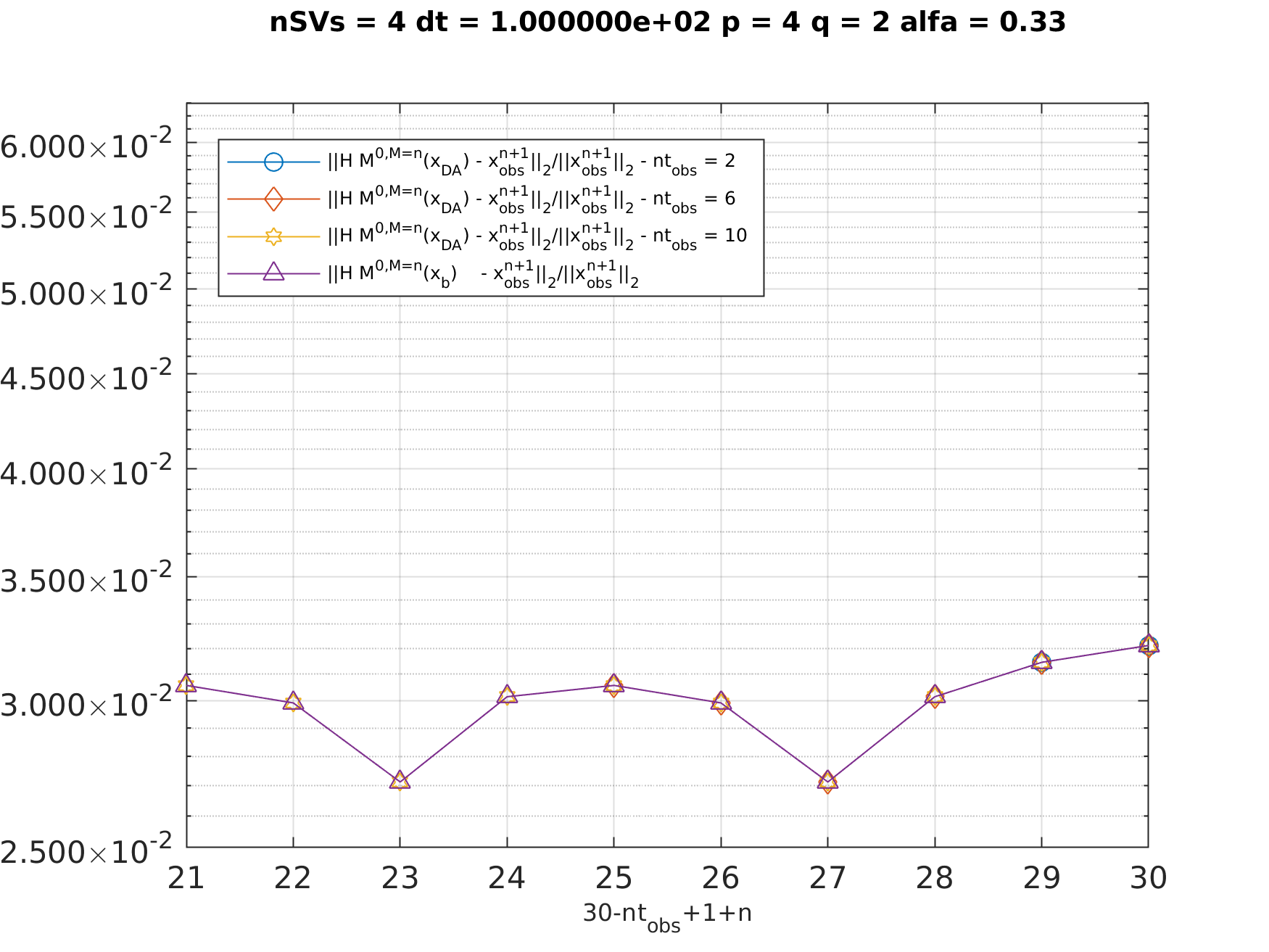}} \\
(b) & (d) \\
\end{tabular}
\end{center}
\caption{\mbox{$\| H \mathbf{M}_{\Delta t}^{0,M_{steps}=n}\left(x^{\Delta t}_{DA}\right) -_{\Delta t}x^{n+1}_{obs}\|_{2}/\left\|_{\Delta t}x^{n+1}_{obs}\right\|_{2}$}
and      \mbox{$\| H \mathbf{M}_{\Delta t}^{0,M_{steps}=n}\left(x^{\Delta t}_{b}\right)  -_{\Delta t}x^{n+1}_{obs}\|_{2}/\left\|_{\Delta t}x^{n+1}_{obs}\right\|_{2}$}
as function of $n=0,\ldots,nt_{obs}-1$ ($\Delta t=100.0$, $nt_{obs}=2,6,10$, $nSVs=4$,
(a)-(b)-labeled plots are related with Problem 1-4 respectively)} 
\label{ConfrontoErrADA.nSVs-04.dt-100.000}
\end{figure}

\begin{figure}
\begin{center}
\begin{tabular}{cc}
{\includegraphics[width=8cm]{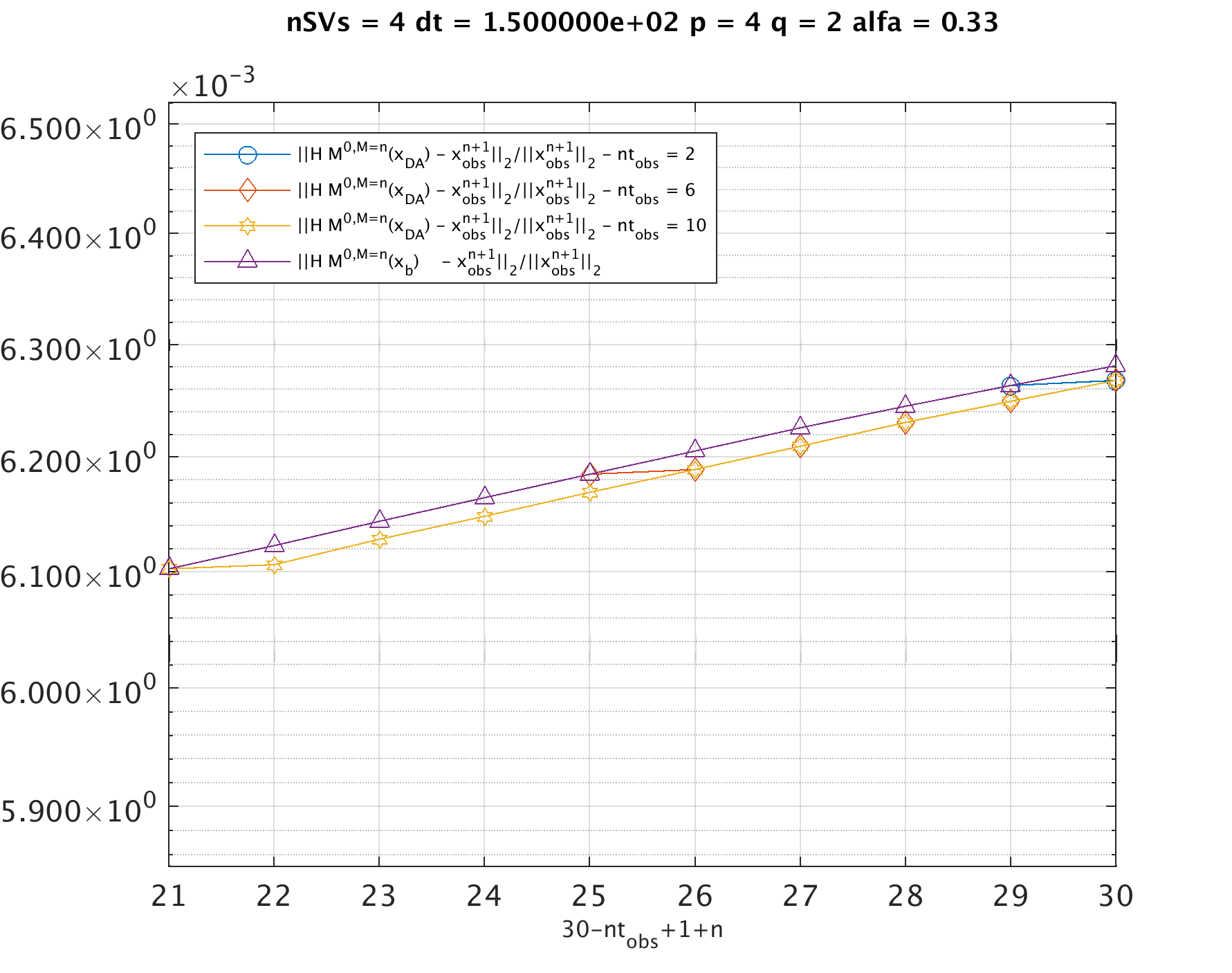}} &
{\includegraphics[width=8cm]{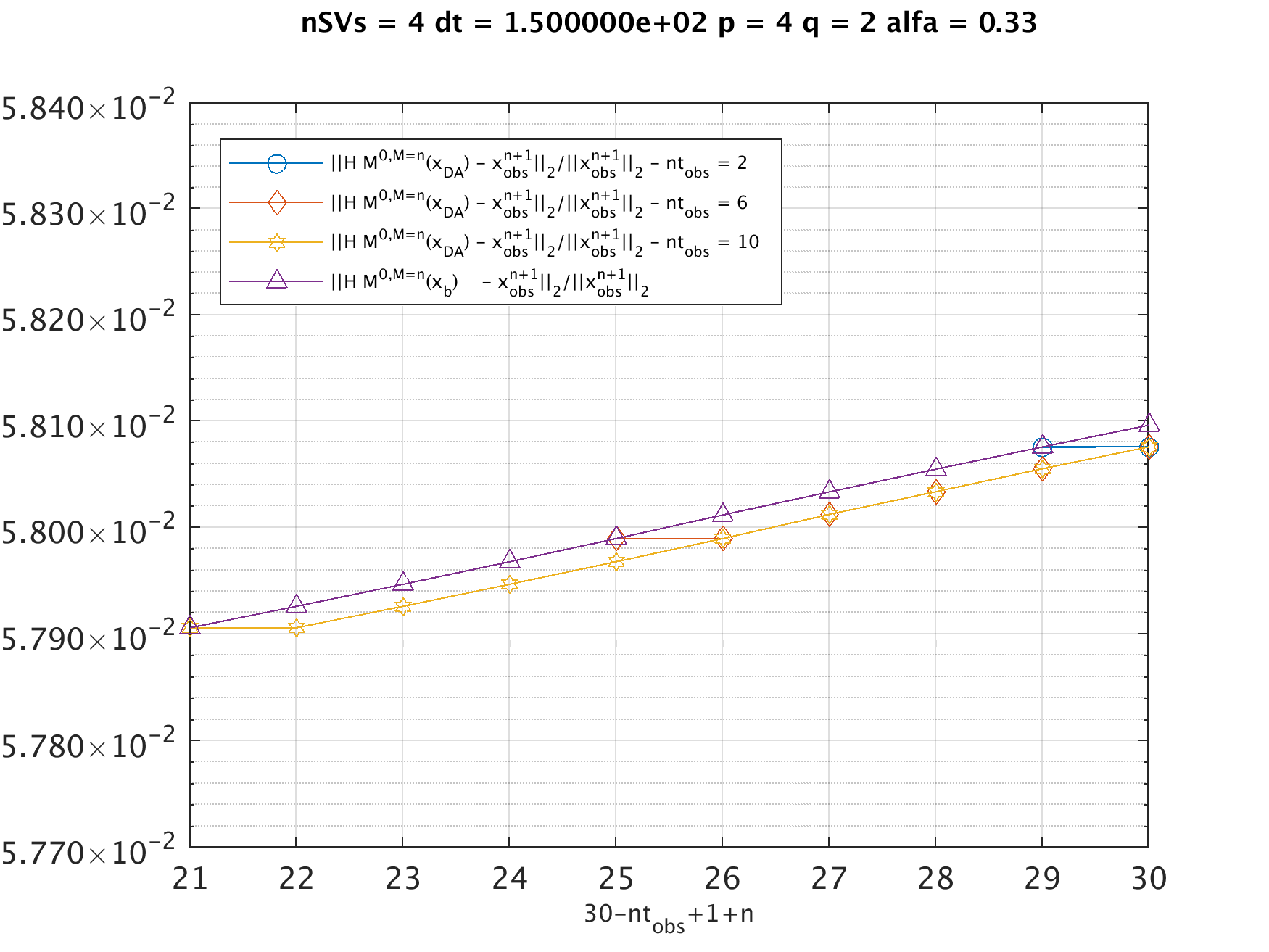}} \\
(a) & (c) \\[4ex]
{\includegraphics[width=8cm]{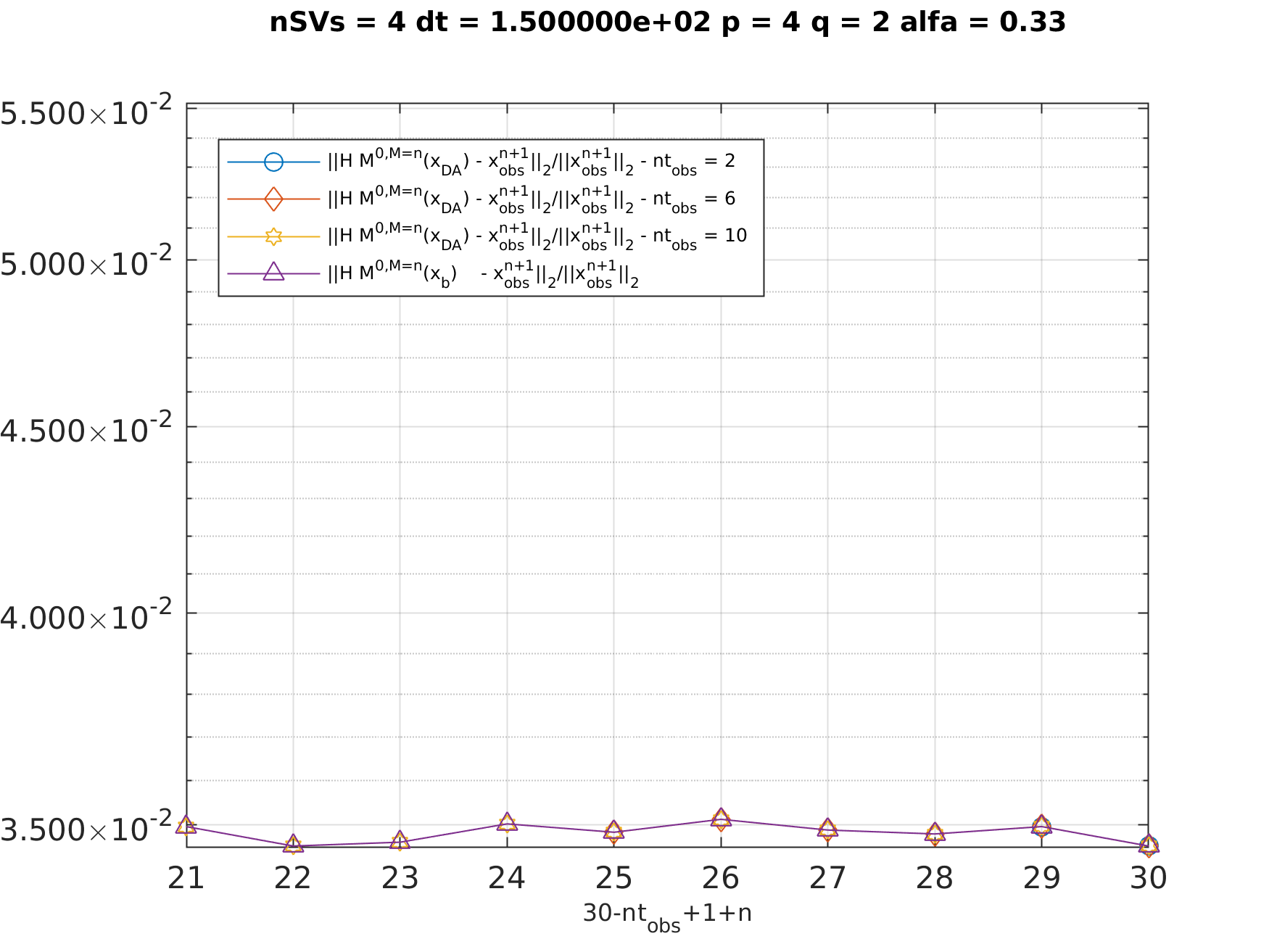}} &
{\includegraphics[width=8cm]{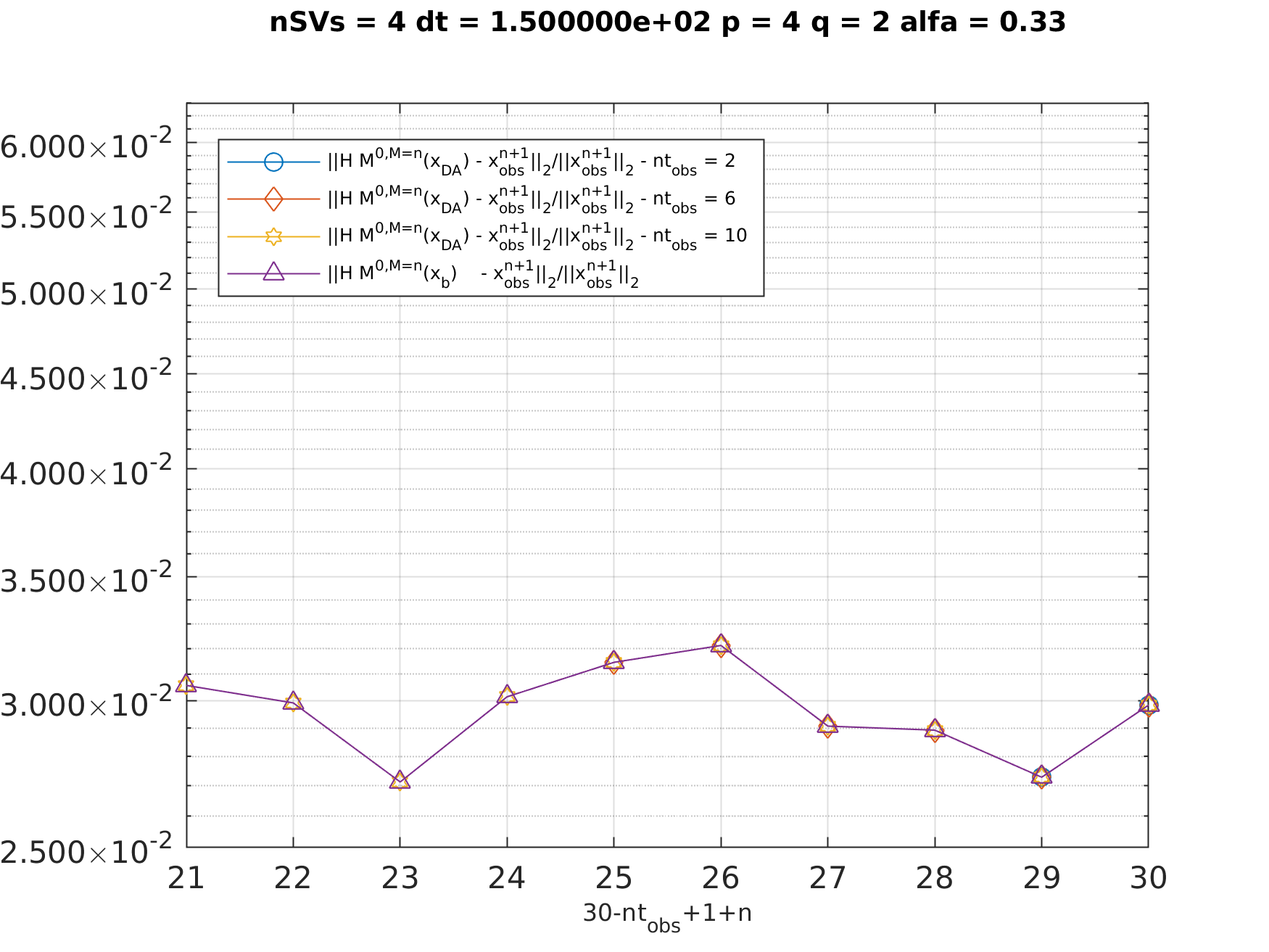}} \\
(b) & (d) \\
\end{tabular}
\end{center}
\caption{\mbox{$\| H \mathbf{M}_{\Delta t}^{0,M_{steps}=n}\left(x^{\Delta t}_{DA}\right) -_{\Delta t}x^{n+1}_{obs}\|_{2}/\left\|_{\Delta t}x^{n+1}_{obs}\right\|_{2}$}
and      \mbox{$\| H \mathbf{M}_{\Delta t}^{0,M_{steps}=n}\left(x^{\Delta t}_{b}\right)  -_{\Delta t}x^{n+1}_{obs}\|_{2}/\left\|_{\Delta t}x^{n+1}_{obs}\right\|_{2}$}
as function of $n=0,\ldots,nt_{obs}-1$ ($\Delta t=150.0$, $nt_{obs}=2,6,10$, $nSVs=4$, 
(a)-(b)-labeled plots are related with Problem 1-4 respectively)} 
\label{ConfrontoErrADA.nSVs-04.dt-150.000}
\end{figure}

\begin{figure}
\begin{center}
\begin{tabular}{cc}
{\includegraphics[width=8cm]{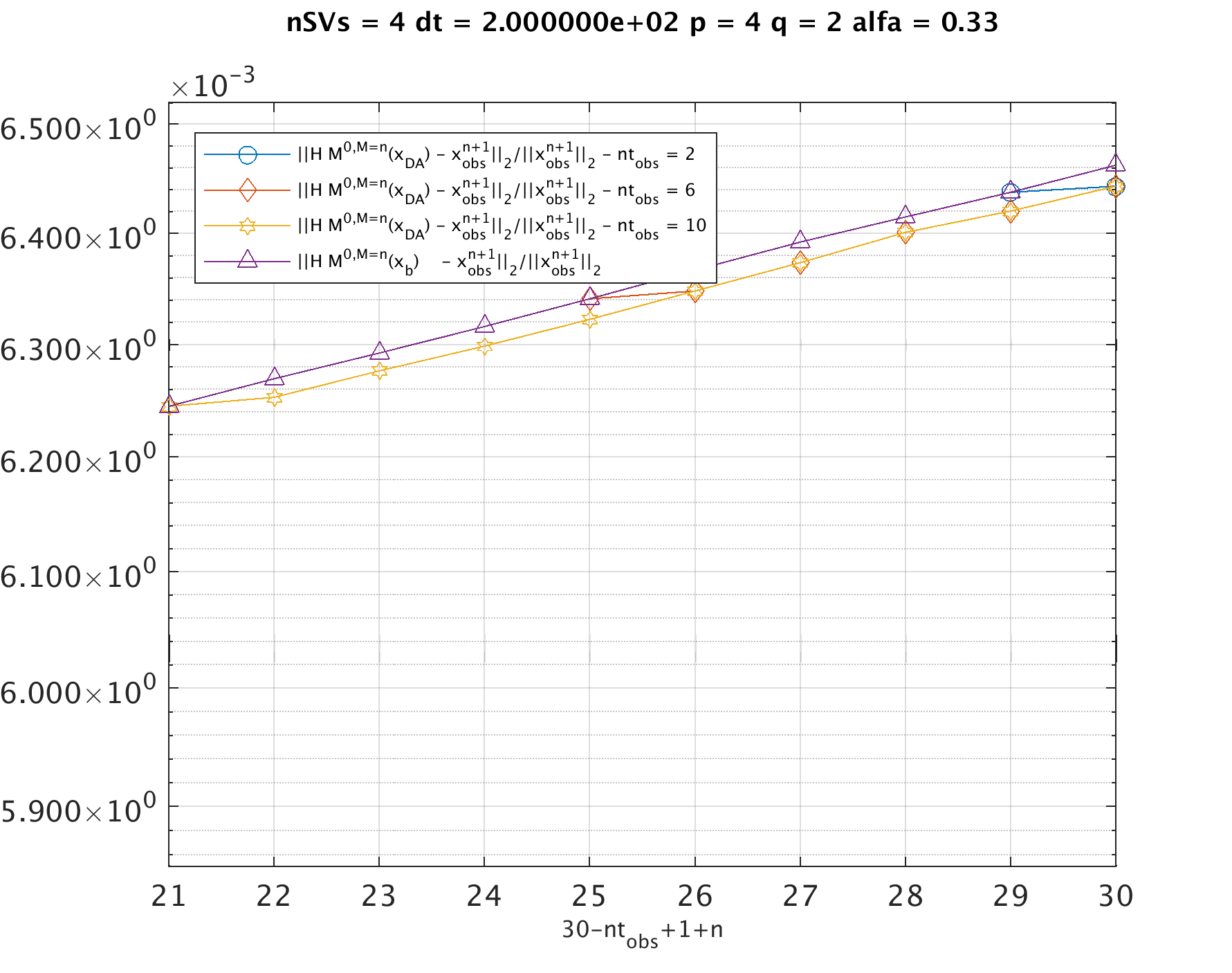}} &
{\includegraphics[width=8cm]{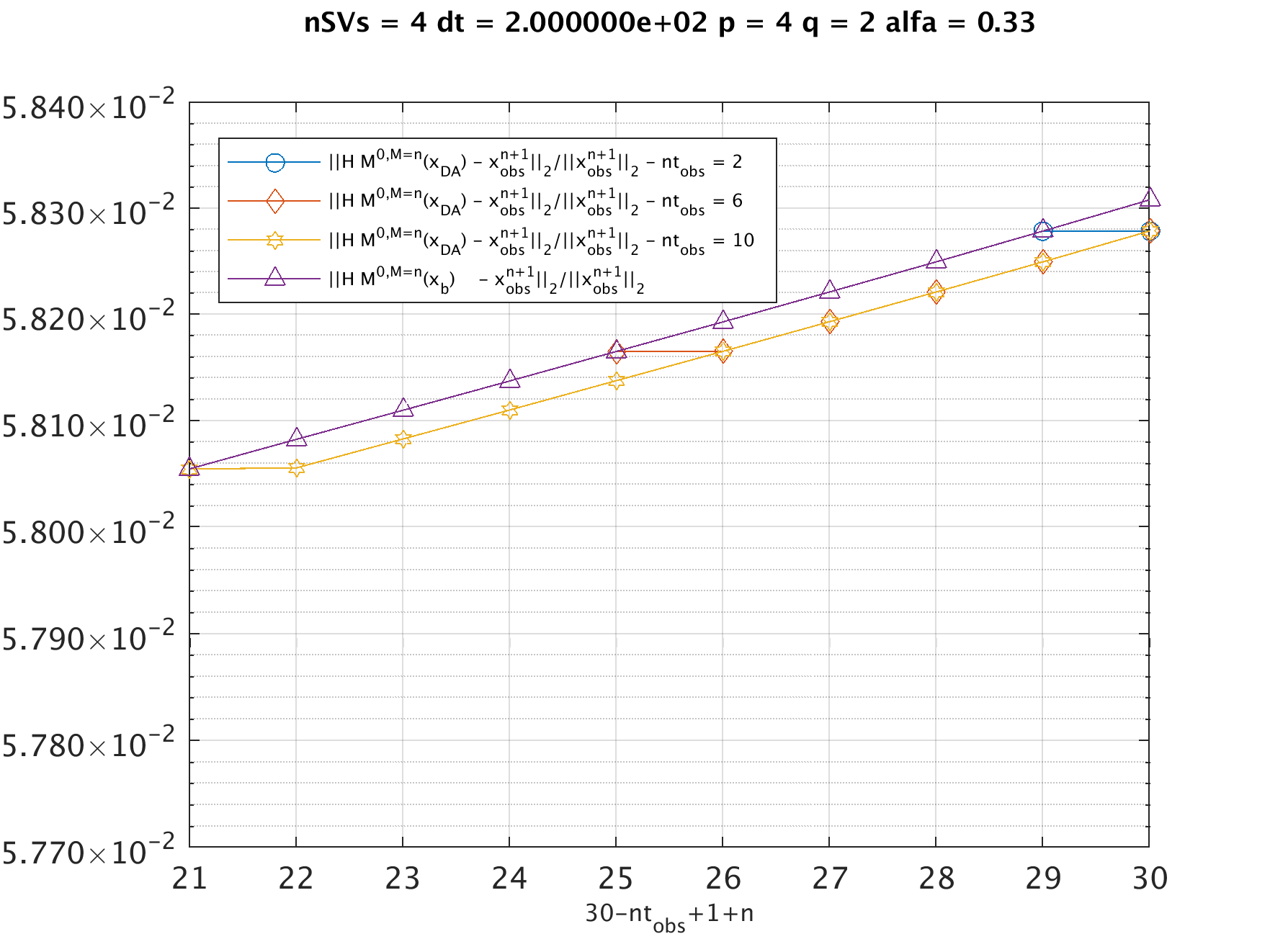}} \\
(a) & (c) \\[4ex]
{\includegraphics[width=8cm]{{Problem2-ECEx-ConfrontoErrADA-Norm2.nSVs-04.dt-150.000.Step030.p-4.q-2.alfa-0.3300}.png}} &
{\includegraphics[width=8cm]{{Problem4-ECEx-ConfrontoErrADA-Norm2.nSVs-04.dt-150.000.Step030.p-4.q-2.alfa-0.3300}.png}} \\
(b) & (d) \\
\end{tabular}
\end{center}
\caption{\mbox{$\| H \mathbf{M}_{\Delta t}^{0,M_{steps}=n}\left(x^{\Delta t}_{DA}\right) -_{\Delta t}x^{n+1}_{obs}\|_{2}/\left\|_{\Delta t}x^{n+1}_{obs}\right\|_{2}$}
and      \mbox{$\| H \mathbf{M}_{\Delta t}^{0,M_{steps}=n}\left(x^{\Delta t}_{b}\right)  -_{\Delta t}x^{n+1}_{obs}\|_{2}/\left\|_{\Delta t}x^{n+1}_{obs}\right\|_{2}$}
as function of $n=0,\ldots,nt_{obs}-1$ ($\Delta t=200.0$, $nt_{obs}=2,6,10$, $nSVs=4$, 
(a)-(b)-labeled plots are related with Problem 1-4 respectively)} 
\label{ConfrontoErrADA.nSVs-04.dt-200.000}
\end{figure}

\begin{figure}
\begin{center}
\begin{tabular}{cc}
{\includegraphics[scale=1.00]{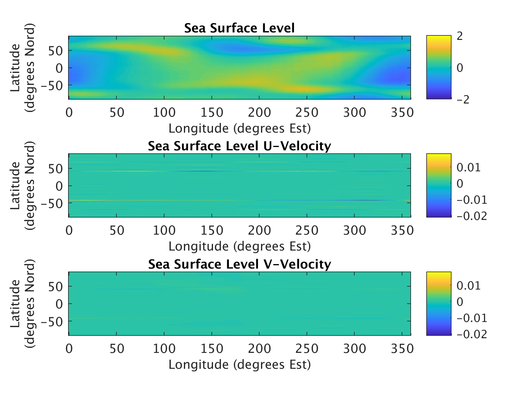}} &
{\includegraphics[scale=1.00]{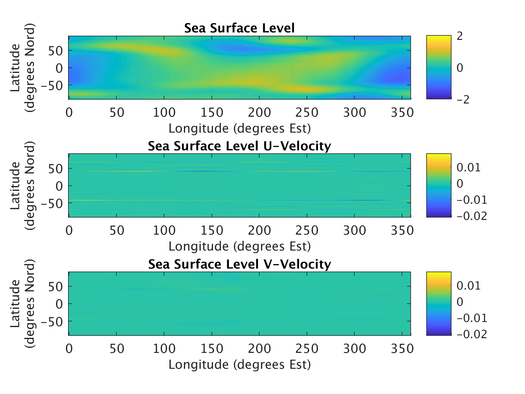}} \\
{(a)} & {(c)} \\[2ex]
{\includegraphics[scale=1.00]{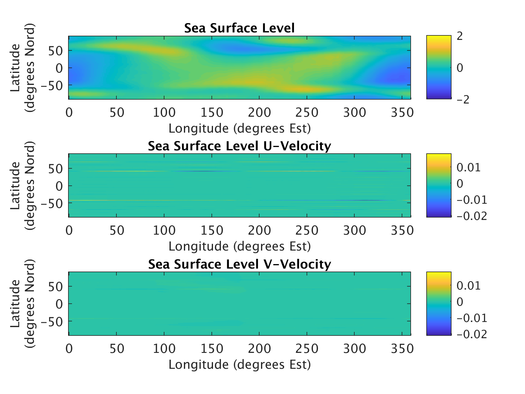}} &
{\includegraphics[scale=1.00]{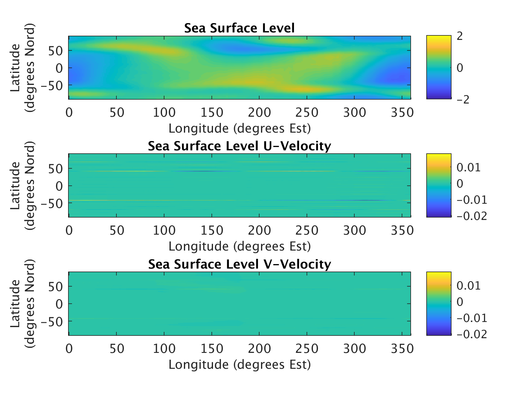}} \\
{(b)} & {(d)} \\
\end{tabular}
\end{center}
\caption{$\mathbf{M}_{\Delta t}^{0,M_{steps}=nt_{obs}-1}\left(x^{\Delta t}_{DA}\right)$ ($\Delta t=50.0$, $nt_{obs}=10$, 
(a)-(b)-labeled plots are related with Problem 1-4 respectively)} 
\label{ModelSolutionAfterDA.nSVs-04.nt_obs-10.dt-50.000}
\end{figure}

\begin{figure}
\begin{center}
\begin{tabular}{cc}
{\includegraphics[scale=1.00]{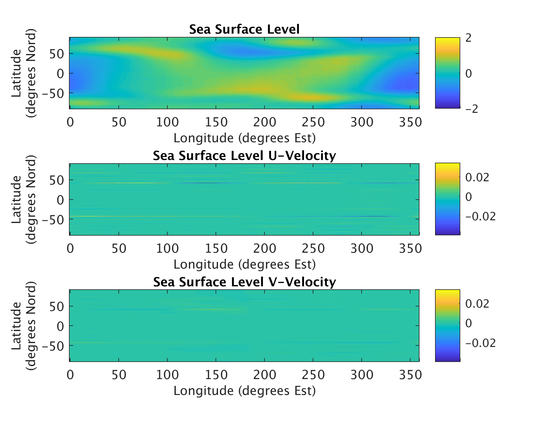}} &
{\includegraphics[scale=1.00]{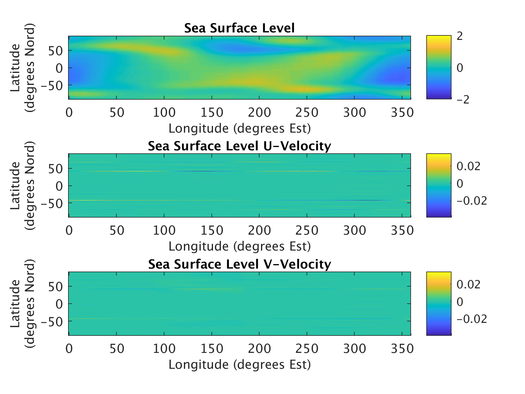}} \\
{(a)} & {(c)} \\[2ex]
{\includegraphics[scale=1.00]{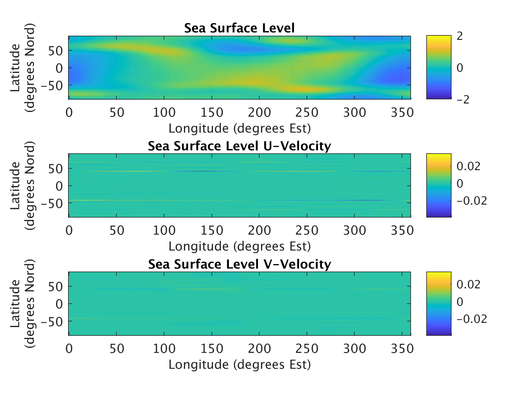}} &
{\includegraphics[scale=1.00]{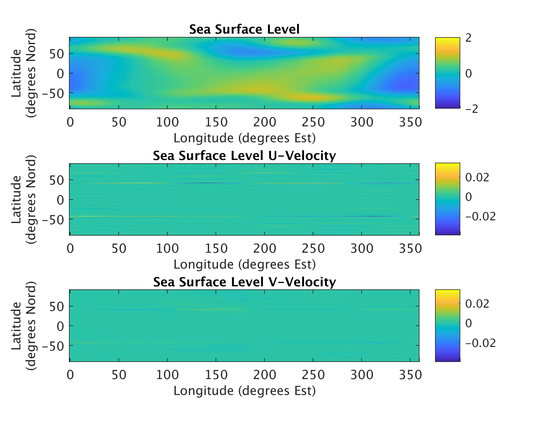}} \\
{(b)} & {(d)} \\
\end{tabular}
\end{center}
\caption{$\mathbf{M}_{\Delta t}^{0,M_{steps}=nt_{obs}-1}\left(x^{\Delta t}_{DA}\right)$ ($\Delta t=100.0$, $nt_{obs}=10$, 
(a)-(b)-labeled plots are related with Problem 1-4 respectively)} 
\label{ModelSolutionAfterDA.nSVs-04.nt_obs-10.dt-100.000}
\end{figure}

\begin{figure}
\begin{center}
\begin{tabular}{cc}
{\includegraphics[scale=1.00]{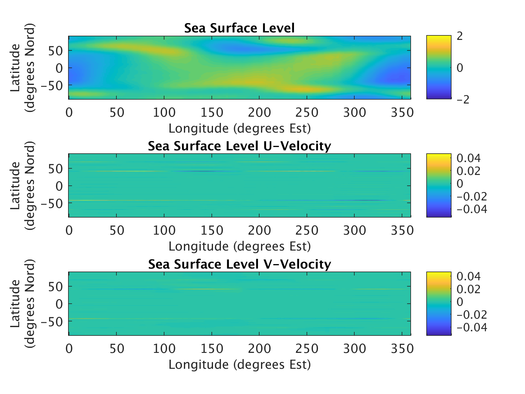}} &
{\includegraphics[scale=1.00]{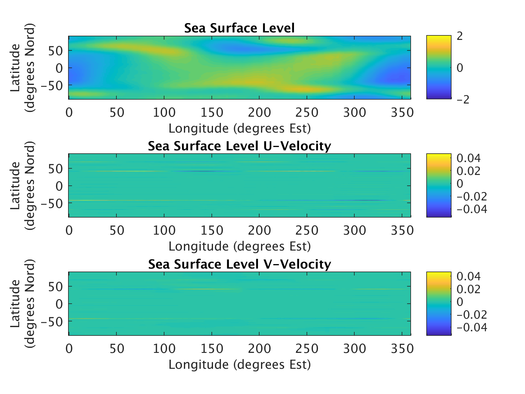}} \\
{(a)} & {(c)} \\[2ex]
{\includegraphics[scale=1.00]{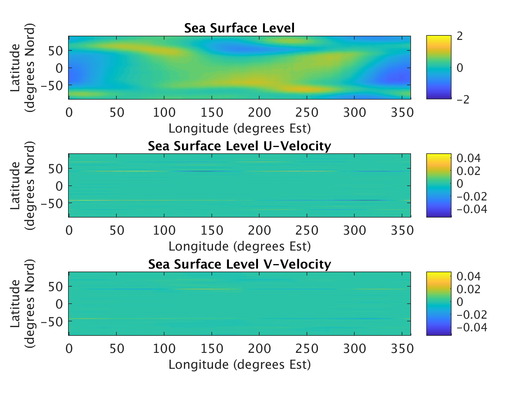}} &
{\includegraphics[scale=1.00]{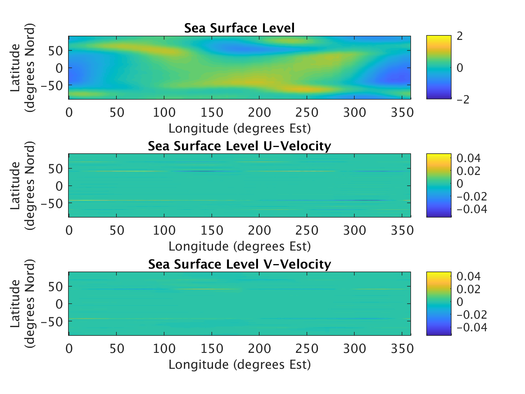}} \\
{(b)} & {(d)} \\
\end{tabular}
\end{center}
\caption{$\mathbf{M}_{\Delta t}^{0,M_{steps}=nt_{obs}-1}\left(x^{\Delta t}_{DA}\right)$ ($\Delta t=150.0$, $nt_{obs}=10$, 
(a)-(b)-labeled plots are related with Problem 1-4 respectively)} 
\label{ModelSolutionAfterDA.nSVs-04.nt_obs-10.dt-150.000}
\end{figure}

\begin{figure}
\begin{center}
\begin{tabular}{cc}
{\includegraphics[scale=1.00]{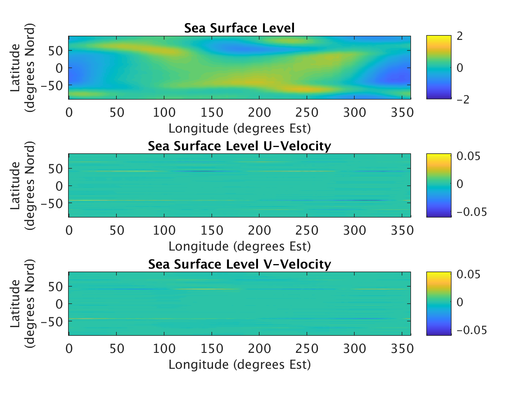}} &
{\includegraphics[scale=1.00]{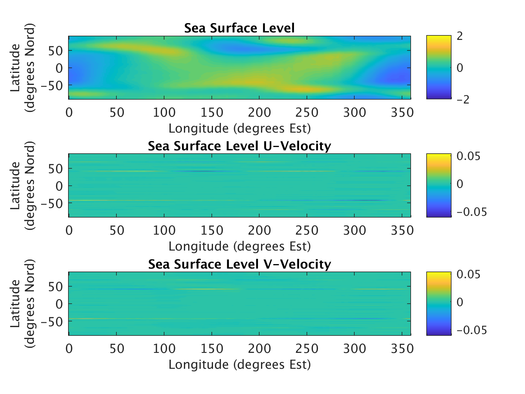}} \\
{(a)} & {(c)} \\[2ex]
{\includegraphics[scale=1.00]{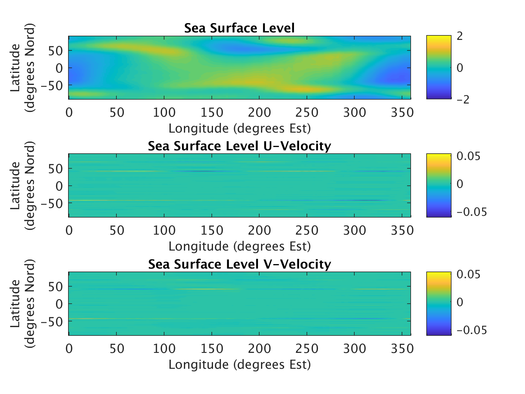}} &
{\includegraphics[scale=1.00]{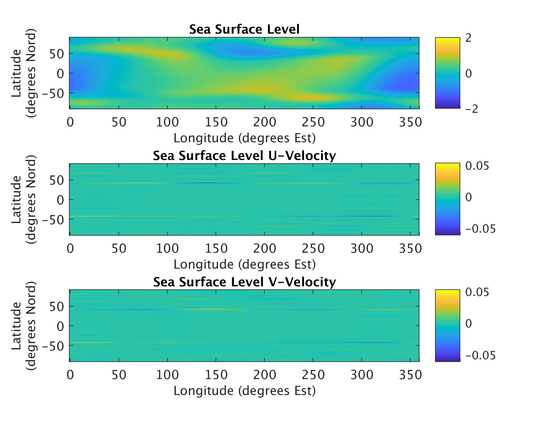}} \\
{(b)} & {(d)} \\
\end{tabular}
\end{center}
\caption{$\mathbf{M}_{\Delta t}^{0,M_{steps}=nt_{obs}-1}\left(x^{\Delta t}_{DA}\right)$ ($\Delta t=200.0$, $nt_{obs}=10$, 
(a)-(b)-labeled plots are related with Problem 1-4 respectively)} 
\label{ModelSolutionAfterDA.nSVs-04.nt_obs-10.dt-200.000}
\end{figure}

In figures \ref{ModelSolutionAfterDA.nSVs-04.nt_obs-10.dt-50.000}, \ref{ModelSolutionAfterDA.nSVs-04.nt_obs-10.dt-100.000}, 
\ref{ModelSolutionAfterDA.nSVs-04.nt_obs-10.dt-150.000} and \ref{ModelSolutionAfterDA.nSVs-04.nt_obs-10.dt-200.000}, 
$\mathbf{M}_{\Delta t}^{0,M_{steps}=nt_{obs}-1}\left(x^{\Delta t}_{DA}\right)$ are showed where
$\Delta t=50.0, 100.0, 150.0, 200.0$ 
(values of $nt_{obs}$ and $nSVs$ are fixed and their values are $nt_{obs}=10$ e $nSVs=4$, (a)-(b)-labeled plots are related with Problem 1-4 respectively).

\section{Conclusions}
In this work are presented and discussed some results of tests performed to validate a software module which implements a DA process.

Such module depends from some parameters such as the number $nt_{obs}$ of observations vectors and the number  $nSVs$ of singular values 
considered for the Truncated SVD of matrix $B$.

The parameter $nSVs$ influences the behavior of DA software module: in particular,
the smaller values for $nSVs$ is (i.e., where $nSVs=4$), more the DA software
module is able to effectively compute the solution of the DA problem.
Also the value of $\Delta t$, the discretization step in time domain used by numerical model, seems to have some influences on the behavior of DA software module:
when the bigger values for $\Delta t$ are used (i.e., when $\Delta t=200$), the DA software module often fails to compute the solution of the DA problem.

About the $nt_{obs}$ parameter, the use of larger values doesn't have significant effects both on the behavior of DA software module and on
use of DA solution in model's application. Furthermore, the use of larger values for $nt_{obs}$ has a bigger computational cost. 
All that said, the use of small values for $nt_{obs}$ 
(i.e., where $nt_{obs}=1$) seems to be more desirable.

%Perhaps the two problems considered for the tests are not able to benefit from the use of a 4DVAR approach (i.e., $nt_{obs}>1$): the first are too close 
%to an ideal problem since take into account full and accurate enough observations vectors, the second instead can be considered unrealistic enough since it use observations very inaccurate.

\end{document}